\documentclass[a4paper,11pt]{article}
\pdfoutput=1 

\usepackage{jheppub} 

\usepackage[T1]{fontenc} 

\usepackage{slashed}
\usepackage{relsize}
\usepackage{tikz}
\usepackage{comment}

\allowdisplaybreaks

\title{\boldmath Towards power corrections in the factorization of baryon quasi-distribution amplitudes in LaMET}

\author[a,1]{Yu-Ji Shi\note{Corresponding author.},}
\author[b,c]{Jun Zeng}

\affiliation[a]{School of Physics, East China University of Science and Technology, Shanghai 200237, China}
\affiliation[b]{College of Physics and Electronic Engineering, Hainan Normal University, Haikou 571158, Hainan, China}
\affiliation[c]{State Key Laboratory of Dark Matter Physics, Key Laboratory for Particle Astrophysics and Cosmology (MOE), Shanghai Key Laboratory for Particle Physics and Cosmology, School of Physics and Astronomy, Shanghai Jiao Tong University, Shanghai 200240, China}

\emailAdd{shiyuji@ecust.edu.cn}
\emailAdd{zengj@hainnu.edu.cn}

\abstract{We study finite-momentum corrections to the quasi-distribution amplitudes of the $\Lambda$ baryon. For the vector and axial amplitudes, the geometric twist-3 projection removes the metric and gamma traces of the local three-quark operators. Their spatial and temporal moments acquire distinct target-mass factors. For the transverse tensor amplitude, the spin projection used on the lattice gives the arithmetic mean of these two factors. Summing the moments gives the relation between each twist-3 quasi-distribution amplitude and its light-cone distribution amplitude to all orders in $m_\Lambda^2/P_z^2$ at tree level. We construct the higher-twist operators in the axial channel and show that they can contribute to fixed moments at order $1/P_z^2$. We also propose a feasible approach to computing their matrix elements on the lattice. We combine the tree-level mass subtraction with leading-power one-loop matching and identify the additional Wilson coefficients required at order $\alpha_s m_\Lambda^2/P_z^2$. The first-order target-mass subtraction is applied to lattice data for the axial $\Lambda$ quasi-distribution amplitude in the $\gamma^t$ channel at several baryon momenta.}

\begin{document} 
\maketitle
\flushbottom

\section{Introduction}

Understanding the internal structure of hadrons in terms of their quark and gluon constituents is a central goal of quantum chromodynamics (QCD). In hard exclusive processes, this structure is encoded in light-cone distribution amplitudes (LCDAs), which describe the longitudinal momentum distribution of partons in the hadron and enter factorization theorems at large momentum transfer~\cite{Lepage:1980fj,Efremov:1978rn}. Baryon LCDAs enter electromagnetic transitions and weak decays of heavy baryons. The first observation of CP violation in a baryon decay, $\Lambda_b^0\to pK^-\pi^+\pi^-$, by LHCb~\cite{LHCb:2025ray} gives further motivation for precise predictions of heavy-baryon decay amplitudes and CP asymmetries. In perturbative QCD these predictions depend on baryon LCDAs~\cite{Shih:1998pb,Keum:2000wi,Lu:2000em,Keum:2000ph,Guo:2005qa,He:2006ud,Lu:2009cm,Han:2022srw,Li:2025rsm,Yang:2025yaw,Han:2024kgz,Han:2025tvc}, whose scale evolution has been calculated through two loops~\cite{Huang:2025bbk}.

Despite their phenomenological importance, baryon LCDAs remain far less understood than their meson counterparts. At leading twist, their definition involves two momentum fractions, multiple Dirac structures, and nontrivial spin-flavor symmetries, which complicate their determination from first principles. Early determinations of baryon LCDAs were primarily based on QCD sum rules or model assumptions~\cite{Chernyak:1987nu,Ball:2008fw}. While these studies provided valuable phenomenological guidance, the resulting picture remained inherently model-dependent. Large-momentum effective theory (LaMET) provides a systematic framework for overcoming these difficulties~\cite{Ji:2013dva,Ji:2014gla}. In this approach, one computes equal-time spatial correlation functions in a boosted hadron state, extracts quasi-distribution amplitudes (quasi-DAs), and then matches them to the physical LCDAs via perturbative factorization. Recent theoretical developments have extended LaMET to baryon quasi-DAs~\cite{Deng:2023csv,Han:2023xbl,Han:2024ucv,Shi:2026mjb,Shi:2026fel}, and lattice studies~\cite{LatticeParton:2024vck,LatticePartonCollaborationLPC:2025vhd} have demonstrated the feasibility of this approach for determining baryon LCDAs.

The LPC Collaboration has obtained the two-dimensional momentum-fraction dependence of the $\Lambda$-baryon LCDAs from lattice quasi-DAs using leading-power matching~\cite{LPC:2026mvw,LPC:2026lcj}. At finite baryon momentum $P_z$, the quasi-DA also receives corrections from the target mass, of order $m_\Lambda^2/P_z^2$, and from higher-twist matrix elements, of order $\Lambda_{\rm QCD}^2/P_z^2$. The lattice analysis uses several momenta and extrapolates in $1/P_z^2$. At the momenta accessible on the lattice, both effects can enter the same fitted slope. For the twist-3 contribution, the light-cone moments fix the reduced matrix elements. This allows us to calculate the dependence on $m_\Lambda$ before fitting the remaining power corrections.

We consider the vector ($V$), axial ($A$) and tensor ($T$) types of baryon DA. The operators for vector and axial DAs each carry one Lorentz index. We use the axial operator to derive the target-mass corrections for the spatial ($\gamma^z$) and temporal ($\gamma^t$) components, and the derivation for the vector operator  is the same.  The symmetric tensor-spinor still contains twist four in its gamma trace after its metric traces have been removed. Subtracting the gamma trace isolates geometric twist-3. For the tensor DA, the lattice operator sums two transverse components of an antisymmetric diquark tensor. We perform this sum after the twist projection and derive the corresponding mass correction to the two-variable distribution.

For the higher-twist terms we give the construction in the axial operator and determine the leading power of their fixed moments. We express the operators in terms of a gauge-invariant nonlocal three-quark operator and derivatives with respect to the two quark separation vectors. Its two straight Wilson lines connect the displaced $u$ and $d$ fields to the $s$ field at the origin. Lattice calculations of leading-twist baryon quasi-DAs already use this type of three-quark operator~\cite{LPC:2026mvw}. We propose a finite-difference prescription for the additional derivatives. Beyond tree level, we use the short-distance expansion to combine the mass subtraction with leading-power one-loop matching and identify the Wilson coefficients needed at order $\alpha_s m_\Lambda^2/P_z^2$.

This paper is organized  as follows.
Sections~\ref{sec:twistprojection} and \ref{sec:targetmassCorr} introduce the three LCDAs and derive the vector and axial mass corrections using the axial operator. Sections~\ref{sec:Highertwist} and \ref{sec:matchingandsubtraction} treat the higher-twist terms and perturbative matching in the same convention. Section~\ref{sec:tensorTMC} derives the tree-level mass correction for the transverse tensor quasi-DA. Section~\ref{sec:numerical} applies the axial temporal subtraction to lattice data, and Sec.~\ref{sec:conclusion} summarizes our results.

\section{Baryon operators and twist projection}
\label{sec:twistprojection}

For a spin-$1/2$ baryon of momentum $P^\mu=(E,0,0,P_z)$ and  $P_z>0$, we introduce the light-cone vectors
\begin{equation}
 n^\mu=\frac{(1,0,0,-1)}{\sqrt{2}},\qquad
 \bar n^\mu=\frac{(1,0,0,1)}{\sqrt{2}} ,
 \label{eq:bdaDirections}
\end{equation}
which obey $n^2=\bar n^2=0$, $n\cdot\bar n=1$ and $p\equiv P\cdot n=(E+P_z)/\sqrt{2}$.
Let $q^{[v]}(t)=W(0,tv)q(tv)$, where the straight Wilson line $W(0,tv)$ transports the quark to the origin.  The three leading-twist LCDAs are defined by~\cite{LPC:2026mvw}
\begin{align}
 \Pi_+\langle0|\epsilon^{abc}(q_{1a}^{T} C\slashed n\,q_{2b})q_{3c}|B(P,\sigma)\rangle
   &=-f_B^{\rm LC}p\,\Phi_B^V\,\gamma_5u_{B,+},\nonumber\\
 \Pi_+\langle0|\epsilon^{abc}(q_{1a}^{T} C\gamma_5\slashed n\,q_{2b})q_{3c}|B(P,\sigma)\rangle
   &= f_B^{\rm LC}p\,\Phi_B^A\,u_{B,+},\nonumber\\
 \Pi_+\langle0|\epsilon^{abc}(q_{1a}^{T}i C\sigma^{\mu\nu}n_\nu q_{2b})\gamma_\mu q_{3c}|B(P,\sigma)\rangle
   &=2f_B^T p\,\Phi_B^T\,\gamma_5u_{B,+}.\label{eq:bdaThreeLCDAs}
\end{align}
Here $q_1,q_2,q_3$ stand for $q_1^{[n]}(t_1),q_2^{[n]}(t_2),q_3(0)$, respectively. The color indices are $a,b,c$. The good component projector is defined as $\Pi_+=\slashed{\bar n}\slashed n/2$ and $u_{B,+}=\Pi_+u_B(P,\sigma)$.  The matrix elements and couplings are defined at the scale $\mu$.  For $X=V,A,T$, the coordinate-space amplitude $\Phi_B^X$ in Eq.~\eqref{eq:bdaThreeLCDAs} is the Fourier transform of the momentum-space LCDA $\phi_B^X$,
 \begin{align}
 \Phi_B^X(pt_1,pt_2;\mu)
 &=\int_{\mathcal S}dx_1\,dx_2\, e^{-ip(x_1t_1+x_2t_2)}\phi_B^X(x_1,x_2;\mu),\label{eq:bdaThreeLCFourier}\\
 \mathcal S&=\{(x_1,x_2):x_1,x_2\ge0,\ x_1+x_2\le1\}.\nonumber
 \end{align}
 
We now take $B=\Lambda$, $(q_1,q_2,q_3)=(u,d,s)$ and suppress the baryon label on the amplitudes. In the isospin limit $\phi^A$ is symmetric under $x_1\leftrightarrow x_2$, while $\phi^V$ and $\phi^T$ are antisymmetric. Thus the local axial matrix element fixes $f_\Lambda^{\rm LC}$ with $\int_{\mathcal S}\phi^A=1$, and the vector and tensor zeroth moments both vanish. 
In Secs.~\ref{sec:twistprojection}--\ref{sec:matchingandsubtraction} we work with the axial operator and write $\phi=\phi^A$. For a diquark index contracted with $w$, the same derivation applies to $V$. Replace the diquark matrix $C\gamma_5\slashed w$ by $C\slashed w$ and multiply the resulting operator by $-\gamma_5$ on its open index. The first line of Eq.~\eqref{eq:bdaThreeLCDAs} then has the same external spinor and sign as the second. Consequently the twist decomposition, power counting and tree-level mass factors are identical for $V$ and $A$. The known leading-power one-loop matching kernel is also common to $V$ and $A$ in the same scheme~\cite{Han:2024ucv,LPC:2026mvw}. The antisymmetric diquark tensor in the third line of Eq.~\eqref{eq:bdaThreeLCDAs} requires a separate treatment, which will be given in Sec.~\ref{sec:tensorTMC}.

\subsection{Light-cone and equal-time operators}
\label{sec:DAdefinitions}

For the $\Lambda$ baryon $E^2=P_z^2+m_\Lambda^2$ and $P^\mu=p\bar n^\mu+m_\Lambda^2 n^\mu/(2p)$. To treat lightlike and equal-time separations in the axial channel, we define
\begin{equation}
 \mathcal O(t_1,t_2;w,v)=
 \epsilon^{ijk}\big[u^{[v]}_i(t_1)\big]^T C\gamma_5\slashed w\,
 d^{[v]}_j(t_2)\,s_k(0).
 \label{eq:bdaThreeQuark}
\end{equation}
Here $i,j,k$ are color indices.
The vector $v$ fixes the two separations, while $w$ selects the Dirac matrix in the diquark.

On the light cone, the leading-twist DA is defined by the three good quark components~\cite{Braun:1999te,Braun:2008ia}.
Write $q_+=\Pi_+q$, $\Pi_-=\slashed n\slashed{\bar n}/2$ and $u_\Lambda=u_\Lambda(P,\sigma)$.
Multiplying $(\slashed P-m_\Lambda)u_\Lambda=0$ by $\slashed n$ gives
\begin{equation}
 2p\,\Pi_-u_\Lambda=m_\Lambda\slashed n u_\Lambda,\qquad
 u_{\Lambda,+}=u_\Lambda-\frac{m_\Lambda}{2p}\slashed n u_\Lambda .
 \label{eq:bdaSpinor}
\end{equation}
For the diquark, $\Pi_+^T C=C\Pi_-$ and $\Pi_-\gamma_5\slashed n\Pi_+=\gamma_5\slashed n$ give
$u_+^T C\gamma_5\slashed n d_+=u^T C\gamma_5\slashed n d$.
The operator obtained by replacing all three quarks in $\mathcal O(t_1,t_2;n,n)$ by good components is therefore
$\mathcal O_{\mathrm{LC},+}(t_1,t_2)=\Pi_+\mathcal O(t_1,t_2;n,n)$, where the projector acts on the open strange-quark spinor index.
The axial definition in Eqs.~\eqref{eq:bdaThreeLCDAs} and \eqref{eq:bdaThreeLCFourier} then reads
\begin{align}
 \langle0|\mathcal O_{\mathrm{LC},+}(t_1,t_2)|\Lambda(P,\sigma)\rangle
 &= f_\Lambda^{\rm LC}\,p\,u_{\Lambda,+}
 \int_{\cal S} dx_1\,dx_2\,
 e^{-ip(x_1t_1+x_2t_2)}\phi(x_1,x_2,\mu),
 \label{eq:bdaLC}\\
 \int_{\cal S} dx_1\,dx_2\,\phi(x_1,x_2,\mu)&=1.\nonumber
 \label{eq:bdaSupport}
\end{align}
The third quark carries momentum fraction $1-x_1-x_2$.
The local limit of Eq.~\eqref{eq:bdaLC} fixes $f_\Lambda^{\rm LC}$ at the same renormalization scale $\mu$.

For the quasi-DAs, the two quark separations are $\ell_1\lambda$ and $\ell_2\lambda$, where
$\lambda^\mu=(n^\mu-\bar n^\mu)/\sqrt{2}=(0,0,0,-1)$. This choice gives
$\lambda^2=-1$, $P\cdot\lambda=P_z$ and $n\cdot\lambda=-1/\sqrt{2}$.
For the temporal Dirac matrix we use the unit vector $v_t^\mu=(1,0,0,0)$. The two channels are defined by
\begin{equation}
 \widetilde{\mathcal O}_\Gamma(\ell_1,\ell_2)
 =\mathcal O(\ell_1,\ell_2;w_\Gamma,\lambda),\qquad
 w_{\gamma^z}=\lambda,\quad w_{\gamma^t}=v_t,\qquad
 \gamma^z=\gamma^3,\quad \gamma^t=\gamma^0 .
 \label{eq:bdaEqualTime}
\end{equation}
Both operators in Eq.~\eqref{eq:bdaEqualTime} contain full quark fields. In this collinear geometry the open spinor
matrix element can be expanded in the two structures $u_{\Lambda,+}$ and $\slashed n u_\Lambda$.
Using the same reference factor $f_\Lambda^{\rm LC}P_z$ for both channels, we write
\begin{align}
 &\langle0|\widetilde{\mathcal O}_\Gamma(\ell_1,\ell_2)|\Lambda(P,\sigma)\rangle
 = f_\Lambda^{\rm LC}P_z\int_{\mathbb R^2}dx_1\,dx_2\,
       e^{-iP_z(x_1\ell_1+x_2\ell_2)}
 \nonumber\\[-1mm]
 &\hspace{18mm}\times
 \big[u_{\Lambda,+}\,\widetilde\phi_\Gamma^{\rm full}(x_1,x_2;P_z)
       +\slashed n u_\Lambda\,\widetilde\psi_\Gamma^{\rm full}(x_1,x_2;P_z)\big] .
 \label{eq:bdaQuasiSpinor}
\end{align}
The superscript $\mathrm{full}$ indicates that all twists are included. We suppress the renormalization scales of the quasi-DAs.
To isolate $\widetilde\phi_\Gamma^{\rm full}$, we contract with $\bar u_\Lambda(P, \sigma)\slashed n$ and sum over $\sigma$.
The second term vanishes by $\slashed n^2=0$. For the first, $\slashed n\Pi_+=\slashed n$ and the spin sum give
$\mathrm{tr}[\slashed n\Pi_+(\slashed P+m_\Lambda)]=4p$. Hence
\begin{equation}
 \frac{\sum_\sigma\bar u_\Lambda\slashed n
  \langle0|\widetilde{\mathcal O}_\Gamma(\ell_1,\ell_2)|\Lambda(P,\sigma)\rangle}
 {4f_\Lambda^{\rm LC}pP_z}
 =\int_{\mathbb R^2}dx_1\,dx_2\,e^{-iP_z(x_1\ell_1+x_2\ell_2)}
    \widetilde\phi_\Gamma^{\rm full}(x_1,x_2;P_z).
 \label{eq:bdaQuasiScalar}
\end{equation}
At tree level the zeroth moment in Eq.~\eqref{eq:bdaQuasiScalar} is the local matrix element divided by $4f_\Lambda^{\rm LC}pP_z$. If the same channel's local matrix element is used as the denominator, the zeroth moment is unity. Beyond tree level the local limit requires a consistent ultraviolet prescription.

\subsection{The twist-3 projector}
\label{sec:localprojector}

Let the nonnegative integers $n_1$ and $n_2$ count the covariant derivatives on the $u$ and $d$ fields, respectively, and set
$N=n_1+n_2$. In the light-cone matrix element these derivatives measure the moment
\begin{equation}
 \mathcal M_{n_1,n_2}(\mu)
 =\int_{\cal S}dx_1\,dx_2\,x_1^{n_1}x_2^{n_2}\phi(x_1,x_2,\mu),
 \qquad \mathcal M_{0,0}=1 .
 \label{eq:bdaMoment}
\end{equation}
To retain the Lorentz indices until the twist projection has been performed, we introduce a dimensionless auxiliary four-vector $z$
and define the local operator
\begin{equation}
 \mathcal O_{n_1,n_2}(z)
 =\epsilon^{ijk}\big[(iz\cdot D)^{n_1}u_i\big]^T C\gamma_5\slashed z
       \big[(iz\cdot D)^{n_2}d_j\big]s_k ,
 \qquad D_\mu=\partial_\mu-ig_s A_\mu .
 \label{eq:bdaLocal}
\end{equation}
The three unprojected quark fields in Eq.~\eqref{eq:bdaLocal} are evaluated at the origin. The operator is homogeneous of degree
$r=N+1$ in $z$. One power comes from the Dirac matrix and the other $N$ powers from the derivatives.
Contraction with the same $z$ selects the fully symmetric part of these $r$ Lorentz indices, including the Dirac matrix index. The two derivative counts remain fixed under this symmetrization.

For a straight line in the direction $v$, the local expansion of $q^{[v]}(t)$ has coefficients
$(v\cdot D)^kq(0)/k!$, as follows from Eq.~\eqref{eq:expoDeriv} in Appendix~\ref{app:bdaTransport}.
At $v=n$ the good-component projector commutes with $n\cdot D$. Applying $(i\partial_{t_1})^{n_1}(i\partial_{t_2})^{n_2}$ to
Eq.~\eqref{eq:bdaLC} thus gives
\begin{align}
 \Pi_+\langle0|\mathcal O_{n_1,n_2}(n)|\Lambda(P,\sigma)\rangle
 &=\left.(i\partial_{t_1})^{n_1}(i\partial_{t_2})^{n_2}
       \langle0|\mathcal O_{\mathrm{LC},+}(t_1,t_2)|\Lambda(P,\sigma)\rangle\right|_{t_1=t_2=0}
       \nonumber\\
 &=f_\Lambda^{\rm LC}\,p^{\,r}\mathcal M_{n_1,n_2}(\mu)\,u_{\Lambda,+}.
 \label{eq:bdaLocalMoment}
\end{align}
We use Eq.~\eqref{eq:bdaLocalMoment} as the light-cone moment relation
with local and light-ray operators defined in the same scheme at scale $\mu$. For an equal-time operator the covariant Taylor expansion gives the
tree-level coefficients of the local operators. Beyond tree level their Wilson coefficients enter through the OPE in Sec.~\ref{sec:renormalizedOPE}.

The $r$ symmetric vector indices in Eq.~\eqref{eq:bdaLocal} are coupled to an open Dirac index. To identify its leading-twist component, we decompose this tensor-spinor into irreducible representations $(j_L,j_R)$ of the complexified Lorentz algebra. The Lorentz spin is $s=j_L+j_R$ and the geometric twist is $\tau=d_{\rm can}-s$~\cite{Braun:2008ia}. For three quark fields and $N$ derivatives, $d_{\rm can}=3(3/2)+N=r+7/2$.
First consider the vector indices alone. Each vector transforms as $(1/2,1/2)$, and the symmetric product decomposes as
\begin{align}
 \operatorname{Sym}^r\!\left(\frac12,\frac12\right)
 =\bigoplus_{k=0}^{\lfloor r/2\rfloor}
      \left(\frac{r-2k}{2},\frac{r-2k}{2}\right).
 \label{eq:bdaTraceSectors}
\end{align}
Here $k$ counts the pairs of vector indices contracted in taking traces. Each trace uses two indices, so the maximum number of traces is $\lfloor r/2\rfloor$, the largest integer less than or equal to $r/2$.

For $k=0$ the vector indices form a traceless rank-$r$ tensor in the representation $(r/2,r/2)$. Coupling it to the left- and right-handed components of the open spinor gives, respectively,
\begin{align}
 \left(\frac r2,\frac r2\right)\otimes\left(\frac12,0\right)
 &=\left(\frac{r+1}{2},\frac r2\right)
    \oplus\left(\frac{r-1}{2},\frac r2\right),\nonumber\\
 \left(\frac r2,\frac r2\right)\otimes\left(0,\frac12\right)
 &=\left(\frac r2,\frac{r+1}{2}\right)
    \oplus\left(\frac r2,\frac{r-1}{2}\right).
 \label{eq:bdaWeylCoupling}
\end{align}
In each line of Eq.~\eqref{eq:bdaWeylCoupling}, the first representation has $s=r+1/2$ and the second has $s=r-1/2$. Their twists are therefore
\begin{align}
 \left(r+\frac72\right)-\left(r+\frac12\right)=3,\qquad
 \left(r+\frac72\right)-\left(r-\frac12\right)=4 .
 \label{eq:bdaTwistThreeFour}
\end{align}
For $k\geq1$, the corresponding term in Eq.~\eqref{eq:bdaTraceSectors} contains $k$ metric tensors multiplying a traceless tensor of rank $r-2k$. The metrics are Lorentz invariant, so the vector indices carry spin $r-2k$. The open spinor adds at most $1/2$. Since taking a trace preserves the number of quark fields and derivatives, the canonical dimension is still $r+7/2$. The largest allowed spin and the corresponding minimum twist in the $k$th term are therefore
\begin{equation}
 s_{\max}=r-2k+\frac12,\qquad
 \tau_{\min}=d_{\rm can}-s_{\max}
 =\left(r+\frac72\right)-\left(r-2k+\frac12\right)=3+2k .
 \label{eq:bdaMetricTwistBound}
\end{equation}
For $r\geq2$, the first metric trace ($k=1$) starts at twist five. Each additional trace raises the minimum twist by two. Removing all metric traces thus leaves just the twist-3 and twist-4 components of Eq.~\eqref{eq:bdaTwistThreeFour}.

To isolate twist-3 in the metric-traceless tensor-spinor $Y_r=Y_r^{(3)}+Y_r^{(4)}$, we use the gamma trace to identify the twist-4 term to subtract.
Eq.~\eqref{eq:bdaWeylCoupling} gives Lorentz spin $r+1/2$ for $Y_r^{(3)}$. A nonzero gamma trace of this component would have to carry the same spin because the contraction is Lorentz covariant.
Contracting one vector index with $\gamma_{\mu_1}$ leaves $r-1$ vector indices and one spinor index. The resulting tensor-spinor has maximum Lorentz spin $(r-1)+1/2=r-1/2$, one unit below the spin of $Y_r^{(3)}$.
Thus $\gamma_{\mu_1}Y_r^{(3)\,\mu_1\cdots\mu_r}=0$, and the gamma trace of $Y_r$ comes entirely from $Y_r^{(4)}$.
\begin{align}
 \gamma_{\mu_1}Y_r^{\mu_1\cdots\mu_r}
 &=\gamma_{\mu_1}\big(Y_r^{(3)\,\mu_1\cdots\mu_r}
                     +Y_r^{(4)\,\mu_1\cdots\mu_r}\big)
  =\gamma_{\mu_1}Y_r^{(4)\,\mu_1\cdots\mu_r}.
 \label{eq:bdaGammaOnFour}
\end{align}
To reconstruct the twist-4 component from its gamma trace in Eq.~\eqref{eq:bdaGammaOnFour}, write $Y_r(z)=z_{\mu_1}\cdots z_{\mu_r}Y_r^{\mu_1\cdots\mu_r}$ and define $\partial_{z^\mu}=\partial/\partial z^\mu$. Acting on this polynomial, $\slashed{\partial}_z=\gamma^\mu\partial_{z^\mu}$ takes the gamma trace and $\Box_z=g^{\mu\nu}\partial_{z^\mu}\partial_{z^\nu}$ takes the metric trace, with factors $r$ and $r(r-1)$ respectively, as in Eq.~\eqref{eq:bdaAppTraces}. Eq.~\eqref{eq:bdaAppGamma} gives the rank-$r$ twist-4 term as $\slashed z\slashed{\partial}_zY_r(z)/[2(r+1)]$. Subtracting this term from $Y_r(z)$ leaves $Y_r^{(3)}(z)$. Applying these two subtractions to the original three-quark operator gives the twist-3 component
\begin{align}
 &\mathcal O^{(3)}_{n_1,n_2}(z)
 = \mathcal P_r \mathcal O_{n_1,n_2}(z),
 \label{eq:bdaProjector}\\
& \mathcal P_r
 = \left(1-\frac{\slashed z\slashed{\partial}_z}{2(r+1)}\right)\mathcal H_r,
 \qquad
 \mathcal H_r = \sum_{k=0}^{\lfloor r/2\rfloor}
    \frac{(-1)^k(r-k)!}{4^k k!\,r!}(z^2)^k\Box_z^k .
 \label{eq:bdaHarmonic}
\end{align}
The harmonic projector $\mathcal H_r$ removes the metric traces. The factor on its left subtracts the twist-4 component, giving $\slashed{\partial}_z\mathcal O^{(3)}_{n_1,n_2}(z)=0$. This last condition already implies that every metric trace vanishes. Indeed, the derivatives with respect to the auxiliary vector commute, so the Clifford algebra gives
\begin{align}
 \slashed{\partial}_z^{\,2}
 =\frac12\{\gamma^\mu,\gamma^\nu\}\partial_{z^\mu}\partial_{z^\nu}
 =\Box_z,\qquad
 \Box_z\mathcal O^{(3)}_{n_1,n_2}
 =\slashed{\partial}_z\big(\slashed{\partial}_z\mathcal O^{(3)}_{n_1,n_2}\big)=0 .
\end{align}
It follows from Eqs.~\eqref{eq:bdaProjector} and \eqref{eq:bdaHarmonic} that the full symmetric operator decomposes as
\begin{equation}
 \mathcal O_{n_1,n_2}(z)
 =\mathcal O^{(3)}_{n_1,n_2}(z)
  +\frac{\slashed z\slashed{\partial}_z}{2(r+1)}\mathcal H_r\mathcal O_{n_1,n_2}(z)
  +(1-\mathcal H_r)\mathcal O_{n_1,n_2}(z).
 \label{eq:bdaTwistDecomposition}
\end{equation}
The middle term above is fixed by the rank-$(r-1)$ gamma trace $\slashed{\partial}_z\mathcal H_r\mathcal O_{n_1,n_2}$. This trace is itself gamma-traceless since $\slashed{\partial}_z^{\,2}\mathcal H_r\mathcal O_{n_1,n_2}=\Box_z\mathcal H_r\mathcal O_{n_1,n_2}=0$. It carries spin $r-1/2$ and determines the twist-4 term. The last term in Eq.~\eqref{eq:bdaTwistDecomposition} contains the metric traces of twist-5 and higher, as follows from Eq.~\eqref{eq:bdaMetricTwistBound}.

To evaluate Eq.~\eqref{eq:bdaTwistDecomposition} on the light cone, we firstly perform the auxiliary derivatives and then set $z=n$. Each term in $(1-\mathcal H_r)\mathcal O_{n_1,n_2}$ contains at least one factor of $z^2$ and vanishes at $n^2=0$. The twist-4 term still has a factor $\slashed n$ on the left. Projecting the open spinor index with $\Pi_+$, as required by the LCDA definition in Eq.~\eqref{eq:bdaLC}, gives
\begin{align}
 \Pi_+\mathcal O^{(3)}_{n_1,n_2}(n)
 &=\Pi_+\mathcal O_{n_1,n_2}(n)
   -\frac{\Pi_+\slashed n}{2(r+1)}
      \big[\slashed{\partial}_z\mathcal H_r\mathcal O_{n_1,n_2}(z)\big]_{z=n}
   \nonumber\\
 &=\Pi_+\mathcal O_{n_1,n_2}(n),
 \label{eq:bdaPreserve}
\end{align}
where the last equality uses $\Pi_+\slashed n=0$. Thus the twist-3 projection preserves the local matrix element that defines the LCDA moment in Eq.~\eqref{eq:bdaLocalMoment}.
For the spatial separation $z=\lambda$, $\lambda^2=-1$ and $\Pi_+\slashed\lambda\ne0$, so both subtractions can affect the spatial moments.

\subsection{Projected matrix elements}
\label{sec:lightconemoments}

In the parity channel of Eq.~\eqref{eq:bdaThreeQuark}, the on-shell matrix element of the unprojected operator in Eq.~\eqref{eq:bdaLocal} has two spinor structures, $u_\Lambda$ and $\slashed z u_\Lambda$. Their coefficients are homogeneous of degrees $r$ and $r-1$ in $z$, respectively. Writing $\chi=P\cdot z$ and $\rho=z^2$, the term $P^{\mu_1}\cdots P^{\mu_r}u_\Lambda$ gives $\chi^r u_\Lambda$ upon contraction with $r$ copies of $z$. Each metric tensor replaces two powers of $\chi$ by one power of $\rho$. For the term proportional to $\slashed z u_\Lambda$, the same counting applies to the remaining $r-1$ indices. At fixed $(n_1,n_2)$ the general matrix element is therefore
\begin{equation}
\begin{aligned}
 &\langle0|\mathcal O_{n_1,n_2}(z)|\Lambda(P,\sigma)\rangle\\
 =& f_\Lambda^{\rm LC}\bigg[
     C_{n_1,n_2}(\mu)\,\chi^r
     +\sum_{k=1}^{\lfloor r/2\rfloor}
          \alpha_k(m_\Lambda^2\rho)^k\chi^{r-2k} +m_\Lambda\slashed z\sum_{k=0}^{\lfloor(r-1)/2\rfloor}
          \beta_k(m_\Lambda^2\rho)^k\chi^{r-1-2k}
     \bigg]u_\Lambda .
\end{aligned}
 \label{eq:bdaFullME}
\end{equation}
The coefficients $C_{n_1,n_2}$, $\alpha_k$ and $\beta_k$ are dimensionless and depend on the derivative assignment and the scale $\mu$. The terms proportional to $\alpha_k$ contain an explicit metric, while those proportional to $\beta_k$ have a gamma matrix inserted into a rank-$(r-1)$ tensor-spinor. Their maximum spins are $r-3/2$ and $r-1/2$, respectively. Applying the spin-$(r+1/2)$ projector to Eq.~\eqref{eq:bdaFullME} therefore gives
\begin{align}
 \langle0|\mathcal O^{(3)}_{n_1,n_2}(z)|\Lambda(P,\sigma)\rangle
 &=f_\Lambda^{\rm LC}\,C_{n_1,n_2}(\mu)\,\mathcal Q_r(z,P)u_\Lambda,\nonumber\\
 \mathcal Q_r(z,P)u_\Lambda
 &\equiv\mathcal P_r[\chi^r u_\Lambda]
   =\big[A_r(\chi,\rho)+m_\Lambda\slashed z B_r(\chi,\rho)\big]u_\Lambda .
 \label{eq:bdaGeneralME}
\end{align}
In terms of the harmonic polynomial $h_r(\chi,\rho)=\mathcal H_r\chi^r$ defined in Eq.~\eqref{eq:bdaAppScalar}, and according to Eq.~\eqref{eq:bdaExplicitQ}, the two spinor coefficients can be expressed as 
\begin{equation}
 A_r(\chi,\rho)=h_r(\chi,\rho)-\frac{\rho}{r+1}\,\partial_\rho h_r(\chi,\rho),
 \qquad
 B_r(\chi,\rho)=-\frac{1}{2(r+1)}\,\partial_\chi h_r(\chi,\rho).
 \label{eq:bdaCovariantCoefficients}
\end{equation}

To determine $C_{n_1,n_2}$ from the light-cone moment, set $z=n$ and so that $\chi=p$ and $\rho=0$. Only the $k=0$ term of Eq.~\eqref{eq:bdaAppScalar} survives in $A_r(p,0)=h_r(p,0)=p^r$. Projecting the open index in Eq.~\eqref{eq:bdaGeneralME} with $\Pi_+$ removes the term proportional to $\slashed n$. Equations~\eqref{eq:bdaPreserve} and \eqref{eq:bdaLocalMoment} give
\begin{align}
 f_\Lambda^{\rm LC} C_{n_1,n_2}p^r u_{\Lambda,+}
 &=\Pi_+\langle0|\mathcal O^{(3)}_{n_1,n_2}(n)|\Lambda(P,\sigma)\rangle\nonumber\\
 &=\Pi_+\langle0|\mathcal O_{n_1,n_2}(n)|\Lambda(P,\sigma)\rangle
 = f_\Lambda^{\rm LC}\mathcal M_{n_1,n_2}p^r u_{\Lambda,+}.
 \label{eq:bdaFixC}
\end{align}
It follows that $C_{n_1,n_2}=\mathcal M_{n_1,n_2}$. Restoring general $z$, the matrix element needed in the following section is
\begin{equation}
 \langle0|\mathcal O^{(3)}_{n_1,n_2}(z)|\Lambda(P,\sigma)\rangle
 = f_\Lambda^{\rm LC}\,\mathcal M_{n_1,n_2}(\mu)\,
     \mathcal Q_r(z,P)u_\Lambda(P,\sigma) .
 \label{eq:bdaMatched}
\end{equation}
The term proportional to $\slashed z$ in $\mathcal Q_r$ contributes to the spatial matrix element at $z=\lambda$.

The two Dirac matrices in Eq.~\eqref{eq:bdaEqualTime} select different components of the same symmetric tensor.
For the spatial channel all $r$ indices are contracted with $\lambda$. In the temporal channel one index is contracted with $v_t$
and the remaining $r-1$ indices with $\lambda$. Differentiating the homogeneous polynomial picks each of the $r$ equivalent indices once, so that
\begin{equation}
 \mathcal O^{(3)}_{z,n_1,n_2}=\mathcal O^{(3)}_{n_1,n_2}(\lambda),\qquad
 \mathcal O^{(3)}_{t,n_1,n_2}
 =\left.\frac{v_t\cdot\partial_z}{r}\mathcal O^{(3)}_{n_1,n_2}(z)\right|_{z=\lambda}.
 \label{eq:bdaChannels}
\end{equation}
We define $\widetilde\phi_\Gamma^{(3,0)}$ by keeping only the twist-3 local operators in the tree-level
expansion of Eq.~\eqref{eq:bdaQuasiScalar}. Taking $n_1$ and $n_2$ distance derivatives as in Eq.~\eqref{eq:bdaLocalMoment} gives
\begin{align}
 \widetilde{\mathcal M}^{(3,0)}_{\Gamma;n_1,n_2}
 &=\int_{\mathbb R^2}dx_1\,dx_2\,x_1^{n_1}x_2^{n_2}
     \widetilde\phi_\Gamma^{(3,0)}(x_1,x_2;P_z)\nonumber\\
 &=\frac{\sum_\sigma \bar u_\Lambda\slashed n
       \langle0|\mathcal O^{(3)}_{\Gamma;n_1,n_2}|\Lambda(P,\sigma)\rangle}
        {4f_\Lambda^{\rm LC}pP_z^{\,r}} .
 \label{eq:bdaQuasiMoment}
\end{align}
For $\Gamma=\gamma^z$ or $\gamma^t$, the numerator contains the $z$ or $t$ component in Eq.~\eqref{eq:bdaChannels}, respectively.
The power $P_z^r$ includes one factor from the reference normalization in Eq.~\eqref{eq:bdaQuasiScalar}
and $N$ factors obtained by differentiating its Fourier phase. Equations~\eqref{eq:bdaMatched} and \eqref{eq:bdaQuasiMoment}
give the starting point for the two moment ratios in Sec.~\ref{sec:momentratio}.
For $N\geq1$, the full temporal operator also contains a mixed-symmetry component, which is treated in Sec.~\ref{sec:Highertwist}.

\section{Tree-level target-mass corrections}
\label{sec:targetmassCorr}

At tree level, the local matrix elements in Eq.~\eqref{eq:bdaQuasiMoment} determine the twist-3 quasi-DA in terms of the LCDA. We derive the moment ratios at finite $P_z$ for $\Gamma=\gamma^z$ and $\gamma^t$ and resum them to obtain the two quasi-DAs. Both channels use the reference normalization $4f_\Lambda^{\rm LC}pP_z$ of Eq.~\eqref{eq:bdaQuasiScalar}. Section~\ref{sec:projectednormalization} gives the conversion to unit normalization.

\subsection{Spatial and temporal moments}
\label{sec:momentratio}

In Eq.~\eqref{eq:bdaMatched}, the LCDA moment multiplies the spinor $\mathcal Q_r(z,P)u_\Lambda(P,\sigma)$. To isolate its kinematic factor in Eq.~\eqref{eq:bdaQuasiMoment}, we contract this spinor with $\bar u_\Lambda(P,\sigma)\slashed n$, sum over $\sigma$, and divide by $\sum_\sigma\bar u_\Lambda\slashed n u_\Lambda=4p$, where $p=P\cdot n$. Inserting the two structures in Eq.~\eqref{eq:bdaGeneralME} and using $\sum_\sigma u_\Lambda\bar u_\Lambda=\slashed P+m_\Lambda$ gives
\begin{align}
 S_r(z)&\equiv\frac{1}{4p}\sum_\sigma
  \bar u_\Lambda(P,\sigma)\slashed n\,\mathcal Q_r(z,P)u_\Lambda(P,\sigma)
  \nonumber\\
 &= A_r(\chi,\rho)+\frac{m_\Lambda^2\,n\cdot z}{p}B_r(\chi,\rho),
 \label{eq:tmSpin}
\end{align}
where $r=n_1+n_2+1$, $\chi=P\cdot z$ and $\rho=z^2$. The functions $A_r$ and $B_r$ are given in Eq.~\eqref{eq:bdaCovariantCoefficients}. For the spatial channel we evaluate $S_r$ at $z=\lambda$. For the temporal channel Eq.~\eqref{eq:bdaChannels} requires one directional derivative before setting $z=\lambda$. Eq.~\eqref{eq:bdaQuasiMoment} then gives
\begin{align}
 \widetilde{\mathcal M}^{(3,0)}_{\gamma^z;n_1,n_2}
 &=\frac{S_r(\lambda)}{P_z^r}\mathcal M_{n_1,n_2}
   \equiv F_r(c)\mathcal M_{n_1,n_2},\nonumber\\
 \widetilde{\mathcal M}^{(3,0)}_{\gamma^t;n_1,n_2}
 &=\frac{[(v_t\cdot\partial_z)S_r(z)]_{z=\lambda}}{rP_z^r}\mathcal M_{n_1,n_2}
   \equiv T_r(c)\mathcal M_{n_1,n_2},\qquad
 c=\frac{m_\Lambda^2}{4P_z^2}.
 \label{eq:tmRatios}
\end{align}
For a given $c$, the mass factors $F_r$ and $T_r$ depend on $n_1$ and $n_2$ only through $r=n_1+n_2+1$. The dependence on how the derivatives are distributed between the two quarks resides in $\mathcal M_{n_1,n_2}$.

At $z=\lambda$, we have $\chi=P_z$, $\rho=-1$ and $n\cdot\lambda=-1/\sqrt{2}$. Let $\Delta=E/P_z=\sqrt{1+4c}$. In the ratio $S_r(\lambda)/P_z^r$ obtained from Eq.~\eqref{eq:tmSpin}, the coefficient multiplying $B_r(P_z,-1)/P_z^{r-1}$ is
\begin{equation}
 \frac{m_\Lambda^2\,n\cdot\lambda}{pP_z}
 =-\frac{m_\Lambda^2}{P_z(E+P_z)}=1-\Delta .
 \label{eq:tmSpatialSign}
\end{equation}
To evaluate $A_r$ and $B_r$, factor $\chi^r$ out of the harmonic polynomial in Eq.~\eqref{eq:bdaAppScalar}. For $\chi\ne0$ this gives
\begin{equation}
 h_r(\chi,\rho)=\chi^r K_r(c_z),\qquad
 c_z=-\frac{m_\Lambda^2\rho}{4\chi^2},\qquad
 K_r(c)=\sum_{k=0}^{\lfloor r/2\rfloor}\binom{r-k}{k}c^k .
 \label{eq:tmK}
\end{equation}
At $z=\lambda$, $c_z=c$. We first differentiate Eq.~\eqref{eq:tmK} at general $z$, obtaining $\rho\,\partial_\rho h_r=\chi^r c_zK_r'(c_z)$ and $\partial_\chi h_r=\chi^{r-1}[rK_r(c_z)-2c_zK_r'(c_z)]$, where the prime denotes differentiation with respect to $c_z$. Inserting these derivatives into Eq.~\eqref{eq:bdaCovariantCoefficients}, setting $z=\lambda$ and using Eq.~\eqref{eq:tmSpatialSign} gives
\begin{align}
 F_r(c)=\frac{S_r(\lambda)}{P_z^r}&=K_r(c)-\frac{c}{r+1}K_r'(c)
           -(1-\Delta)\frac{rK_r(c)-2cK_r'(c)}{2(r+1)}\nonumber\\
       &=\frac{(r+2+r\Delta)K_r(c)-2c\Delta K_r'(c)}{2(r+1)} ,
 \label{eq:tmF}
\end{align}
where the term proportional to $1-\Delta$ is the contribution of the $B_r$ spinor structure in Eq.~\eqref{eq:tmSpin}.

For the temporal contraction in Eq.~\eqref{eq:tmRatios}, take $z_\epsilon=\lambda+\epsilon v_t$ at fixed $P$ and $m_\Lambda$. Since $d z_\epsilon^\mu/d\epsilon=v_t^\mu$, we have
\begin{equation}
 T_r(c)
 =\frac{1}{rP_z^r}\left.(v_t\cdot\partial_z)S_r(z)\right|_{z=\lambda}
 =\frac{1}{rP_z^r}\left.\frac{d}{d\epsilon}S_r(z_\epsilon)\right|_{\epsilon=0}.
 \label{eq:reorgtimepathderivative}
\end{equation}
To take the derivative in Eq.~\eqref{eq:reorgtimepathderivative}, we need $S_r$ along the path $z_\epsilon$. Evaluating the spin projection in Eq.~\eqref{eq:tmSpin} with the coefficients of Eq.~\eqref{eq:bdaCovariantCoefficients} gives
\begin{equation}
 \frac{S_r(z_\epsilon)}{\chi_\epsilon^r}
 =K_r(c_\epsilon)-\frac{c_\epsilon}{r+1}K_r'(c_\epsilon)
 -\frac{m_\Lambda^2\,n\cdot z_\epsilon}{p\,\chi_\epsilon}\,
   \frac{rK_r(c_\epsilon)-2c_\epsilon K_r'(c_\epsilon)}{2(r+1)} , \label{eq:SdivideChi}
\end{equation}
where $\chi_\epsilon=P\cdot z_\epsilon$, and $c_\epsilon$ denotes $c_z$ of Eq.~\eqref{eq:tmK} evaluated at $z_\epsilon$. Along this path, $\chi_\epsilon=P_z(1+\Delta\epsilon)$, $z_\epsilon^2=\epsilon^2-1$ and $n\cdot z_\epsilon=(\epsilon-1)/\sqrt{2}$. With $p=(E+P_z)/\sqrt{2}$ and $m_\Lambda^2=(E-P_z)(E+P_z)$, these contractions give
\begin{align}
 c_\epsilon\equiv c_{z_\epsilon}
 &=-\frac{m_\Lambda^2 z_\epsilon^2}{4(P\cdot z_\epsilon)^2}
   =c\,\frac{1-\epsilon^2}{(1+\Delta\epsilon)^2},\nonumber\\
 \frac{m_\Lambda^2\,n\cdot z_\epsilon}{p\,\chi_\epsilon}
 &=\frac{m_\Lambda^2}{P_z(E+P_z)}\frac{\epsilon-1}{1+\Delta\epsilon}
   =(\Delta-1)\frac{\epsilon-1}{1+\Delta\epsilon}.
 \label{eq:tmPath}
\end{align}
At $z=\lambda$, Eq.~\eqref{eq:tmSpatialSign} gives $1-\Delta$ for the spinor coefficient. To express the coefficient along $z_\epsilon$ in the same form, let $\Delta_\epsilon=\sqrt{1+4c_\epsilon}$. Using $\Delta^2-1=4c$ in the first line of Eq.~\eqref{eq:tmPath} and taking the positive square root near $\epsilon=0$, we find
\begin{align}
 1+4c_\epsilon
 &=\frac{(1+\Delta\epsilon)^2+(\Delta^2-1)(1-\epsilon^2)}
         {(1+\Delta\epsilon)^2}
  =\frac{(\Delta+\epsilon)^2}{(1+\Delta\epsilon)^2},\nonumber\\
 1-\sqrt{1+4c_\epsilon}
 &=1-\frac{\Delta+\epsilon}{1+\Delta\epsilon}
  =(\Delta-1)\frac{\epsilon-1}{1+\Delta\epsilon}
  =\frac{m_\Lambda^2\,n\cdot z_\epsilon}{p\,(P\cdot z_\epsilon)} .
 \label{eq:tmPathSpinFactor}
\end{align}
With the spinor coefficient $1-\Delta_\epsilon$ from Eq.~\eqref{eq:tmPathSpinFactor}, the right-hand side of Eq.~\eqref{eq:SdivideChi} equals $F_r(c_\epsilon)$ by the first line of Eq.~\eqref{eq:tmF}. Multiplying by $\chi_\epsilon^r$ gives
\begin{align}
 S_r(z_\epsilon)=(P\cdot z_\epsilon)^r F_r(c_\epsilon)
                =P_z^r(1+\Delta\epsilon)^r F_r(c_\epsilon).
 \label{eq:tmPathScalar}
\end{align}
At $\epsilon=0$, Eq.~\eqref{eq:tmPath} gives $c_\epsilon=c$ and $d c_\epsilon/d\epsilon|_0=-2c\Delta$, while $d(1+\Delta\epsilon)^r/d\epsilon|_0=r\Delta$. Inserting Eq.~\eqref{eq:tmPathScalar} into Eq.~\eqref{eq:reorgtimepathderivative} and applying the product rule, we obtain
\begin{align}
 T_r(c)
 &=\frac{1}{r}\left.\frac{d}{d\epsilon}
     \big[(1+\Delta\epsilon)^r F_r(c_\epsilon)\big]\right|_{\epsilon=0}=\Delta\left[F_r(c)-\frac{2c}{r}F_r'(c)\right].
 \label{eq:tmT}
\end{align}
Equations~\eqref{eq:tmF} and \eqref{eq:tmT} retain all powers of $m_\Lambda^2/P_z^2$. To first order at fixed $r\geq1$, $K_r(c)=1+(r-1)c+O(c^2)$ and $\Delta=1+2c+O(c^2)$, giving
\begin{equation}
 F_r(c)=1+\frac{r^2}{r+1}\,c+O(c^2),\qquad
 T_r(c)=1+\frac{r^2+2}{r+1}\,c+O(c^2).
 \label{eq:tmSmall}
\end{equation}

\subsection{Resummation in momentum space}
\label{sec:Target-mass}

We now use the moment factors in Eq.~\eqref{eq:tmRatios} to reconstruct the two twist-3 quasi-DAs. We extend $\phi(x_1,x_2,\mu)$ by zero outside the simplex $\mathcal S$ defined in Eq.~\eqref{eq:bdaThreeLCFourier} and suppress the common scale $\mu$. To express the moment factors in terms of rescalings of $x_1$ and $x_2$, introduce the two roots
\begin{equation}
 a=\frac{1+\Delta}{2},\qquad b=\frac{1-\Delta}{2},\qquad
 a+b=1,\quad a-b=\Delta,\quad ab=-c .
 \label{eq:tmRoots}
\end{equation}
The polynomial in Eq.~\eqref{eq:tmK} satisfies $K_r=K_{r-1}+cK_{r-2}$ with $K_0=K_1=1$. Since $a$ and $b$ are the roots of the characteristic equation, $K_r=(a^{r+1}-b^{r+1})/\Delta$. Inserting this expression into Eqs.~\eqref{eq:tmF} and \eqref{eq:tmT} gives
\begin{align}
 F_r&=\frac{a^{r+1}}{\Delta}
       -\frac{b}{\Delta^2}\frac{a^{r+1}-b^{r+1}}{r+1},\nonumber\\
 T_r&=\frac{a^{r+1}}{\Delta}
       -\frac{b(\Delta+2)}{\Delta^3}\frac{a^{r+1}-b^{r+1}}{r+1}
       +\frac{4ab^2}{\Delta^3}\frac{a^r-b^r}{r}.
 \label{eq:tmRootFactors}
\end{align}
Appendix~\ref{app:tmRoots} gives the derivation of Eq.~\eqref{eq:tmRootFactors}. To relate the powers of $a$ and $b$ to rescalings of the LCDA, consider the $(n_1,n_2)$ moment of $\phi(x_1/s,x_2/s)$ for real $s\ne0$. Changing variables to $x_i=s y_i$ and using $N=n_1+n_2=r-1$, we obtain
\begin{align}
 \int_{\mathbb R^2}\!dx_1\,dx_2\,x_1^{n_1}x_2^{n_2}
     \phi\!\left(\frac{x_1}{s},\frac{x_2}{s}\right)
 &=s^N |s|^2\int_{\mathcal S}\!dy_1\,dy_2\,y_1^{n_1}y_2^{n_2}\phi(y_1,y_2)
 =s^{r+1}\mathcal M_{n_1,n_2}.
 \label{eq:tmDilation}
\end{align}
The two differences of powers in Eq.~\eqref{eq:tmRootFactors} are the scale integrals
\begin{equation}
 \frac{a^{r+1}-b^{r+1}}{r+1}=\int_b^a\frac{ds}{s}\,s^{r+1},
 \qquad
 \frac{a^r-b^r}{r}=\int_b^a\frac{ds}{s^2}\,s^{r+1}.
 \label{eq:tmScaleWeights}
\end{equation}
Combining Eqs.~\eqref{eq:tmDilation} and \eqref{eq:tmScaleWeights} term by term yields the spatial quasi-DA,
\begin{equation}
 \widetilde\phi_{\gamma^z}^{(3,0)}(x_1,x_2;P_z)
 =\frac{1}{\Delta}\phi\!\left(\frac{x_1}{a},\frac{x_2}{a}\right)
  -\frac b{\Delta^2}\int_b^a\frac{ds}{s}\,
     \phi\!\left(\frac{x_1}{s},\frac{x_2}{s}\right),
 \label{eq:tmSpatialDA}
\end{equation}
and the temporal quasi-DA,
\begin{align}
 \widetilde\phi_{\gamma^t}^{(3,0)}(x_1,x_2;P_z)
 &=\frac{1}{\Delta}\phi\!\left(\frac{x_1}{a},\frac{x_2}{a}\right)
   -\frac{b(\Delta+2)}{\Delta^3}\int_b^a\frac{ds}{s}\,
       \phi\!\left(\frac{x_1}{s},\frac{x_2}{s}\right)\nonumber\\
 &\quad+\frac{4ab^2}{\Delta^3}\int_b^a\frac{ds}{s^2}\,
       \phi\!\left(\frac{x_1}{s},\frac{x_2}{s}\right).
 \label{eq:tmTemporalDA}
\end{align}
The $ds/s^2$ term in Eq.~\eqref{eq:tmTemporalDA} comes from the term proportional to $(a^r-b^r)/r$ in Eq.~\eqref{eq:tmRootFactors}. For $c>0$, the scale $s$ ranges over both $b<s<0$ and $0<s<a$. We can integrate both contributions over $0<u<1$ by setting $s=bu$ on the negative interval and $s=au$ on the positive interval. Writing $f_+=1+\Delta=2a$ and $f_-=\Delta-1=-2b$, we obtain for the spatial channel
\begin{align}
 \widetilde\phi_{\gamma^z}^{(3,0)}(x_1,x_2;P_z)
 &=\frac{1}{\Delta}\phi\!\left(\frac{2x_1}{f_+},\frac{2x_2}{f_+}\right)\nonumber\\
 &\quad+\frac{f_-}{2\Delta^2}\int_0^1\frac{du}{u}
 \left[\phi\!\left(\frac{2x_1}{u f_+},\frac{2x_2}{u f_+}\right)
      -\phi\!\left(-\frac{2x_1}{u f_-},-\frac{2x_2}{u f_-}\right)\right].
 \label{eq:tmSpatialDAu}
\end{align}
For the temporal channel, Eq.~\eqref{eq:tmTemporalDA} gives
\begin{align}
 \widetilde\phi_{\gamma^t}^{(3,0)}(x_1,x_2;P_z)
 &=\frac{1}{\Delta}\phi\!\left(\frac{2x_1}{f_+},\frac{2x_2}{f_+}\right)\nonumber\\
 &\quad+\frac{f_-^2}{\Delta^3}\int_0^1\frac{du}{u^2}
       \phi\!\left(\frac{2x_1}{u f_+},\frac{2x_2}{u f_+}\right)\nonumber\\
 &\quad+\frac{f_-f_+}{\Delta^3}\int_0^1\frac{du}{u^2}
       \phi\!\left(-\frac{2x_1}{u f_-},-\frac{2x_2}{u f_-}\right)\nonumber\\
 &\quad+\frac{f_-(\Delta+2)}{2\Delta^3}\int_0^1\frac{du}{u}
 \left[\phi\!\left(\frac{2x_1}{u f_+},\frac{2x_2}{u f_+}\right)
      -\phi\!\left(-\frac{2x_1}{u f_-},-\frac{2x_2}{u f_-}\right)\right].
 \label{eq:tmTemporalDAu}
\end{align}
Appendix~\ref{app:tmBranches} derives Eqs.~\eqref{eq:tmSpatialDAu} and \eqref{eq:tmTemporalDAu} and shows that each branch converges absolutely at $u=0$ for almost every $(x_1,x_2)$. The forms involving $f_-$ apply for $c>0$. At $c=0$, $a=1$ and $b=0$, and Eqs.~\eqref{eq:tmSpatialDA} and \eqref{eq:tmTemporalDA} reduce directly to $\widetilde\phi_{\Gamma}^{(3,0)}(x_1,x_2;P_z)=\phi(x_1,x_2)$ in both channels.

The leading mass correction follows by expanding Eqs.~\eqref{eq:tmSpatialDA} and \eqref{eq:tmTemporalDA} to first order in $c=m_\Lambda^2/(4P_z^2)$. In the first term of each equation, $1/\Delta=1-2c+O(c^2)$ changes the overall normalization, while $a=1+c+O(c^2)$ shifts both arguments of $\phi$. Since $b=-c+O(c^2)$, the $ds/s$ term already carries a factor of $c$ and its integration limits can be set to $0$ and $1$ at this order. The $ds/s^2$ term in the temporal channel starts at $O(c^2)$. With $\kappa_{\gamma^z}=1$ and $\kappa_{\gamma^t}=3$, the result is
\begin{align}
 \widetilde\phi_\Gamma^{(3,0)}(x_1,x_2;P_z)
 ={}&\phi(x_1,x_2)-c(2+x_1\partial_{x_1}+x_2\partial_{x_2})\phi(x_1,x_2)\nonumber\\
 &+c\,\kappa_\Gamma\int_0^1\frac{ds}{s}\,
       \phi\!\left(\frac{x_1}{s},\frac{x_2}{s}\right)+O(c^2).
 \label{eq:tmDAFirstOrder}
\end{align}
For the $(n_1,n_2)$ moment, write $N=n_1+n_2$. The derivative contribution $x_1\partial_{x_1}+x_2\partial_{x_2}$ gives $cN\mathcal M_{n_1,n_2}$. For the scale integral, Eq.~\eqref{eq:tmDilation} gives $s^{N+2}\mathcal M_{n_1,n_2}$ before integration with $ds/s$. Adding both corrections to the LCDA moment gives
\[
 \begin{aligned}
 \widetilde{\mathcal M}^{(3,0)}_{\Gamma;n_1,n_2}
 &=\left[1+cN+c\kappa_\Gamma\int_0^1s^{N+1}\,ds\right]\mathcal M_{n_1,n_2}
   +O(c^2)\\
 &=\left[1+c\left(N+\frac{\kappa_\Gamma}{N+2}\right)\right]\mathcal M_{n_1,n_2}
   +O(c^2).
 \end{aligned}
\]
With $r=N+1$, the coefficient of $c$ is $r^2/(r+1)$ for $\gamma^z$ and $(r^2+2)/(r+1)$ for $\gamma^t$, as in Eq.~\eqref{eq:tmSmall}.

\subsection{Local normalization}
\label{sec:projectednormalization}

Integrating the quasi-DA over both momentum fractions selects $n_1=n_2=0$ and hence $r=1$. The corresponding LCDA moment is $\mathcal M_{0,0}=1$. In the common $f_\Lambda^{\rm LC}$ normalization, setting $K_1=1$ in Eqs.~\eqref{eq:tmF} and \eqref{eq:tmT} gives
\begin{align}
 \int_{\mathbb R^2}\!dx_1\,dx_2\,\widetilde\phi_{\gamma^z}^{(3,0)}(x_1,x_2;P_z)
 &= F_1(c)=\frac{3+\Delta}{4},\nonumber\\
 \int_{\mathbb R^2}\!dx_1\,dx_2\,\widetilde\phi_{\gamma^t}^{(3,0)}(x_1,x_2;P_z)
 &= T_1(c)=\frac{1+3\Delta}{4}.
 \label{eq:tmLocalFactors}
\end{align}
To normalize the projected axial quasi-DAs to unit integral, we define a local prefactor for each channel using Eq.~\eqref{eq:tmLocalFactors},
\begin{equation}
 f_{\Lambda,\gamma^z}^{(3,0)}(P_z)=F_1(c)f_\Lambda^{\rm LC},\qquad
 f_{\Lambda,\gamma^t}^{(3,0)}(P_z)=T_1(c)f_\Lambda^{\rm LC}.
 \label{eq:tmEffectiveF}
\end{equation}
Using Eq.~\eqref{eq:tmEffectiveF} in the definition of the projected quasi-DAs then gives
\begin{align}
 \widehat{\widetilde\phi}_{\gamma^z}^{(3,0)}(x_1,x_2;P_z)
 &=\widetilde\phi_{\gamma^z}^{(3,0)}(x_1,x_2;P_z)/F_1(c),\nonumber\\
 \widehat{\widetilde\phi}_{\gamma^t}^{(3,0)}(x_1,x_2;P_z)
 &=\widetilde\phi_{\gamma^t}^{(3,0)}(x_1,x_2;P_z)/T_1(c).
 \label{eq:tmUnitDA}
\end{align}
The moment factors of the unit-normalized DAs in Eq.~\eqref{eq:tmUnitDA} are therefore $F_r/F_1$ and $T_r/T_1$. In each case the local prefactor comes from the same twist-3 projection used in the nonlocal numerator. A lattice calculation can also normalize the full nonlocal matrix element of Eq.~\eqref{eq:bdaQuasiScalar} by its own value at $\ell_1=\ell_2=0$. That local matrix element can contain higher-twist contributions, so the conversion factors in Eq.~\eqref{eq:tmEffectiveF} apply specifically to the twist-3 projected DAs.

\section{Higher-twist operators}
\label{sec:Highertwist}

The factors $F_r$ and $T_r$ in Eq.~\eqref{eq:tmRatios} multiply the twist-3 LCDA moment $\mathcal M_{n_1,n_2}$. The full lattice matrix element in Eq.~\eqref{eq:bdaQuasiScalar} also contains higher-twist contributions. We identify the corresponding local operators in the spatial and temporal channels, sum them to obtain nonlocal operators, and describe how to construct their matrix elements on the lattice. We then determine how their contributions to the normalized moments scale with $P_z$, using the tree-level expansion and the light-cone normalization of Eq.~\eqref{eq:bdaQuasiMoment}.

\subsection{Local twist decomposition}
\label{sec:higheroperators}

For fixed derivative counts $(n_1,n_2)$, let $r=n_1+n_2+1$. We leave the Lorentz index of the diquark gamma matrix in Eq.~\eqref{eq:bdaLocal} open so that we can contract it with either the spatial or the temporal direction. The resulting operator and its contraction with $z^\mu$ are
\begin{align}
 V_\mu(z)&=\epsilon^{ijk}\big[(iz\cdot D)^{n_1}u_i\big]^T C\gamma_5\gamma_\mu
                 \big[(iz\cdot D)^{n_2}d_j\big]s_k,\nonumber\\
 \mathcal O_r(z)&=z^\mu V_\mu(z)=\mathcal O_{n_1,n_2}(z).
 \label{eq:ht4vector}
\end{align}
All quark fields in Eq.~\eqref{eq:ht4vector} are evaluated at the origin. Contracting the gamma-matrix index and the $r-1$ derivative indices with the same vector $z$ selects the fully symmetric part of the coefficient tensor. Differentiating $\mathcal O_r(z)$ with respect to $z^\mu$ leaves the index $\mu$ in each of the $r$ positions in turn. Averaging these $r$ terms gives the symmetric part of $V_\mu$. With $\partial_\mu=\partial/\partial z^\mu$, we write $V_\mu=V_\mu^{\rm S}+V_\mu^{\rm M}$, where
\begin{align}
 V_\mu^{\rm S}(z)&=\frac{1}{r}\partial_\mu\mathcal O_r(z),\nonumber\\
 V_\mu^{\rm M}(z)&=V_\mu-\frac{1}{r}\partial_\mu\mathcal O_r(z)
 =\frac{z^\nu}{r} B_{\mu\nu}(z),\qquad
 B_{\mu\nu}(z)=\partial_\nu V_\mu(z)-\partial_\mu V_\nu(z).
 \label{eq:ht4curl}
\end{align}
The last expression for $V_\mu^{\rm M}$ in Eq.~\eqref{eq:ht4curl} follows from homogeneity, $z\cdot\partial_z V_\mu=(r-1)V_\mu$, and the product rule, $\partial_\mu\mathcal O_r=V_\mu+z^\nu\partial_\mu V_\nu$. In the spatial channel, $\lambda^\mu\lambda^\nu B_{\mu\nu}(\lambda)=0$ by antisymmetry. In the temporal channel, $v_t^\mu\lambda^\nu B_{\mu\nu}(\lambda)$ can contribute. The two local operators are therefore
\begin{align}
 \mathcal O^{\rm full}_{\gamma^z;n_1,n_2}&=\mathcal O_r(\lambda),\nonumber\\
 \mathcal O^{\rm full}_{\gamma^t;n_1,n_2}
 &=\left.\frac{v_t\cdot\partial_z}{r}\mathcal O_r(z)\right|_{z=\lambda}
   +\frac{v_t^\mu\lambda^\nu}{r}B_{\mu\nu}(\lambda).
 \label{eq:ht4raw}
\end{align}
The superscript $\mathrm{full}$ denotes the local operator before twist projection. At tree level, we obtain this operator by applying $(i\partial_{\ell_1})^{n_1}(i\partial_{\ell_2})^{n_2}$ to $\widetilde{\mathcal O}_\Gamma$ in Eq.~\eqref{eq:bdaEqualTime} and setting $\ell_1=\ell_2=0$. In the spatial channel this gives precisely $\mathcal O_{n_1,n_2}(\lambda)$ as defined in Eq.~\eqref{eq:bdaLocal}.

The symmetric parts of both operators in Eq.~\eqref{eq:ht4raw} follow from the same polynomial $\mathcal O_r(z)$. Write $\rho=z^2$. Each pair of indices carried by a metric tensor contributes one factor of $\rho$ upon contraction with $z$, leaving two fewer indices on the remaining tensor. Starting from $R_r=\mathcal O_r$, denote the metric-traceless part of each remainder $R_j$ by $X_j$. For $r\ge4$ the first two steps give
\begin{align}
 \mathcal O_r
 &=X_r+\rho R_{r-2}
  =X_r+\rho\big(X_{r-2}+\rho R_{r-4}\big)
  =X_r+\rho X_{r-2}+\rho^2 R_{r-4}=\cdots ,
 \label{eq:ht4firstlayers}
\end{align}
where $\Box_z X_j=0$. At each degree $j$, we use the harmonic projector $\mathcal H_j$ in Eq.~\eqref{eq:bdaHarmonic} to extract $X_j$. Subtracting $X_j$ from $R_j$ and dividing by $\rho$ gives the remainder of degree $j-2$. The recursion is
\begin{align}
 X_j&=\mathcal H_j R_j,\qquad
 R_{j-2}=\frac{R_j-X_j}{\rho}
 =\sum_{k=1}^{\lfloor j/2\rfloor}
   \frac{(-1)^{k+1}(j-k)!}{4^k k!\,j!}\,\rho^{k-1}\Box_z^k R_j
 \qquad (j\ge2).
 \label{eq:ht4recur}
\end{align}
The $k=0$ term of $\mathcal H_jR_j$ cancels $R_j$ in the numerator. Each remaining term contains a factor of $\rho$, so the quotient in Eq.~\eqref{eq:ht4recur} is a polynomial. For the first two steps in Eq.~\eqref{eq:ht4firstlayers} this gives
\begin{align}
 X_r&=\mathcal H_r\mathcal O_r,&
 R_{r-2}&=\frac{\mathcal O_r-X_r}{\rho},\nonumber\\
 X_{r-2}&=\mathcal H_{r-2}R_{r-2},&
 R_{r-4}&=\frac{R_{r-2}-X_{r-2}}{\rho}
          =\frac{\mathcal O_r-X_r-\rho X_{r-2}}{\rho^2}.
 \label{eq:ht4twosteps}
\end{align}
The recursion stops at degree zero or one, where every homogeneous polynomial of that degree is harmonic. Substituting the successive remainders back into Eq.~\eqref{eq:ht4firstlayers} gives
\begin{equation}
 \mathcal O_r(z)=\sum_{k=0}^{\lfloor r/2\rfloor}\rho^k X_{r-2k}(z),
 \label{eq:ht4harmonicexp}
\end{equation}
where the term $\rho^k X_{r-2k}$ contains $k$ metric tensors and has minimum twist $3+2k$ by Eq.~\eqref{eq:bdaMetricTwistBound}. Thus only $X_r$ contains twists three and four. Its gamma trace determines the twist-4 component through Eq.~\eqref{eq:bdaTwistDecomposition}. The terms with $k\ge1$ in Eq.~\eqref{eq:ht4harmonicexp} give the complete symmetric contribution of twist five and higher. Within this sum, twist five occurs only in the $k=1$ term and is selected by applying $\mathcal P_{r-2}$ to $X_{r-2}$. These projections give
\begin{align}
 \mathcal O_r^{(4)}
 &=\frac{\slashed z\slashed{\partial}_z}{2(r+1)}X_r
   =\frac{\slashed z\slashed{\partial}_z}{2(r+1)}\mathcal H_r\mathcal O_r,
 \nonumber\\
 \mathcal O_r^{(\ge5)}
 &=\sum_{k=1}^{\lfloor r/2\rfloor}\rho^kX_{r-2k}
   =\mathcal O_r-X_r=(1-\mathcal H_r)\mathcal O_r,
 \nonumber\\
 \mathcal O_r^{(5)}
 &=\rho\,\mathcal P_{r-2}X_{r-2}
   =\frac{\rho}{4r}\mathcal P_{r-2}\Box_z\mathcal O_r
     \qquad(r\ge2).
 \label{eq:ht4symmetric}
\end{align}
We can also extract $X_{r-2k}$ directly from Eq.~\eqref{eq:ht4harmonicexp}. Taking $k$ metric traces and projecting the remaining indices onto their metric-traceless part gives
\begin{equation}
 X_{r-2k}
 =\frac{(r-2k+1)!}{4^k k!\,(r-k+1)!}\,
    \mathcal H_{r-2k}\Box_z^k\mathcal O_r ,
 \qquad 0\le k\le\lfloor r/2\rfloor .
 \label{eq:ht4layer}
\end{equation}
Let $j=r-2k$. Within $\rho^k X_j$, the gamma-traceless part of $X_j$ gives twist $3+2k$, and its gamma trace fixes the twist-$(4+2k)$ component when $j\ge1$. Appendix~\ref{app:ht4project} derives the operators with both kinds of twists
\begin{align}
 \mathcal O_r^{(3+2k)}
 &=\rho^k\mathcal P_jX_j
 =\frac{\rho^k(j+1)!}{4^k k!(r-k+1)!}\mathcal P_j\Box_z^k\mathcal O_r,
 \nonumber\\
 \mathcal O_r^{(4+2k)}
 &=\rho^k\frac{\slashed z\slashed{\partial}_z}{2(j+1)}X_j
 =\frac{\rho^k j!}{2\cdot4^k k!(r-k+1)!}
    \slashed z\slashed{\partial}_z\mathcal H_j\Box_z^k\mathcal O_r .
\end{align}
The first line holds for $0\le k\le\lfloor r/2\rfloor$ and the second for $0\le k\le\lfloor(r-1)/2\rfloor$, since $X_0$ has no vector index to contract with a gamma matrix.

To evaluate the the $\mathcal O_r^{(4)}$ and $\mathcal O_r^{(\ge5)}$ in Eq.~\eqref{eq:ht4symmetric}, we expand $\mathcal H_r$ using Eq.~\eqref{eq:bdaHarmonic}. In the twist-4 term, $\slashed{\partial}_z$ differentiates both $\rho^k$ and $\Box_z^k\mathcal O_r$. Using $\slashed{\partial}_z\rho^k=2k\rho^{k-1}\slashed z$ for $k\ge1$ and $\slashed z\slashed z=\rho$, we obtain
\begin{equation}
 \frac{\slashed z\slashed{\partial}_z}{2}
     \big[\rho^k\Box_z^k\mathcal O_r\big]
 =\rho^k\left(k+\frac{\slashed z\slashed{\partial}_z}{2}\right)
       \Box_z^k\mathcal O_r .
 \label{eq:ht4gammaproduct}
\end{equation}
Using Eq.~\eqref{eq:ht4gammaproduct} for the twist-4 component and expanding $1-\mathcal H_r$ for the metric traces, we obtain at arbitrary $z$
\begin{align}
 \mathcal O_r^{(4)}(z)
 &=\sum_{k=0}^{\lfloor r/2\rfloor}
    \frac{(r-k)!\,(-\rho)^k}{4^k k!\,(r+1)!}
    \left(k+\frac{\slashed z\slashed{\partial}_z}{2}\right)\Box_z^k\mathcal O_r(z),
 \nonumber\\
 \mathcal O_r^{(\ge5)}(z)
 &=-\sum_{k=1}^{\lfloor r/2\rfloor}
    \frac{(r-k)!\,(-\rho)^k}{4^k k!\,r!}\,\Box_z^k\mathcal O_r(z).
 \label{eq:ht4finitepolys}
\end{align}
For the spatial channel, we take the derivatives in Eq.~\eqref{eq:ht4finitepolys} and then set $z=\lambda$. For the symmetric part of the temporal operator, Eq.~\eqref{eq:ht4raw} calls for one more derivative, $\partial_{v_t}=v_t\cdot\partial_z$, and a factor of $1/r$. At $z=\lambda$, $(-\rho)^k=1$ and $\partial_{v_t}(-\rho)^k=0$ because $v_t\cdot\lambda=0$. The derivative acts on $\Box_z^k\mathcal O_r$ in both terms and also on the explicit $\slashed z$ in the twist-4 term, where it gives $\slashed v_t$. Suppressing the fixed indices $(n_1,n_2)$, we obtain
\begin{align}
 \mathcal O_{\gamma^z,{\rm S}}^{(4)}
 &=\sum_{k=0}^{\lfloor r/2\rfloor}\frac{(r-k)!}{4^k k!\,(r+1)!}
   \left.\left(k+\frac{\slashed\lambda\slashed{\partial}_z}{2}\right)
   \Box_z^k\mathcal O_r\right|_{z=\lambda},\label{eq:ht4localzfour}\\
 \mathcal O_{\gamma^t,{\rm S}}^{(4)}
 &=\sum_{k=0}^{\lfloor r/2\rfloor}\frac{(r-k)!}{4^k k!\,r(r+1)!}
   \left.\left(k\partial_{v_t}+\frac{\slashed v_t\slashed{\partial}_z}{2}
      +\frac{\slashed\lambda\slashed{\partial}_z\partial_{v_t}}{2}\right)
       \Box_z^k\mathcal O_r\right|_{z=\lambda},\label{eq:ht4localtfour}\\
 \mathcal O_{\gamma^z,{\rm S}}^{(\ge5)}
 &=-\sum_{k=1}^{\lfloor r/2\rfloor}\frac{(r-k)!}{4^k k!\,r!}
   \left.\Box_z^k\mathcal O_r\right|_{z=\lambda},\label{eq:ht4localzmetric}\\
 \mathcal O_{\gamma^t,{\rm S}}^{(\ge5)}
 &=-\sum_{k=1}^{\lfloor r/2\rfloor}\frac{(r-k)!}{4^k k!\,r\,r!}
   \left.\partial_{v_t}\Box_z^k\mathcal O_r\right|_{z=\lambda}.
   \label{eq:ht4localtmetric}
\end{align}
Equations~\eqref{eq:bdaChannels} and \eqref{eq:ht4localzfour}--\eqref{eq:ht4localtmetric} give the symmetric terms in both channels. Including the mixed-symmetry term from Eq.~\eqref{eq:ht4raw} in the temporal channel gives the full local operators
\begin{equation}
 \begin{aligned}
 \mathcal O_{\gamma^z}^{\rm full}
 &=\mathcal O_{\gamma^z}^{(3)}+\mathcal O_{\gamma^z,{\rm S}}^{(4)}
         +\mathcal O_{\gamma^z,{\rm S}}^{(\ge5)},\\
 \mathcal O_{\gamma^t}^{\rm full}
 &=\mathcal O_{\gamma^t}^{(3)}+\mathcal O_{\gamma^t,{\rm S}}^{(4)}
         +\mathcal O_{\gamma^t,{\rm S}}^{(\ge5)}+\mathcal O_{\gamma^t,{\rm M}},
 \qquad
 \mathcal O_{\gamma^t,{\rm M}}=\frac{v_t^\mu\lambda^\nu}{r}B_{\mu\nu}(\lambda).
 \end{aligned}
 \label{eq:ht4channelbalance}
\end{equation}
Equation~\eqref{eq:ht4mixedproject} in Appendix~\ref{app:ht4mixed} separates the mixed temporal operator $\mathcal O_{\gamma^t,{\rm M}}$ into its twist-4 component and the sum of its components of twist five and higher.

\subsection{Nonlocal operators}
\label{sec:ht4gen}

We sum the local operators in Eq.~\eqref{eq:ht4channelbalance} over $(n_1,n_2)$ to reconstruct the nonlocal operators at quark separations $\ell_1$ and $\ell_2$. Let $\mathcal V_\mu(\ell_1,\ell_2;z)$ denote the operator in Eq.~\eqref{eq:bdaThreeQuark} with the diquark matrix $\slashed w$ replaced by $\gamma_\mu$ and the separation direction set to $v=z$. The component of twist $\tau$ in $\mathcal O(\ell_1,\ell_2;z)=z^\mu\mathcal V_\mu(\ell_1,\ell_2;z)$ is
\begin{equation}
 \mathcal O^{(\tau)}(\ell_1,\ell_2;z)
 =\sum_{n_1,n_2\ge0}
  \frac{(-i\ell_1)^{n_1}(-i\ell_2)^{n_2}}{n_1!n_2!}\,
  \mathcal O_{n_1,n_2}^{(\tau)}(z).
 \label{eq:ht4gen}
\end{equation}
To perform the sum in Eq.~\eqref{eq:ht4gen}, we write the factorial ratios in Eq.~\eqref{eq:ht4finitepolys} as integrals. Set $N=n_1+n_2=r-1$. The identity $\int_0^1d\alpha\,\alpha^p(1-\alpha)^q=p!q!/(p+q+1)!$, with $p=r-k$ and $q=k$ or $k-1$, gives
\begin{align}
 \frac{(r-k)!}{k!(r+1)!}
 &=\frac{1}{(k!)^2}\int_0^1 d\alpha\,\alpha^{N}\alpha^{1-k}(1-\alpha)^k,
 \nonumber\\
 \frac{(r-k)!}{k!r!}
 &=\frac{1}{k!(k-1)!}\int_0^1 d\alpha\,\alpha^{N}\alpha^{1-k}(1-\alpha)^{k-1}
 \qquad(k\ge1).
 \label{eq:ht4beta}
\end{align}
Multiplying the weight in Eq.~\eqref{eq:ht4gen} by $\alpha^N$ rescales both quark separations by the same factor, since
\begin{equation}
 \alpha^N\frac{(-i\ell_1)^{n_1}(-i\ell_2)^{n_2}}{n_1!n_2!}
 =\frac{(-i\alpha\ell_1)^{n_1}(-i\alpha\ell_2)^{n_2}}{n_1!n_2!}.
 \label{eq:ht4distweights}
\end{equation}
The weights in Eq.~\eqref{eq:ht4distweights} are independent of $z$, so $\Box_z^k$ commutes with the sum over $(n_1,n_2)$. At fixed $k$ the sum gives $\Box_z^k\mathcal O(\alpha\ell_1,\alpha\ell_2;z)$. Equation~\eqref{eq:ht4finitepolys} then gives
\begin{align}
 \mathcal O^{(4)}(\ell_1,\ell_2;z)
 &=\sum_{k\ge0}\frac{(-z^2)^k}{4^k(k!)^2}
   \int_0^1d\alpha\,\alpha^{1-k}(1-\alpha)^k
   \left(k+\frac{\slashed z\slashed{\partial}_z}{2}\right)\Box_z^k
   \mathcal O(\alpha\ell_1,\alpha\ell_2;z),\nonumber\\
 \mathcal O^{(\ge5)}(\ell_1,\ell_2;z)
 &=-\sum_{k\ge1}\frac{(-z^2)^k}{4^kk!(k-1)!}
   \int_0^1d\alpha\,\alpha^{1-k}(1-\alpha)^{k-1}\Box_z^k
   \mathcal O(\alpha\ell_1,\alpha\ell_2;z).
 \label{eq:ht4gensym}
\end{align}
We now take the spatial and temporal contractions in Eq.~\eqref{eq:ht4raw}. Both terms of the temporal operator carry $1/r=\int_0^1d\beta\,\beta^N$. As in Eq.~\eqref{eq:ht4distweights}, multiplying the weight by $\beta^N$ rescales both separations by $\beta$. The nonlocal operators are
\begin{align}
 \widetilde{\mathcal O}_{\gamma^z,{\rm S}}^{(\tau)}(\ell_1,\ell_2)
 &=\mathcal O^{(\tau)}(\ell_1,\ell_2;\lambda),\nonumber\\
 \widetilde{\mathcal O}_{\gamma^t,{\rm S}}^{(\tau)}(\ell_1,\ell_2)
 &=\left.\int_0^1d\beta\, (v_t\cdot\partial_z)
                   \mathcal O^{(\tau)}(\beta\ell_1,\beta\ell_2;z)\right|_{z=\lambda},\nonumber\\
 \widetilde{\mathcal O}_{\gamma^t,{\rm M}}(\ell_1,\ell_2)
 &=v_t^\mu\lambda^\nu\int_0^1d\beta\,
   \left[\partial_\nu\mathcal V_\mu(\beta\ell_1,\beta\ell_2;z)
        -\partial_\mu\mathcal V_\nu(\beta\ell_1,\beta\ell_2;z)\right]_{z=\lambda}.
 \label{eq:ht4genchannels}
\end{align}
In the first two lines of Eq.~\eqref{eq:ht4genchannels}, the superscripts $4$ and $\ge5$ denote the twist-4 component and the sum over twists five and higher, respectively. The derivatives with respect to $z$ are taken at fixed separations and fixed integration variables before setting $z=\lambda$. Substituting Eq.~\eqref{eq:ht4gensym} into the second line gives quark separations $(\alpha\beta\ell_1,\alpha\beta\ell_2)$. The mixed term in the third line involves only $(\beta\ell_1,\beta\ell_2)$ and includes all its twist components. Appendix~\ref{app:ht4mixed} separates its twist-4 component at each rank $r$.

Equation~\eqref{eq:wcfiveterms} in Appendix~\ref{app:ht4convolutions} expresses $\tfrac12\Box_z\mathcal O(L_1,L_2;z)$ as five Wilson-line integrals at arbitrary $z$. For the terms with $k\ge1$ in Eq.~\eqref{eq:ht4gensym}, we set $L_i=\alpha\ell_i$ and use
\[
 \Box_z^k\mathcal O(L_1,L_2;z)
 =2\Box_z^{k-1}\!\left[\tfrac12\Box_z\mathcal O(L_1,L_2;z)\right],
\]
with Eq.~\eqref{eq:wcfiveterms} in the brackets. The $k=0$ twist-4 term requires only the first derivative in Eq.~\eqref{eq:wcudsfirst}. For the symmetric temporal operator in Eq.~\eqref{eq:ht4genchannels}, we use $L_i=\alpha\beta\ell_i$ and take the temporal derivative before setting $z=\lambda$. For the mixed temporal operator, we apply Eq.~\eqref{eq:wconeline} to the two displaced quark lines of $\mathcal V_\mu$ at $L_i=\beta\ell_i$ and take the antisymmetric combination in Eq.~\eqref{eq:ht4genchannels}. Similar Wilson-line insertions occur in the nucleon quasi-PDF and pion quasi-DA~\cite{Chen:2016utp,Zhang:2017bzy}. Each additional trace introduces more integrals over insertion positions, making a term-by-term lattice calculation impractical. Section~\ref{sec:higherlattice} therefore constructs the same derivatives from finite differences of full three-quark correlators.

\subsection{Lattice construction}
\label{sec:higherlattice}

To evaluate the five matrix elements in Eq.~\eqref{eq:ht4genchannels} on the lattice, we use the same three-quark correlator and shift one quark endpoint at a time.  Eqs.~\eqref{eq:ht4gensym} and \eqref{eq:ht4genchannels} give the signed quark separations
\[
 L_i=\begin{cases}
       \alpha\ell_i       & \text{for }\widetilde{\mathcal O}_{\gamma^z,\mathrm S}^{(\tau)},\\
       \alpha\beta\ell_i  & \text{for }\widetilde{\mathcal O}_{\gamma^t,\mathrm S}^{(\tau)},\\
       \beta\ell_i        & \text{for }\widetilde{\mathcal O}_{\gamma^t,\mathrm M},
      \end{cases}
 \qquad i=1,2,\qquad 0\le\alpha,\beta\le1 ,
\]
The label $\mathrm S$ means that the Lorentz index of the diquark gamma matrix is symmetrized with all the derivative indices, as in Eq.~\eqref{eq:ht4curl}. The label $\mathrm M$ denotes the remaining part of the temporal operator. Here $\tau=4$ or $\ge5$. On a lattice of spacing $a$, we measure the correlator at endpoints $X_i=L_i\lambda_E$, with $\lambda_E=-e_3$ and integer $L_i/a$, and at neighboring sites for the finite differences. Intermediate lengths in the $\alpha,\beta$ integrals can be obtained by interpolation.

We use Euclidean indices $A,B=1,2,3,4$ and the continuation $z_M^0=-iz_E^4$, $z_M^j=z_E^j$, $\gamma_E^4=\gamma_M^0$ and $\gamma_E^j=-i\gamma_M^j$ for $j=1,2,3$. The index $B$ labels the gamma matrix in the diquark. The Euclidean operator is
\begin{equation}
 V^E_B(X_1,X_2)=\epsilon^{ijk}[W(0,X_1)u(X_1)]_i^T C_E\gamma_{5,E}\gamma_E^B
                  [W(0,X_2)d(X_2)]_j s_k(0).
 \label{eq:ht4euV}
\end{equation}
We extract the vacuum-to-baryon matrix element from the ground-state contribution to the correlator of $V_B^E$ with a baryon source of momentum $P_z$, retaining the open spinor index. Each Wilson line transports its quark from the displaced endpoint to the origin. Varying $z_E$ at fixed $L_1$ and $L_2$ moves both endpoints. The chain rule therefore gives
\begin{align}
 \frac{\partial}{\partial z_E^A}V^E_B(L_1z_E,L_2z_E)
 &=\left.\left(L_1\frac{\partial}{\partial X_1^A}
           +L_2\frac{\partial}{\partial X_2^A}\right)V^E_B(X_1,X_2)
     \right|_{X_i=L_i z_E}.
 \label{eq:ht4chain}
\end{align}
We take the spacing $a$ to be the same in all four Euclidean directions. For a displacement along $e_A$, choose $h_A=m_A a$ with $m_A$ a positive integer. Replacing each endpoint derivative in Eq.~\eqref{eq:ht4chain} by a centered difference gives
\begin{align}
 \nabla_A V^E_B(X_1,X_2)
 ={}&\frac{L_1}{2h_A}\big[V^E_B(X_1+h_Ae_A,X_2)-V^E_B(X_1-h_Ae_A,X_2)\big]\nonumber\\
 &+\frac{L_2}{2h_A}\big[V^E_B(X_1,X_2+h_Ae_A)-V^E_B(X_1,X_2-h_Ae_A)\big].
 \label{eq:ht4difference}
\end{align}
When a quark endpoint moves, its Wilson line must connect the origin to the new site. In the continuum, the straight segment to an endpoint $X$ is $x(u)=uX$, $0\le u\le1$. Shifting $X$ by $\delta X$ displaces the point $x(u)$ by $u\,\delta X$. With the convention $[D_\mu,D_\nu]=-ig_sG_{\mu\nu}$, the resulting derivative of the transported field is
\begin{align}
 \partial_{X^\mu}[W(0,X)q(X)]
 ={}&W(0,X)D_\mu q(X)\nonumber\\
 &-ig_s\int_0^1du\,u\,W(0,uX)X^\nu G_{\mu\nu}(uX)W(uX,X)q(X).
 \label{eq:ht4wilson}
\end{align}
The first term in Eq.~\eqref{eq:ht4wilson} contains the covariant derivative of the quark field at the endpoint. The field-strength integral accounts for the change along the line. Each term in Eq.~\eqref{eq:ht4difference} therefore uses the Wilson line appropriate to its shifted endpoint.

\begin{figure}
\centering
\begin{tikzpicture}[x=.60cm,y=.60cm,font=\small,
   gridline/.style={draw=black!22,line width=.26pt},
   links/.style={draw=blue!60!teal,line width=1.35pt,line join=round,line cap=round},
   target/.style={draw=black!85,dash pattern=on 3pt off 2pt,line width=.7pt}]
 \foreach \panel/\spacing/\count in {0/1/2,1/.5/4}{
   \begin{scope}[shift={({11*\panel},0)}]
    \foreach \j in {0,...,16}{
      \pgfmathsetmacro{\xx}{\j*\spacing}
      \ifdim\xx pt<8.01pt \draw[gridline] (\xx,-2) -- (\xx,2); \fi
    }
    \foreach \j in {0,...,8}{
      \pgfmathsetmacro{\yy}{-2+\j*\spacing}
      \ifdim\yy pt<2.01pt \draw[gridline] (0,\yy) -- (8,\yy); \fi
    }
    \draw[black!48,line width=.5pt] (0,0) -- (8,0);
    \foreach \sgn in {-1,1}{
      \draw[links] (0,0)
       \foreach \j in {1,...,\count}{
          -- ({8*(\j-.5)/\count},{\sgn*(\j-1)*\spacing})
          -- ({8*(\j-.5)/\count},{\sgn*\j*\spacing})}
       -- (8,{\sgn*2});
      \draw[target] (0,0) -- (8,{\sgn*2});
      \fill[black] (8,{\sgn*2}) circle[radius=1.6pt];
    }
    \fill[black] (0,0) circle[radius=1.6pt] node[left=2pt] {$0$};
    \draw[fill=white] (8,0) circle[radius=1.6pt] node[right=2pt] {$X$};
    \node[right=2pt] at (8,2) {$X_+$};
    \node[right=2pt] at (8,-2) {$X_-$};
    \ifnum\panel=0
      \draw[<->,thin] (9.55,.04) -- (9.55,1.96) node[midway,right=1pt] {$h$};
      \node at (4,2.72) {(a) \quad $a,\quad h=2a$};
    \else
      \node at (4,2.72) {(b) \quad $a'=a/2,\quad h=4a'$};
    \fi
   \end{scope}
 }
 \draw[links] (3.2,-3.05) -- (4.4,-3.05)
      node[right=3pt,text=black] {lattice links};
 \draw[target] (10.6,-3.05) -- (11.8,-3.05)
      node[right=3pt,text=black] {continuum straight line};
\end{tikzpicture}
\caption{Lattice Wilson lines for a transverse endpoint derivative. The blue paths connect the origin to $X_\pm=X\pm h e_A$, with $X=L_i\lambda_E$ and $h=h_A$. The dashed lines mark the corresponding straight segments. Both panels show the same physical endpoints, with the other quark held fixed. The transverse displacement $h$ spans two links at spacing $a$ in panel (a) and four at spacing $a/2$ in panel (b). The two panels illustrate the continuum extrapolation at fixed $h$.}
\label{fig:ht4offaxis}
\end{figure}
Figure~\ref{fig:ht4offaxis} shows the two transverse displacements $X_\pm=X\pm h_A e_A$ of one endpoint, with the other endpoint held fixed. We approximate each straight segment from $0$ to $X_\pm$ by a staircase of $M$ elementary links. The successive lattice sites satisfy
\begin{align}
 x_0=0,\qquad x_M=X_\pm,\qquad
 x_{j+1}-x_j=\sigma_j a e_{\mu_j},
 \qquad \mu_j\in\{1,2,3,4\},\quad \sigma_j=\pm1 .
\end{align}
Each link has length $a$, while $h_A=m_A a$ measures the displacement from the unshifted endpoint $X$.
Let $U_\mu(x)$ denote the link variable on the bond from $x$ to $x+a e_\mu$. The factors for steps in the positive and negative directions are
\begin{align}
 U_{+\mu}(x)=U_\mu(x),\qquad
 U_{-\mu}(x)=U_\mu^\dagger(x-a e_\mu).
\end{align}
Their ordered product gives the lattice Wilson line
\begin{align}
 W_a(0,X_\pm)
 =U_{\sigma_0\mu_0}(x_0)\,U_{\sigma_1\mu_1}(x_1)\cdots
  U_{\sigma_{M-1}\mu_{M-1}}(x_{M-1}).
\end{align}
A Bresenham-type rule selects the sites along each segment. For opposite transverse displacements, one may use paths related by reflection~\cite{Musch:2010ka}. At fixed physical endpoints, reducing $a$ gives a staircase that follows the straight segment more closely.

Applying the centered difference in Eq.~\eqref{eq:ht4difference} successively gives higher derivatives. For fixed directions $A$ and $B$, write $V_{p,q}=V_B^E(X_1+p h_Ae_A,X_2+q h_Ae_A)$ for integer offsets $p,q$. Two applications in direction $A$ give
\begin{align}
 (\nabla_A^2 V_B^E)(X_1,X_2)
 ={}&\frac{L_1^2}{4h_A^2}(V_{2,0}-2V_{0,0}+V_{-2,0})\nonumber\\
 &+\frac{L_1 L_2}{2h_A^2}(V_{1,1}-V_{1,-1}-V_{-1,1}+V_{-1,-1})\nonumber\\
 &+\frac{L_2^2}{4h_A^2}(V_{0,2}-2V_{0,0}+V_{0,-2}) .
 \label{eq:ht4multisecond}
\end{align}

Let $M^R_{B,a}(X_1,X_2)$ denote the ground-state matrix element of Eq.~\eqref{eq:ht4euV} after matching the lattice operator to the continuum straight-line operator in a common renormalization scheme. This matching must account for Wilson-line self-energies and operator mixing for the chosen paths. Even for a quark bilinear, the mass counterterm for a diagonal staircase can differ from its on-axis value~\cite{Green:2020xco}. We first take the continuum limit at fixed $P_z$, $L_i$, and physical displacement $h_A$, as illustrated in Fig.~\ref{fig:ht4offaxis}. We then repeat the extrapolation at smaller $h_A$ to obtain the derivative,
\begin{equation}
 \left.\partial_{z_E^A}^2 M_B^R(L_1z_E,L_2z_E)\right|_{z_E=\lambda_E}
 =
 \lim_{h_A\to0}\,
 \lim_{\substack{a\to0\\m_A a\to h_A}}
 \left.(\nabla_A^2 M^R_{B,a})(X_1,X_2)\right|_{X_i=L_i\lambda_E},
 \label{eq:ht4multilimits}
\end{equation}
where $M_B^R$ denotes the continuum straight-line matrix element. In the inner limit, the displacement on each lattice is $m_A a$, chosen to approach the prescribed $h_A$. After this limit, the centered difference has an $O(h_A^2)$ truncation error for a sufficiently smooth matrix element. The outer limit removes this error, provided the derivative exists after any required contact subtractions. For mixed derivatives, all displacement lengths are held fixed in physical units during the continuum extrapolation.

For a transverse shift, the staircase in Fig.~\ref{fig:ht4offaxis} contains $|L_i|/a$ longitudinal links and $h_A/a$ transverse links. When only a few longitudinal links span $|L_i|$, there are few sites at which to place the transverse steps. Resolving such short separations requires finer lattices. At fixed physical four-volume the number of lattice sites grows as $a^{-4}$, and Monte Carlo autocorrelations can further increase the cost per independent measurement~\cite{Schaefer:2010hu}. With a fixed computational budget, the finest lattices may therefore have insufficient statistics to resolve the differences in Eqs.~\eqref{eq:ht4difference} and \eqref{eq:ht4multisecond}. Their statistical errors depend on the correlations among the shifted matrix elements.

\subsection{Power counting}
\label{sec:higherpowercounting}

We take the large-$P_z$ limit at fixed derivative counts $(n_1,n_2)$ and fixed renormalization scale, so the rank $r=n_1+n_2+1$ is fixed. Consider first the symmetric operators in Eq.~\eqref{eq:ht4symmetric}. Factoring out $\slashed z$ from the twist-4 operator leaves a gamma-traceless tensor-spinor of rank $r-1$. Factoring out $z^2$ from the twist-5 operator leaves one of rank $r-2$. Restoring these factors gives the vacuum-to-baryon matrix elements of the two operators,
\begin{align}
 \langle0|\mathcal O_r^{(4)}(z)|\Lambda(P,\sigma)\rangle
  &=f_\Lambda^{\rm LC}h^{(4)}\slashed z\,\mathcal Q_{r-1}(z,P)u_\Lambda(P,\sigma),\nonumber\\
 \langle0|\mathcal O_r^{(5)}(z)|\Lambda(P,\sigma)\rangle
  &=f_\Lambda^{\rm LC}h^{(5)}z^2\mathcal Q_{r-2}(z,P)u_\Lambda(P,\sigma)
   \qquad(r\ge2).
 \label{eq:ht4me}
\end{align}
The polynomials $\mathcal Q_{r-1}$ and $\mathcal Q_{r-2}$ in Eq.~\eqref{eq:ht4me} follow from the arbitrary-rank expression derived in Eqs.~\eqref{eq:bdaGeneralME} and \eqref{eq:bdaCovariantCoefficients}. The reduced moments $h^{(4)}=h^{(4)}_{n_1,n_2}$ and $h^{(5)}=h^{(5)}_{n_1,n_2}$ have mass dimensions one and two, respectively. At $z=n$, the twist-4 term vanishes under the light-cone projection because $\Pi_+\slashed n=0$, and the twist-5 term vanishes because $n^2=0$. The light-cone moment therefore fixes neither reduced moment.

For twist four, the spin contraction in Eq.~\eqref{eq:bdaQuasiMoment} includes the explicit $\slashed z$ from Eq.~\eqref{eq:ht4me}. Using $\slashed z\slashed z=z^2$ and the spin sums $\sum_\sigma\bar u_\Lambda\slashed n u_\Lambda=4p$ and $\sum_\sigma\bar u_\Lambda\slashed n\slashed z u_\Lambda=4m_\Lambda(n\cdot z)$, we obtain
\begin{equation}
 \frac{1}{4p}\sum_\sigma\bar u_\Lambda\slashed n\slashed z\,
     \mathcal Q_{r-1}(z,P)u_\Lambda
 =m_\Lambda\left[\frac{n\cdot z}{p}A_{r-1}(P\cdot z,z^2)
                      +z^2 B_{r-1}(P\cdot z,z^2)\right].
 \label{eq:ht4spin}
\end{equation}
For the spatial channel, we set $z=\lambda$ in Eq.~\eqref{eq:ht4spin}. For the symmetric part of the temporal channel, we use the directional derivative in Eq.~\eqref{eq:reorgtimepathderivative}. We set $z_\epsilon=\lambda+\epsilon v_t$ at fixed $P$, differentiate Eq.~\eqref{eq:ht4spin} with respect to $\epsilon$, and divide by $r$. With the kinematics of Eq.~\eqref{eq:tmPath}, $P\cdot z_\epsilon=P_z+\epsilon E$, $z_\epsilon^2=-1+\epsilon^2$, and $(n\cdot z_\epsilon)/p=(\epsilon-1)/(E+P_z)$. Their derivatives at $\epsilon=0$ are $E$, $0$, and $1/(E+P_z)$, respectively.  With the normalization of Eq.~\eqref{eq:bdaQuasiMoment}, the two contributions are
\begin{align}
 \widetilde{\mathcal M}_{\gamma^z,{\rm S}}^{(4,0)}
 &=\frac{m_\Lambda h^{(4)}}{P_z^r}
    \left[-\frac{A_{r-1}(P_z,-1)}{E+P_z}-B_{r-1}(P_z,-1)\right],\nonumber\\
 \widetilde{\mathcal M}_{\gamma^t,{\rm S}}^{(4,0)}
 &=\frac{m_\Lambda h^{(4)}}{rP_z^r}
    \left[\frac{A_{r-1}(P_z,-1)-E A'_{r-1}(P_z,-1)}{E+P_z}
          -E B'_{r-1}(P_z,-1)\right].
 \label{eq:ht4finite}
\end{align}
The second superscript, $0$, denotes tree level. A prime on $A_{r-1}$ or $B_{r-1}$ denotes differentiation with respect to its first argument at fixed second argument. At fixed rank, Eq.~\eqref{eq:bdaCovariantCoefficients} gives
$A_{r-1}(P_z,-1)=P_z^{r-1}+O(m_\Lambda^2 P_z^{r-3})$ and
$B_{r-1}(P_z,-1)=-(r-1)P_z^{r-2}/(2r)+O(m_\Lambda^2 P_z^{r-4})$.
Substituting these expansions and $E=P_z+O(m_\Lambda^2/P_z)$ into Eq.~\eqref{eq:ht4finite} yields
\begin{align}
 \widetilde{\mathcal M}_{\gamma^z,{\rm S}}^{(4,0)}
 &=\frac{m_\Lambda h^{(4)}}{P_z^2}
     \left[-\frac12+\frac{r-1}{2r}\right]+O(P_z^{-4})
  =-\frac{m_\Lambda h^{(4)}}{2rP_z^2}+O(P_z^{-4}),\nonumber\\
 \widetilde{\mathcal M}_{\gamma^t,{\rm S}}^{(4,0)}
 &=\frac{m_\Lambda h^{(4)}}{rP_z^2}
     \left[\frac{2-r}{2}+\frac{(r-1)(r-2)}{2r}\right]+O(P_z^{-4})
  =\frac{(2-r)m_\Lambda h^{(4)}}{2r^2P_z^2}+O(P_z^{-4}).
 \label{eq:ht4brackets}
\end{align}

For twist five, the spin contraction in Eq.~\eqref{eq:tmSpin} reduces the second line of Eq.~\eqref{eq:ht4me} to $h^{(5)}z^2 S_{r-2}(z)$ after division by $4f_\Lambda^{\rm LC}p$. At $z=\lambda$ we have $z^2=-1$ and $(v_t\cdot\partial_z)z^2=0$, so the temporal derivative acts only on $S_{r-2}$. Equation~\eqref{eq:tmRatios} then gives
\begin{align}
 \widetilde{\mathcal M}_{\gamma^z,{\rm S}}^{(5,0)}
 &=-\frac{h^{(5)}}{P_z^r}S_{r-2}(\lambda)
  =-\frac{h^{(5)}}{P_z^2}F_{r-2}(c) \qquad(r\ge2),
 \label{eq:ht4fivez}\\
 \widetilde{\mathcal M}_{\gamma^t,{\rm S}}^{(5,0)}
 &=-\frac{h^{(5)}}{rP_z^r}
    \left.\frac{d S_{r-2}(z_\epsilon)}{d\epsilon}\right|_{\epsilon=0}
  =-\frac{r-2}{r}\frac{h^{(5)}}{P_z^2}T_{r-2}(c) \qquad(r\ge3).
 \label{eq:ht4fivet}
\end{align}
At $r=2$, $S_0=F_0=1$ and the temporal twist-5 contribution vanishes. For $r\ge3$, $F_{r-2}(0)=T_{r-2}(0)=1$, so Eqs.~\eqref{eq:ht4fivez} and \eqref{eq:ht4fivet} give the leading $1/P_z^2$ coefficients in both channels. Each additional metric trace in Eq.~\eqref{eq:ht4families} reduces the remaining tensor rank by two and supplies a further factor of $1/P_z^2$ in the normalized moment.

The temporal channel also contains the mixed-symmetry term in Eq.~\eqref{eq:ht4raw}. For $r\ge2$, Appendix~\ref{app:ht4mixedpower} derives the large-$P_z$ expansion of its normalized moment,
\begin{equation}
 \widetilde{\mathcal M}^{\rm M,0}_{\gamma^t;n_1,n_2}
 =\frac{H^{\rm M}_{n_1,n_2}}{P_z^2}+O(P_z^{-4}),\qquad
 H^{\rm M}_{n_1,n_2}=\lim_{P_z\to\infty}P_z^2
                 \widetilde{\mathcal M}^{\rm M,0}_{\gamma^t;n_1,n_2}.
 \label{eq:ht4mixedpower}
\end{equation}
Equation~\eqref{eq:ht4mixedrestcoef} expresses $H^{\rm M}_{n_1,n_2}$ in terms of an independent rest-frame matrix element of $B_{\mu\nu}(n)$. If this coefficient vanishes, the mixed moment is $O(P_z^{-4})$.

Combining Eqs.~\eqref{eq:ht4brackets}--\eqref{eq:ht4mixedpower} with the exact twist-3 factors in Eq.~\eqref{eq:tmRatios}, we find the full tree-level moments in the light-cone normalization,
\begin{align}
 \widetilde{\mathcal M}_{\gamma^z;n_1,n_2}^{\rm full,0}
 &=F_r(c)\mathcal M_{n_1,n_2}
   -\frac{1}{P_z^2}\left(\frac{m_\Lambda h^{(4)}}{2r}+h^{(5)}\right)
   +O(P_z^{-4}),\nonumber\\
 \widetilde{\mathcal M}_{\gamma^t;n_1,n_2}^{\rm full,0}
 &=T_r(c)\mathcal M_{n_1,n_2}
   +\frac{1}{P_z^2}\left[\frac{(2-r)m_\Lambda h^{(4)}}{2r^2}
          -\frac{r-2}{r}h^{(5)}+H^{\rm M}_{n_1,n_2}\right]+O(P_z^{-4}).
 \label{eq:ht4leading}
\end{align}
The terms proportional to $\mathcal M_{n_1,n_2}$ include all twist-3 target-mass corrections. The coefficients of the explicit $1/P_z^2$ terms come from higher-twist matrix elements.

To determine the full local normalization, we set $r=1$ and use $\mathcal M_{0,0}=1$. At this rank the metric trace and the mixed operator are absent, so we set $h^{(5)}_{0,0}=H^{\rm M}_{0,0}=0$ in Eq.~\eqref{eq:ht4leading}. Adding the exact twist-4 moments in Eq.~\eqref{eq:ht4finite}, evaluated at $A_0=1$ and $B_0=0$, to the projected zeroth moments of Sec.~\ref{sec:projectednormalization} gives
\begin{equation}
 \widetilde{\mathcal M}^{\rm full,0}_{\gamma^z;0,0}
 =F_1(c)-\frac{m_\Lambda h^{(4)}_{0,0}}{P_z(E+P_z)},\qquad
 \widetilde{\mathcal M}^{\rm full,0}_{\gamma^t;0,0}
 =T_1(c)+\frac{m_\Lambda h^{(4)}_{0,0}}{P_z(E+P_z)} .
 \label{eq:ht4localnorm}
\end{equation}
To obtain unit normalization for the axial amplitudes, we divide each full moment in Eq.~\eqref{eq:ht4leading} by the zeroth moment of the same channel in Eq.~\eqref{eq:ht4localnorm}.

\section{Matching and mass subtraction}
\label{sec:matchingandsubtraction}

At finite $P_z$, the twist-3 quasi-DA receives the tree-level target-mass corrections derived in Sec.~\ref{sec:targetmassCorr} for the spatial and temporal channels. To extract the LCDA from lattice data, we combine these corrections with perturbative matching and account for higher-twist contributions to the measured quasi-DA. We retain all powers of $c=m_\Lambda^2/(4P_z^2)$ at tree level and the leading power in $c$ at one loop. We use $a_s=\alpha_s(\mu)/(4\pi)$, where $\mu$ is the factorization scale of the LCDA. The equal-time operator is renormalized at scale $\mu_R$.

\subsection{Renormalized twist-3 operators}
\label{sec:renormalizedOPE}

At short separations the twist-3 component of the equal-time operator in Eq.~\eqref{eq:bdaEqualTime} has the expansion
\begin{equation}
 \widetilde{\mathcal O}_{\Gamma,R}^{(3)}(\ell_1,\ell_2;\mu_R)
 \sim \sum_{N=0}^{\infty}\sum_I
 \left[
   \mathcal C_{\Gamma,NI}^{(0)\,\rho_1\cdots\rho_r}
   +a_s\mathcal C_{\Gamma,NI}^{(1)\,\rho_1\cdots\rho_r}
   +O(a_s^2)
 \right]
 \mathcal O^{(3)}_{NI,R,\rho_1\cdots\rho_r}(\mu).
 \label{eq:match5ope}
\end{equation}
Here $N$ is the total number of derivatives, $r=N+1$ is the tensor rank, and $I$ labels the local operators at fixed $N$. The subscript $R$ denotes a renormalized operator. The Wilson tensors $\mathcal C_{\Gamma,NI}^{(i)}$ depend on $\ell_1,\ell_2,\mu_R,\mu$ and may act on the open spinor index. In the derivative-assignment basis, $I=(k_1,k_2,k_3)$ counts the derivatives on the $u$, $d$ and $s$ fields, respectively, with $k_1+k_2+k_3=N$. Since the strange quark is at the origin, only assignments of the form $I=(n_1,n_2,0)$ occur at tree level. Their Wilson tensors are
\begin{equation}
 \mathcal C_{\Gamma,NI}^{(0)\,\rho_1\cdots\rho_r}
 =\frac{(-i\ell_1)^{n_1}(-i\ell_2)^{n_2}}{n_1!\,n_2!}\,
     w_\Gamma^{\rho_1}\lambda^{\rho_2}\cdots\lambda^{\rho_r},
 \qquad w_{\gamma^z}=\lambda,\quad w_{\gamma^t}=v_t .
 \label{eq:match5tree}
\end{equation}
In a Lorentz-covariant, mass-independent continuum scheme, physical operators can mix if they have the same canonical dimension, Lorentz representation and internal quantum numbers. The canonical dimension $9/2+N$ fixes $N$ but allows derivatives on any of the three quark fields. Note that total derivatives of lower-rank operators can also contribute when projected onto the same final rank $r$. The sum over $I$ therefore runs over a complete basis at fixed $N$.

Consider first the matrix element of a single basis operator. Contract all $r$ indices with $z$ and write $u_\Lambda=u_\Lambda(P,\sigma)$. In the parity channel of Eq.~\eqref{eq:bdaGeneralME}, the matrix element is
\begin{equation}
 \langle0|\mathcal O^{(3)}_{NI,R}(z;\mu)|\Lambda(P,\sigma)\rangle
 = [A_{NI}+m_\Lambda\slashed z B_{NI}]u_\Lambda
 = a_{NI}\mathcal Q_r(z,P)u_\Lambda .
 \label{eq:match5ab}
\end{equation}
Homogeneity fixes the leading term $A_{NI}(\chi,0)=a_{NI}\chi^r$, where $\chi=P\cdot z$. The gamma-trace condition then fixes the remaining coefficients through the recursion in Eq.~\eqref{eq:bdaAppRecurrence}. This gives the same polynomial $\mathcal Q_r$ as in Eq.~\eqref{eq:bdaGeneralME}, up to the overall coefficient $a_{NI}$. To determine it, set $z=n$ and apply $\Pi_+$. With $p=P\cdot n$ and $\Pi_+\slashed n=0$, the light-cone component defines the reduced moment $\mathcal M_{N;I}(\mu)$ through
\begin{align}
 \Pi_+\langle0|\mathcal O^{(3)}_{NI,R}(n;\mu)|\Lambda(P,\sigma)\rangle
 &=a_{NI}p^r u_{\Lambda,+}
 \equiv f_\Lambda^{\rm LC}(\mu)\mathcal M_{N;I}(\mu)p^r u_{\Lambda,+}.
 \label{eq:match5lcfix}
\end{align}
Thus for arbitrary $z$ Eq.~\eqref{eq:match5ab} gives the matrix element
\begin{equation}
 \langle0|\mathcal O^{(3)}_{NI,R}(z;\mu)|\Lambda(P,\sigma)\rangle
 =f_\Lambda^{\rm LC}(\mu)\mathcal M_{N;I}(\mu)\mathcal Q_r(z,P)u_\Lambda .
 \label{eq:match5moment}
\end{equation}

At fixed $N$, all operators in the mixing sum have the same $\mathcal Q_r$. Let $Z_{N,IJ}(\mu,\mu_0)$ evolve the local basis from $\mu_0$ to $\mu$. Substituting Eq.~\eqref{eq:match5moment} at scale $\mu_0$ gives
\begin{align}
 \langle0|\mathcal O^{(3)}_{NI,R}(z;\mu)|\Lambda(P,\sigma)\rangle
 &=\sum_J Z_{N,IJ}(\mu,\mu_0)
       \langle0|\mathcal O^{(3)}_{NJ,R}(z;\mu_0)|\Lambda(P,\sigma)\rangle \nonumber\\
 &=\left[\sum_J Z_{N,IJ}(\mu,\mu_0)f_\Lambda^{\rm LC}(\mu_0)
              \mathcal M_{N;J}(\mu_0)\right]\mathcal Q_r(z,P)u_\Lambda .
 \label{eq:match5mix}
\end{align}
Taking the light-cone component of Eq.~\eqref{eq:match5mix} identifies the bracket with $f_\Lambda^{\rm LC}(\mu)\mathcal M_{N;I}(\mu)$ by Eq.~\eqref{eq:match5lcfix}. The same tensor $\mathcal Q_r$ thus determines the spatial and temporal contractions at either scale. Using the spin sum in Eq.~\eqref{eq:tmSpin}, we obtain
\begin{align}
 &\frac{w_\Gamma^{\rho_1}\lambda^{\rho_2}\cdots\lambda^{\rho_r}}{4pP_z^r}
   \sum_\sigma\bar u_\Lambda\slashed n\,
        \mathcal Q_{r,\rho_1\cdots\rho_r}(P)u_\Lambda \nonumber\\
 &\qquad=\begin{cases}
    S_r(\lambda)/P_z^r=F_r(c),&\Gamma=\gamma^z,\\[3pt]
    \left.(v_t\cdot\partial_z)S_r(z)\right|_{z=\lambda}/(rP_z^r)=T_r(c),
       &\Gamma=\gamma^t .
 \end{cases}
 \label{eq:match5geom}
\end{align}
The two contractions in Eq.~\eqref{eq:match5geom} give $F_r(c)$ and $T_r(c)$ even when operators with no derivatives on the strange quark mix with operators that have such derivatives. To evaluate the one-loop term in Eq.~\eqref{eq:match5ope}, one must contract the same local tensor with the corresponding one-loop Wilson tensor.

\subsection{The one-loop matching kernel}
\label{sec:match5kernel}

To express each reduced moment in Eq.~\eqref{eq:match5moment} as an integral over the LCDA, place the three quarks at $\xi_j n$ and write $y_3=1-y_1-y_2$. Translation invariance gives the Fourier phase $\exp[-ip\sum_{j=1}^3 y_j\xi_j]$. Each derivative $i\partial_{\xi_j}$ contributes a factor $p y_j$, so an operator with derivative assignment $(k_1,k_2,k_3)$ on the three quarks has weight $y_1^{k_1}y_2^{k_2}y_3^{k_3}$. At fixed $N$ and $\mu$, define $\beta_{I;k_1k_2k_3}$ by expanding the light-cone component of basis operator $I$ in operators with definite derivative assignments,
\begin{equation}
 \Pi_+\mathcal O^{(3)}_{NI,R}(n;\mu)
 =\sum_{\substack{k_1,k_2,k_3\geq 0\\ k_1+k_2+k_3=N}}
   \beta_{I;k_1k_2k_3}\,
   \Pi_+\mathcal O^{(3)}_{N;(k_1,k_2,k_3),R}(n;\mu).
 \label{eq:match5betabasis}
\end{equation}
For $I=(n_1,n_2,n_3)$, $\beta_{I;k_1k_2k_3}=\delta_{k_1n_1}\delta_{k_2n_2}\delta_{k_3n_3}$. If $I$ denotes a linear combination of operators, $\beta_{I;k_1k_2k_3}$ gives the coefficient of the operator with assignment $(k_1,k_2,k_3)$. Taking the vacuum-to-baryon matrix element of Eq.~\eqref{eq:match5betabasis} and factoring out $f_\Lambda^{\rm LC}(\mu)p^{N+1}u_{\Lambda,+}$ as in Eq.~\eqref{eq:match5lcfix} gives
\begin{align}
 \mathcal M_{N;I}(\mu)
 &=\sum_{k_1+k_2+k_3=N}\beta_{I;k_1k_2k_3}\,
              \mathcal M_{N;(k_1,k_2,k_3)}(\mu)\nonumber\\
 &=\int_{\mathcal S}dy_1\,dy_2\, w_{NI}(y_1,y_2)\phi(y_1,y_2,\mu),\nonumber\\
 w_{NI}(y_1,y_2)
 &=\sum_{k_1+k_2+k_3=N}\beta_{I;k_1k_2k_3}\,
         y_1^{k_1}y_2^{k_2}(1-y_1-y_2)^{k_3}.
 \label{eq:match5weight}
\end{align}
Appendix~\ref{sec:match5weights} derives Eq.~\eqref{eq:match5weight} from the anchored LCDA of Eq.~\eqref{eq:bdaLC} and gives its equivalent form in terms of two-index moments.

For $\Gamma=\gamma^z,\gamma^t$, we apply the spin contraction of Eq.~\eqref{eq:bdaQuasiScalar} to the renormalized twist-3 operator in Eq.~\eqref{eq:match5ope}. This defines the scalar coordinate-space amplitude
\begin{align}
 \mathcal A_\Gamma(\ell_1,\ell_2;P_z,\mu_R,\mu)
 &\equiv \frac{\displaystyle\sum_\sigma\bar u_\Lambda(P,\sigma)\slashed n\,
   \langle0| \widetilde{\mathcal O}_{\Gamma,R}^{(3)}(\ell_1,\ell_2;\mu_R)
                 |\Lambda(P,\sigma)\rangle}
        {4f_\Lambda^{\rm LC}(\mu)\,pP_z} \nonumber\\
 &=\mathcal A_\Gamma^{(0)}(\ell_1,\ell_2)
   +a_s\mathcal A_\Gamma^{(1)}(\ell_1,\ell_2)+O(a_s^2).
 \label{eq:match5amplitudeDef}
\end{align}
The superscripts $(0)$ and $(1)$ label the tree-level and one-loop terms in the Wilson expansion, both evaluated with the LCDA at scale $\mu$. We suppress the arguments $P_z,\mu_R,\mu$ on these two terms.

To evaluate this matrix element, first keep $N$ fixed in Eq.~\eqref{eq:match5ope}. Substituting Eqs.~\eqref{eq:match5moment} and \eqref{eq:match5weight} puts the finite sum over $I$ inside the light-cone integral,
\begin{align}
 &\sum_I\mathcal C_{\Gamma,NI}^{\rho_1\cdots\rho_r}
   \langle0|\mathcal O^{(3)}_{NI,R,\rho_1\cdots\rho_r}(\mu)|\Lambda(P,\sigma)\rangle
   \nonumber\\
=&\  f_\Lambda^{\rm LC}(\mu)\int_{\mathcal S}dy_1\,dy_2\,
     \phi(y_1,y_2,\mu)
     \left[\sum_I\mathcal C_{\Gamma,NI}^{\rho_1\cdots\rho_r}
                      w_{NI}(y_1,y_2)\right]
         \mathcal Q_{r,\rho_1\cdots\rho_r}(P)u_\Lambda .
 \label{eq:match5fold}
\end{align}
We now sum Eq.~\eqref{eq:match5fold} over $N$ and contract the spinor index as in Eq.~\eqref{eq:match5amplitudeDef}. The factor $f_\Lambda^{\rm LC}(\mu)$ cancels against the reference normalization. At each order $j=0,1$ we obtain
\begin{align}
 \mathcal A_\Gamma^{(j)}(\ell_1,\ell_2)
 &=\int_{\mathcal S}dy_1\,dy_2\,\mathcal W_{\Gamma,c}^{(j)}(\ell_1,\ell_2;y_1,y_2)
                         \phi(y_1,y_2,\mu),
 \label{eq:match5matrixW}
\end{align}
where the coordinate-space kernel is
\begin{align}
 &\mathcal W_{\Gamma,c}^{(j)}(\ell_1,\ell_2;y_1,y_2)\nonumber\\
 =&\frac{1}{4pP_z}\sum_{N,I,\sigma}w_{NI}(y_1,y_2)\,
 \bar u_\Lambda\slashed n\,\mathcal C_{\Gamma,NI}^{(j)\,\rho_1\cdots\rho_r}
        (\ell_1,\ell_2;\mu_R,\mu)\mathcal Q_{r,\rho_1\cdots\rho_r}(P)u_\Lambda .
 \label{eq:match5wilsonW}
\end{align}
In Eq.~\eqref{eq:match5wilsonW} the input fractions $y_1,y_2$ remain fixed. The baryon dependence in $\mathcal Q_r$ is kinematic, and the LCDA enters only through the integral in Eq.~\eqref{eq:match5matrixW}.

At tree level only $I=(n_1,n_2,0)$ contributes. Contracting the directions in Eq.~\eqref{eq:match5tree} with the local tensor gives $P_z^r F_r(c)$ or $P_z^r T_r(c)$ after division by $4p$, as in Eq.~\eqref{eq:match5geom}. The remaining $1/P_z$ in Eq.~\eqref{eq:match5wilsonW} leaves $P_z^{r-1}=P_z^{n_1+n_2}$, and Eq.~\eqref{eq:match5matrixW} reduces to
\begin{equation}
 \mathcal A_\Gamma^{(0)}(\ell_1,\ell_2)
 =\sum_{n_1,n_2\ge0}\frac{(-iP_z\ell_1)^{n_1}(-iP_z\ell_2)^{n_2}}{n_1!\,n_2!}\,
  \mathcal M_{n_1,n_2}(\mu)
  \begin{cases}
     F_{n_1+n_2+1}(c),&\Gamma=\gamma^z,\\
     T_{n_1+n_2+1}(c),&\Gamma=\gamma^t .
  \end{cases}
 \label{eq:match5treeamplitude}
\end{equation}
At $\ell_1=\ell_2=0$, only the zeroth moment $\mathcal M_{0,0}=1$ survives in Eq.~\eqref{eq:match5treeamplitude}. The spatial and temporal tree amplitudes at the origin are therefore $F_1(c)$ and $T_1(c)$, respectively. Beyond tree level, unit normalization requires the separately renormalized local twist-3 matrix element in the same channel. Using the reference normalization of Eq.~\eqref{eq:match5amplitudeDef}, we define this local value by
\begin{align}
 d_\Gamma(c)
 &\equiv\frac{\sum_\sigma\bar u_\Lambda\slashed n
  \langle0|\widetilde{\mathcal O}^{(3) \rm loc}_{\Gamma,R}(\mu_R)
                        |\Lambda(P,\sigma)\rangle}
       {4f_\Lambda^{\rm LC}(\mu)pP_z}
 =d_\Gamma^{(0)}(c)+a_s d_\Gamma^{(1)}(c)+O(a_s^2),\nonumber\\
 d_{\gamma^z}^{(0)}(c)&=F_1(c),\qquad d_{\gamma^t}^{(0)}(c)=T_1(c).
 \label{eq:match5locald}
\end{align}
Only the zeroth light-cone moment enters $d_\Gamma$, so this denominator is independent of the shape of $\phi$. Write $\mathcal W_{\Gamma,c}=\mathcal W_{\Gamma,c}^{(0)}+a_s\mathcal W_{\Gamma,c}^{(1)}+O(a_s^2)$. Taking the Fourier transform of the ratio $\mathcal A_\Gamma/d_\Gamma$ and inserting Eq.~\eqref{eq:match5matrixW}, we obtain the factorization formula for the unit-normalized twist-3 quasi-DA,
\begin{align}
 &\widehat{\widetilde\phi}^{(3)}_\Gamma(x_1,x_2;P_z,\mu_R)\nonumber\\
 &\quad=P_z^2\int_{\mathbb R^2}\frac{d\ell_1\,d\ell_2}{(2\pi)^2}\,
     e^{iP_z(x_1\ell_1+x_2\ell_2)}
     \frac{\mathcal A_\Gamma(\ell_1,\ell_2;P_z,\mu_R,\mu)}{d_\Gamma(c)}
     \nonumber\\
 &\quad=\int_{\mathcal S}dy_1\,dy_2\,
      \widehat K_{\Gamma,c}(x_1,x_2;y_1,y_2;P_z,\mu_R,\mu)\phi(y_1,y_2,\mu)\nonumber\\
 &\quad \equiv [\widehat K_{\Gamma,c}\otimes\phi](x_1,x_2;P_z,\mu_R,\mu),\label{eq:match5conv}\\
 &\widehat K_{\Gamma,c}(x_1,x_2;y_1,y_2;P_z,\mu_R,\mu)
 =P_z^2\int_{\mathbb R^2}\frac{d\ell_1\,d\ell_2}{(2\pi)^2}
       e^{iP_z(x_1\ell_1+x_2\ell_2)}
       \frac{\mathcal W_{\Gamma,c}(\ell_1,\ell_2;y_1,y_2)}{d_\Gamma(c)} .
 \label{eq:match5fourier}
\end{align}
The kernel $\widehat K_{\Gamma,c}$ in Eq.~\eqref{eq:match5fourier} maps the LCDA to the twist-3 quasi-DA at finite $P_z$.
At $c=0$, $F_r(0)=T_r(0)=1$ for every $r$. Inserting the moments from Eq.~\eqref{eq:bdaMoment} into Eq.~\eqref{eq:match5treeamplitude} and summing the two series gives the coordinate kernel $\mathcal W_{\Gamma,0}^{(0)}(\ell_1,\ell_2;y_1,y_2)=e^{-iP_z(y_1\ell_1+y_2\ell_2)}$. Integrating this phase against $\phi$ gives the tree-level amplitude in Eq.~\eqref{eq:match5matrixW}. Since $d_\Gamma^{(0)}(0)=1$, the Fourier transform of the kernel in Eq.~\eqref{eq:match5fourier} gives $\delta(x_1-y_1)\delta(x_2-y_2)$.

At finite $c$, we expand the ratio $\mathcal W_{\Gamma,c}/d_\Gamma$ in Eq.~\eqref{eq:match5fourier} using the local denominator of Eq.~\eqref{eq:match5locald}. Through one loop this gives
\begin{align}
 \frac{\mathcal W_{\Gamma,c}}{d_\Gamma}
 &=\big(\mathcal W_{\Gamma,c}^{(0)}+a_s\mathcal W_{\Gamma,c}^{(1)}\big)
    \left[\frac{1}{d_\Gamma^{(0)}}-
          \frac{a_s d_\Gamma^{(1)}}{(d_\Gamma^{(0)})^2}\right]+O(a_s^2)
          \nonumber\\
 &=\frac{\mathcal W_{\Gamma,c}^{(0)}}{d_\Gamma^{(0)}}
 +a_s\left[\frac{\mathcal W_{\Gamma,c}^{(1)}}{d_\Gamma^{(0)}}
          -\frac{d_\Gamma^{(1)}\mathcal W_{\Gamma,c}^{(0)}}{(d_\Gamma^{(0)})^2}
      \right]+O(a_s^2).
 \label{eq:match5quot}
\end{align}
Fourier transforming the coefficient of $a_s$ in Eq.~\eqref{eq:match5quot} gives the one-loop momentum-space matching kernel,
\begin{align}
 &\widehat K_{\Gamma,c}^{(1)}(x_1,x_2;y_1,y_2;P_z,\mu_R,\mu)\nonumber\\
 = &\  P_z^2\int_{\mathbb R^2}\frac{d\ell_1\,d\ell_2}{(2\pi)^2}\,
 e^{iP_z(x_1\ell_1+x_2\ell_2)}
 \left[\frac{\mathcal W_{\Gamma,c}^{(1)}(\ell_1,\ell_2;y_1,y_2)}{d_\Gamma^{(0)}(c)}
 -\frac{d_\Gamma^{(1)}(c)\mathcal W_{\Gamma,c}^{(0)}(\ell_1,\ell_2;y_1,y_2)}
             {(d_\Gamma^{(0)}(c))^2}\right].
 \label{eq:match5oneW}
\end{align}
The first term comes from the nonlocal matrix element. The second comes from expanding the inverse local denominator. Substitution of Eq.~\eqref{eq:match5wilsonW} into both terms gives the momentum-space kernel directly in terms of the Wilson tensors,
\begin{align}
 &\widehat K_{\Gamma,c}^{(1)}(x_1,x_2;y_1,y_2;P_z,\mu_R,\mu)\nonumber\\
 =&\ \frac{P_z}{4p\,d_\Gamma^{(0)}(c)}
   \int_{\mathbb R^2}\frac{d\ell_1\,d\ell_2}{(2\pi)^2}
          e^{iP_z(x_1\ell_1+x_2\ell_2)}
    \sum_{N,I,\sigma}w_{NI}(y_1,y_2)\,
    \bar u_\Lambda\slashed n \nonumber\\
 &\ \times\left[
   \mathcal C_{\Gamma,NI}^{(1)\,\rho_1\cdots\rho_r}(\ell_1,\ell_2;\mu_R,\mu)
  -\frac{d_\Gamma^{(1)}(c)}{d_\Gamma^{(0)}(c)}
   \mathcal C_{\Gamma,NI}^{(0)\,\rho_1\cdots\rho_r}(\ell_1,\ell_2;\mu_R,\mu)
       \right]\mathcal Q_{r,\rho_1\cdots\rho_r}(P)u_\Lambda .
 \label{eq:match5oneWilson}
\end{align}
Equation~\eqref{eq:match5oneWilson} specifies which one-loop coefficients are needed at finite $c$.

\subsection{Matching in the two channels}
\label{sec:match5channelresults}

Let $\widehat{\mathcal T}_{\Gamma,c}^{(0)}$ denote the linear map defined by the tree-level kernel in Eq.~\eqref{eq:match5fourier}. Acting on $\phi$, it gives the unit-normalized tree-level quasi-DA of Eq.~\eqref{eq:tmUnitDA},
\begin{equation}
\widehat{\widetilde\phi}^{(3,0)}_\Gamma(x_1,x_2;P_z)=[\widehat{\mathcal T}_{\Gamma,c}^{(0)}\phi](x_1,x_2;\mu). \label{eq:matchT}
\end{equation} 
At $c=0$ this map is the identity. At one loop, we compare Eq.~\eqref{eq:match5conv} with leading-power matching using the same spin projection, renormalization scheme and local normalization. For arbitrary light-cone input the coefficient of $a_s c^0$ in both convolutions must agree, which identifies the leading-power kernel as
\begin{equation}
 \left.\widehat K_{\Gamma,c}^{(1)}(x_1,x_2;y_1,y_2;P_z,\mu_R,\mu)\right|_{c^0}
 =C_{\Gamma,\mathrm{LP}}^{(1)}(x_1,x_2;y_1,y_2;P_z,\mu_R,\mu).
 \label{eq:match5LP}
\end{equation}
Here $C_{\Gamma,\mathrm{LP}}^{(1)}$ is the leading-power matching kernel at one loop. It has been obtained by Ref.~\cite{Deng:2023csv} using a direct matching between the quasi-DAs and LCDAs in the $\overline{\mathrm{MS}}$ scheme with $\mu_R=\mu$. For quasi-DAs renormalized in the hybrid scheme, the corresponding kernels are given in Ref.~\cite{Han:2024ucv}.  Adding and subtracting $a_s[C_{\Gamma,\mathrm{LP}}^{(1)}\otimes\phi]$ in Eq.~\eqref{eq:match5conv} gives
\begin{align}
 \widehat{\widetilde\phi}^{(3)}_\Gamma(x_1,x_2;P_z,\mu_R)
 ={}&[\widehat{\mathcal T}_{\Gamma,c}^{(0)}\phi](x_1,x_2;\mu)
       +a_s[C_{\Gamma,\mathrm{LP}}^{(1)}\otimes\phi](x_1,x_2;P_z,\mu_R,\mu)
       \nonumber\\
 &+\mathcal R_{\Gamma,\mathrm{mix}}(x_1,x_2;P_z,\mu_R,\mu)+O(a_s^2),\nonumber\\
 \mathcal R_{\Gamma,\mathrm{mix}}(x_1,x_2;P_z,\mu_R,\mu)
 \equiv{}&a_s[(\widehat K_{\Gamma,c}^{(1)}-C_{\Gamma,\mathrm{LP}}^{(1)})
                         \otimes\phi](x_1,x_2;P_z,\mu_R,\mu).
 \label{eq:match5forward}
\end{align}
At fixed local rank, or for a suitably regulated distribution, the twist-3 remainder $\mathcal R_{\Gamma,\mathrm{mix}}$ starts at $O(a_s c)$. Its power dependence can be accompanied by logarithms.

Inserting the resummed tree contributions of Eqs.~\eqref{eq:tmSpatialDA} and \eqref{eq:tmTemporalDA} into Eq.~\eqref{eq:match5forward} gives the two quasi-DAs with all tree-level mass corrections and the leading-power one-loop term. With $\Delta=\sqrt{1+4c}$, $a=(1+\Delta)/2$ and $b=(1-\Delta)/2$, the results are
\begin{align}
 &\widehat{\widetilde\phi}^{(3)}_{\gamma^z}(x_1,x_2;P_z,\mu_R)
   =\frac{1}{F_1(c)\Delta}\,\phi\!\left(\frac{x_1}{a},\frac{x_2}{a},\mu\right)
   -\frac{b}{F_1(c)\Delta^2}\int_b^a\frac{ds}{s}\,
         \phi\!\left(\frac{x_1}{s},\frac{x_2}{s},\mu\right)\nonumber\\
 &\hspace{22mm}
   +a_s[C_{\gamma^z,\mathrm{LP}}^{(1)}\otimes\phi](x_1,x_2;P_z,\mu_R,\mu)
   +O(a_s c)+O(a_s^2),\label{eq:match5spaceDA}\\
 &\widehat{\widetilde\phi}^{(3)}_{\gamma^t}(x_1,x_2;P_z,\mu_R)
   =\frac{1}{T_1(c)\Delta}\,\phi\!\left(\frac{x_1}{a},\frac{x_2}{a},\mu\right)
   -\frac{b(\Delta+2)}{T_1(c)\Delta^3}\int_b^a\frac{ds}{s}\,
         \phi\!\left(\frac{x_1}{s},\frac{x_2}{s},\mu\right)\nonumber\\
 &\hspace{22mm}
   +\frac{4ab^2}{T_1(c)\Delta^3}\int_b^a\frac{ds}{s^2}\,
         \phi\!\left(\frac{x_1}{s},\frac{x_2}{s},\mu\right)\nonumber\\
 &\hspace{22mm}
   +a_s[C_{\gamma^t,\mathrm{LP}}^{(1)}\otimes\phi](x_1,x_2;P_z,\mu_R,\mu)
   +O(a_s c)+O(a_s^2).
 \label{eq:match5timeDA}
\end{align}
Here $F_1(c)=(3+\Delta)/4$ and $T_1(c)=(1+3\Delta)/4$ are the local factors in Eq.~\eqref{eq:tmLocalFactors}, and $\phi$ is extended by zero outside $\mathcal S$. 

The quasi-DA calculated on the lattice also contains the higher-twist matrix elements of Sec.~\ref{sec:Highertwist}. Consider lattice data normalized by the full local matrix element in the measured channel. Denote this local value by $d_{\Gamma,\rm full}$ in the same units as $d_\Gamma$ in Eq.~\eqref{eq:match5locald}. Starting from the full and twist-3 quasi-DAs in the common reference normalization of Eq.~\eqref{eq:bdaQuasiScalar} and dividing each by its own local value gives the difference
\begin{align}
 \mathcal R_{\Gamma,\mathrm{HT}}
 &\equiv \widehat{\widetilde\phi}^{\rm full}_\Gamma-\widehat{\widetilde\phi}^{(3)}_\Gamma
 =\frac{\widetilde\phi^{\rm full}_\Gamma}{d_{\Gamma,\rm full}}
  -\frac{\widetilde\phi^{(3)}_\Gamma}{d_\Gamma}\nonumber\\
 &=\frac{\widetilde\phi^{\rm full}_\Gamma-\widetilde\phi^{(3)}_\Gamma
       -(d_{\Gamma,\rm full}-d_\Gamma)\widehat{\widetilde\phi}^{(3)}_\Gamma}
            {d_{\Gamma,\rm full}} .
 \label{eq:match5ht}
\end{align}
The last term in the numerator of Eq.~\eqref{eq:match5ht} accounts for the higher-twist contribution to the local normalization. Once all twist-3 contributions for this projection have been included, the difference $\mathcal R_{\Gamma,\mathrm{HT}}$ contains only higher twist. For the spin contraction used here, Eqs.~\eqref{eq:ht4leading} and \eqref{eq:ht4localnorm} show that the tree-level corrections to the higher-twist moments start at order $1/P_z^2$.

Combining Eqs.~\eqref{eq:match5forward} and \eqref{eq:match5ht} gives the forward relation for the locally normalized lattice quasi-DA,
\begin{align}
 \widehat{\widetilde\phi}^{\rm full}_\Gamma(x_1,x_2;P_z,\mu_R)
 ={}&[\widehat{\mathcal T}^{(0)}_{\Gamma,c}\phi](x_1,x_2;\mu)
 +a_s[C_{\Gamma,\mathrm{LP}}^{(1)}\otimes\phi](x_1,x_2;P_z,\mu_R,\mu)\nonumber\\
 &+\mathcal R_{\Gamma,\mathrm{mix}}(x_1,x_2;P_z,\mu_R,\mu)
  +\mathcal R_{\Gamma,\mathrm{HT}}(x_1,x_2;P_z,\mu_R)+O(a_s^2).
 \label{eq:match5fullforward}
\end{align}
The tree-level mass shift in Eq.~\eqref{eq:match5fullforward} is $[\widehat{\mathcal T}_{\Gamma,c}^{(0)}\phi]-\phi$. The second term is fixed by the leading-power matching kernel in the chosen scheme. Determining $\mathcal R_{\Gamma,\mathrm{mix}}$ through $O(a_s c)$ requires the mass-dependent part of the one-loop kernel in Eq.~\eqref{eq:match5oneWilson}. Determining $\mathcal R_{\Gamma,\mathrm{HT}}$ requires independent higher-twist matrix elements.

\subsection{Subtraction and inverse matching}
\label{sec:subtracttargetmass}

The tree-level inverse $(\widehat{\mathcal T}_{\Gamma,c}^{(0)})^{-1}$ removes the mass dependence of the first term in Eq.~\eqref{eq:match5fullforward}. Applied to the measured quasi-DA, it also acts on the one-loop term and both remainders.  For a subtraction through order $1/P_z^2$, we expand the inverse to first order in $c=m_\Lambda^2/(4P_z^2)$. Appendix~\ref{app:match5inverse} gives a full inverse  and its range of convergence.

For a distribution $g(x_1,x_2)$ on $\mathbb R^2$, Eqs.~\eqref{eq:tmDAFirstOrder} and \eqref{eq:tmLocalFactors} give the expansion of the normalized tree-level map,
\begin{align}
 \widehat{\mathcal T}_{\Gamma,c}^{(0)}g
 &=g+c\,\mathcal D_\Gamma g+O(c^2),\nonumber\\
 [\mathcal D_\Gamma g](x_1,x_2)
 &=-\left(2+\frac{\kappa_\Gamma}{2}
                +x_1\partial_{x_1}+x_2\partial_{x_2}\right)g(x_1,x_2)
   +\kappa_\Gamma\int_0^1\frac{ds}{s}\,
              g\!\left(\frac{x_1}{s},\frac{x_2}{s}\right),
 \label{eq:match5D}
\end{align}
where $\kappa_{\gamma^z}=1$ and $\kappa_{\gamma^t}=3$. The term $-\kappa_\Gamma g/2$ in $\mathcal D_\Gamma g$ comes from expanding the local normalization.
The inverse to this order is $(\widehat{\mathcal T}_{\Gamma,c}^{(0)})^{-1}=1-c\mathcal D_\Gamma+O(c^2)$. Applying it to the measured quasi-DA gives the first-order subtraction
\begin{align}
 \widehat{\widetilde\phi}^{[1]}_{\Gamma,\rm sub}(x_1,x_2;P_z,\mu_R)
 ={}&\widehat{\widetilde\phi}^{\rm full}_\Gamma(x_1,x_2;P_z,\mu_R)\nonumber\\
 &+c\left(2+\frac{\kappa_\Gamma}{2}
                +x_1\partial_{x_1}+x_2\partial_{x_2}\right)
       \widehat{\widetilde\phi}^{\rm full}_\Gamma(x_1,x_2;P_z,\mu_R)\nonumber\\
 &-c\,\kappa_\Gamma\int_0^1\frac{ds}{s}\,
       \widehat{\widetilde\phi}^{\rm full}_\Gamma
         \!\left(\frac{x_1}{s},\frac{x_2}{s};P_z,\mu_R\right).
 \label{eq:match5datasub}
\end{align}
The superscript $[1]$ denotes truncation at first order in $c$. The scale integral samples the full quasi-DA at $(x_1/s,x_2/s)$, including fractions outside the light-cone simplex. For lattice data on a finite grid, evaluating the integral requires an estimate of the unmeasured tails and their uncertainty.  

Substituting Eq.~\eqref{eq:match5fullforward} into Eq.~\eqref{eq:match5datasub} relates the subtracted quasi-DA at momentum $P_z$ to the LCDA at scale $\mu$,
\begin{align}
 \widehat{\widetilde\phi}^{[1]}_{\Gamma,\rm sub}
 &=\phi+a_s[C_{\Gamma,\mathrm{LP}}^{(1)}\otimes\phi]+\delta_\Gamma,\nonumber\\
 \delta_\Gamma
 &=\mathcal R_{\Gamma,\mathrm{mix}}
   -a_sc\,\mathcal D_\Gamma[C_{\Gamma,\mathrm{LP}}^{(1)}\otimes\phi]
   +(1-c\mathcal D_\Gamma)\mathcal R_{\Gamma,\mathrm{HT}}\nonumber\\
   &\quad\quad+O(c^2)+O(a_s c^2)+O(a_s^2).
 \label{eq:match5residual}
\end{align}
The first two terms of $\delta_\Gamma$ are both of order $a_s c$. The first depends on the finite-mass one-loop kernel in Eq.~\eqref{eq:match5oneWilson}. The second is fixed by the tree-level subtraction acting on the known leading-power kernel. If a higher-twist moment starts at $\Lambda_{\rm h}^2/P_z^2$, where $\Lambda_{\rm h}$ is a hadronic scale, the change induced in that moment by $-c\mathcal D_\Gamma$ starts at $m_\Lambda^2\Lambda_{\rm h}^2/P_z^4$. Thus the tree subtraction can be applied before the higher-twist extrapolation at this accuracy.

Equation~\eqref{eq:match5residual} still contains $a_s[C_{\Gamma,\mathrm{LP}}^{(1)}\otimes\phi]$. In this convolution we may replace $\phi$ by the subtracted data to one-loop accuracy at leading power. The resulting inverse matching on the physical simplex reads
\begin{align}
&\bar\phi^{[1]}_\Gamma(x_1,x_2;P_z,\mu_R,\mu)
 =\ \widehat{\widetilde\phi}^{[1]}_{\Gamma,\rm sub}(x_1,x_2;P_z,\mu_R)\nonumber\\
 &\quad\quad\quad\quad\quad\quad-a_s\int_{\mathcal S}dy_1\,dy_2\,
       C_{\Gamma,\mathrm{LP}}^{(1)}(x_1,x_2;y_1,y_2;P_z,\mu_R,\mu)\,
       \widehat{\widetilde\phi}^{[1]}_{\Gamma,\rm sub}(y_1,y_2;P_z,\mu_R).
 \label{eq:match5inverse}
\end{align}
The convolution in Eq.~\eqref{eq:match5inverse} integrates over the input fractions $(y_1,y_2)$ in the light-cone simplex $\mathcal S$ of Eq.~\eqref{eq:bdaThreeLCFourier}. Substituting Eq.~\eqref{eq:match5residual} into both terms gives
\begin{align}
 \bar\phi^{[1]}_\Gamma
 &=\phi+a_s[C_{\Gamma,\mathrm{LP}}^{(1)}\otimes\phi]+\delta_\Gamma
    -a_s C_{\Gamma,\mathrm{LP}}^{(1)}\otimes
       \big(\phi+a_s[C_{\Gamma,\mathrm{LP}}^{(1)}\otimes\phi]+\delta_\Gamma\big)
       \nonumber\\
 &=\phi+\delta_\Gamma-a_s[C_{\Gamma,\mathrm{LP}}^{(1)}\otimes\delta_\Gamma]
      +O(a_s^2).
 \label{eq:match5cancel}
\end{align}
The leading-power one-loop term cancels. Acting on the higher-twist part of $\delta_\Gamma$, inverse matching can generate a correction of order $a_s\Lambda_{\rm h}^2/P_z^2$. The omitted two-loop kernel can contribute at leading power.

\section{Target-mass corrections to the tensor DA}
\label{sec:tensorTMC}

For a $\Lambda$ baryon moving along the third axis, the tensor sink operator used in the lattice study~\cite{LPC:2026mvw}, translated so that the strange quark is at the origin, reads
\begin{equation}
 \widetilde{\mathcal O}^{T}_{\perp,t}(\ell_1,\ell_2)
 =\sum_{i=1}^2\epsilon^{abc}
     [u_a^{[\lambda]}(\ell_1)]^T C\,\frac{[\gamma^t,\gamma^i]}{2}\,
      d_b^{[\lambda]}(\ell_2)\,\gamma_5\gamma_i\,s_c(0).
 \label{eq:tensorLatticeSink}
\end{equation}
Here $i=1,2$ selects the $x,y$ components transverse to the baryon momentum.  We keep the tensor indices open until after the twist-3 projection. Taking the transverse sum then relates the tensor moment to the two axial moment factors in Eq.~\eqref{eq:tmRatios}. We work at tree level and use the  external spin projection of Eq.~\eqref{eq:bdaQuasiScalar}.

\subsection{The transverse tensor operator}
\label{sec:tensorOperator}

To implement the projection, write Eq.~\eqref{eq:tensorLatticeSink} as a contraction of an antisymmetric tensor,
 \begin{align}
 \mathcal B^{\mu\nu}(\ell_1,\ell_2;z)
 &=\epsilon^{abc}[u_a^{[z]}(\ell_1)]^T
       C\,\frac{[\gamma^\mu,\gamma^\nu]}{2}\,d_b^{[z]}(\ell_2)\,\gamma_5s_c(0),\nonumber\\
 \widetilde{\mathcal O}^{T}_{\perp,t}(\ell_1,\ell_2)
 &=\sum_{i=1}^2\gamma_i(v_t)_\nu\mathcal B^{i\nu}(\ell_1,\ell_2;\lambda).\label{eq:tensorSink}
 \end{align}
The second line reproduces Eq.~\eqref{eq:tensorLatticeSink} because $[\gamma^i,\gamma^t]=-[\gamma^t,\gamma^i]$ and $\gamma_i\gamma_5=-\gamma_5\gamma_i$. In our conventions the spin projector $\gamma^t+\gamma^z$ used by lattice calculation \cite{LPC:2026mvw} is $\sqrt{2}\slashed n$.

The tensor coupling $f_\Lambda^T$ was defined in Eq.~\eqref{eq:bdaThreeLCDAs}. Multiplication of its tensor matrix element by $\gamma_5$ gives the spinor convention in Eq.~\eqref{eq:tensorSink} and the reference factor $2f_\Lambda^T$. We use this factor to define the full quasi-DA before imposing the lattice local normalization
\begin{equation}
 \frac{\sum_\sigma\bar u_\Lambda\slashed n
       \langle0|\widetilde{\mathcal O}^{T}_{\perp,t}(\ell_1,\ell_2)|\Lambda\rangle}
      {8f_\Lambda^T pP_z}
 =\int_{\mathbb R^2}dx_1\,dx_2\,e^{-iP_z(x_1\ell_1+x_2\ell_2)}
       \widetilde\phi_{\perp,t}^{T,\mathrm{full}}(x_1,x_2;P_z).
 \label{eq:tensorFourier}
\end{equation}
For $n_1$ derivatives on $u$ and $n_2$ on $d$, let $N=n_1+n_2$, $r=N+1$ and $I=\rho_1\cdots\rho_N$, with $z_I=z_{\rho_1}\cdots z_{\rho_N}$. The local operator corresponding to  Eq.~\eqref{eq:tensorSink} is
\begin{equation}
 z_I B_{n_1,n_2}^{\mu\nu;I}
 =\epsilon^{abc}[(iz\cdot D)^{n_1}u_a]^T C\,\frac{[\gamma^\mu,\gamma^\nu]}2\,
             [(iz\cdot D)^{n_2}d_b]\,\gamma_5s_c .
 \label{eq:tensorLocal}
\end{equation}
The indices in $I$ are symmetric. We suppress $n_1,n_2$ on $B$ when the derivative assignment is fixed. On the light cone Eq.~\eqref{eq:tensorLocal} gives the moment of the tensor LCDA $\phi^T$,
\begin{align}
 &\frac{1}{4p}\sum_\sigma\bar u_\Lambda\slashed n\,
   \langle0|\gamma_\mu n_\nu n_I B^{\mu\nu;I}|\Lambda\rangle
 =2f_\Lambda^T p^r\mathcal M^T_{n_1,n_2}.\nonumber\\
 & \mathcal M^T_{n_1,n_2}
 =\int_{\mathcal S}dx_1\,dx_2\,x_1^{n_1}x_2^{n_2}\phi^T(x_1,x_2),
 \label{eq:tensorLCmoment}
\end{align}
where $z=n$ is used. In the first line the longitudinal terms of the $\mu$ contraction vanish by $n_\mu n_\nu B^{\mu\nu;I}=0$ and $\slashed n^2=0$. For the equal-time moment, acting on Eq.~\eqref{eq:tensorFourier} with $(i\partial_{\ell_1})^{n_1}(i\partial_{\ell_2})^{n_2}$ and setting both distances to zero gives the quasi moment
\begin{equation}
 \widetilde{\mathcal M}^{T,\mathrm{full},0}_{\perp,t;n_1,n_2}
 =\frac{\sum_\sigma\bar u_\Lambda\slashed n
       \langle0|\sum_{i=1}^2\gamma_i(v_t)_\nu\lambda_I B^{i\nu;I}|\Lambda\rangle}
       {8 f_\Lambda^T pP_z^r}.
 \label{eq:tensorQuasiMoment}
\end{equation}

\subsection{The twist-3 moment}
\label{sec:tensorMoment}

For fixed $(n_1,n_2)$, the local operator in Eq.~\eqref{eq:tensorLocal} has dimension $N+9/2$, so twist three requires the maximal Lorentz spin $j_L+j_R=N+3/2$. Its $N$ symmetric derivative indices must then be traceless and carry $(N/2,N/2)$. Couple these indices first to the antisymmetric pair $\mu,\nu$ and then to the open spinor. For the $(1,0)$ part of the antisymmetric tensor, the two choices of spinor chirality give
\begin{equation}
\begin{aligned}
 (\tfrac N2,\tfrac N2)\otimes(1,0)\otimes(\tfrac12,0)
    &\longrightarrow(\tfrac N2+\tfrac32,\tfrac N2),\\
 (\tfrac N2,\tfrac N2)\otimes(1,0)\otimes(0,\tfrac12)
    &\longrightarrow(\tfrac N2+1,\tfrac N2+\tfrac12).
\end{aligned}
\label{eq:coupleTspinor}
\end{equation}
The arrows retain the maximal spin in each product. The $(0,1)$ part gives the left--right conjugate of each line. 
In the $\Lambda$ rest frame, a representation $(j_L,j_R)$ contains each rotation spin from $|j_L-j_R|$ to $j_L+j_R$ once. The first line of Eq.~\eqref{eq:coupleTspinor} and its conjugate start at spin $3/2$, so their vacuum-to-$\Lambda$ matrix elements vanish. The second line and its conjugate each contain one spin-$1/2$ multiplet. Each therefore admits one coupling to the baryon. 

To construct the matrix element of the  twist-3 component $B^{(3) \mu\nu;I}$, let $Y^{\nu I}$ denote the symmetric, traceless tensor-spinor of rank $r=N+1$ defined by Eq.~\eqref{eq:bdaGeneralME}. It obeys
\begin{equation}
 z_\nu z_IY^{\nu I}=\mathcal Q_r(z,P)u_\Lambda
  =[A_r(P\cdot z,z^2)+m_\Lambda\slashed z B_r(P\cdot z,z^2)]u_\Lambda,
 \qquad \gamma_\nu Y^{\nu I}=0 .
 \label{eq:tensorY}
\end{equation}
The gamma-trace condition in Eq.~\eqref{eq:tensorY} removes the $j_L+j_R=r-1/2$ terms in Eq.~\eqref{eq:bdaWeylCoupling}. The left-handed component of $Y$ thus transforms as $((r+1)/2,r/2)$ and the right-handed component as its conjugate. With $r=N+1$, these are precisely the representations selected by the second line of Eq.~\eqref{eq:coupleTspinor} and its conjugate. To supply the antisymmetric indices of $B^{(3) \mu\nu;I}$, we construct a tensor $\gamma^\mu Y^{\nu I}-\gamma^\nu Y^{\mu I}$  which obeys
\edef\tensorYBaseTag{\theequation}
\begin{equation}
 \gamma_\mu(\gamma^\mu Y^{\nu I}-\gamma^\nu Y^{\mu I})
 =4Y^{\nu I}-(2Y^{\nu I}-\gamma^\nu\gamma_\mu Y^{\mu I})
 =2Y^{\nu I}.
 \label{eq:tensorEmbeddingTrace}
\end{equation}
Contraction with $\gamma_\mu/2$ therefore recovers $Y^{\nu I}$ from the antisymmetric combination. To fix its overall coefficient, contract Eq.~\eqref{eq:tensorEmbeddingTrace} with $n_\nu n_I$ and insert Eq.~\eqref{eq:tensorY} to obtain
\[
 \frac{1}{4p}\sum_\sigma\bar u_\Lambda\slashed n\,n_\nu n_I
 \gamma_\mu(\gamma^\mu Y^{\nu I}-\gamma^\nu Y^{\mu I})
 =\frac{2}{4p}\sum_\sigma\bar u_\Lambda\slashed n\mathcal Q_r(n,P)u_\Lambda
 =2p^r ,
\]
where $A_r(p,0)=p^r$. Matching to Eq.~\eqref{eq:tensorLCmoment} fixes the coefficient,
\begin{equation}
 \langle0|B^{(3)\,\mu\nu;I}_{n_1,n_2}|\Lambda(P,\sigma)\rangle
 = f_\Lambda^T\mathcal M^T_{n_1,n_2}
       (\gamma^\mu Y^{\nu I}-\gamma^\nu Y^{\mu I}).
 \label{eq:tensorEmbedding}
\end{equation}

To evaluate the transverse sum in Eq.~\eqref{eq:tensorQuasiMoment}, contract the derivative indices in Eq.~\eqref{eq:tensorEmbedding} with $\lambda_I$ and write $Y^\nu=\lambda_IY^{\nu I}$. For a vector $w$ in the $t$--$z$ plane, let $Y(w)=w_\nu Y^\nu$. The lattice operator selects $w=v_t$. Substituting Eq.~\eqref{eq:tensorEmbedding} for the twist-3 part of the matrix element in Eq.~\eqref{eq:tensorQuasiMoment} gives
\begin{align}
 \widetilde{\mathcal M}^{T,(3,0)}_{\perp,t;n_1,n_2}
 &=\frac{\mathcal M^T_{n_1,n_2}}{4pP_z^r}
     \sum_\sigma\bar u_\Lambda\slashed n V(v_t),\nonumber\\
 V(w)&\equiv\frac12\sum_{i=1}^2\gamma_i\big[\gamma^iY(w)-\slashed wY^i\big]
       =Y(w)+\frac12\slashed w\sum_{i=1}^2\gamma_iY^i .\label{eq:tensorTransverse}
\end{align}
The transverse components $Y^i$ contribute even at $P^i=\lambda^i=0$. Differentiating Eq.~\eqref{eq:tensorY} with respect to $z_i$ and then setting $z=\lambda$ leaves only the derivative of $\slashed z$ on the right. Symmetry of the $r$ indices on the left gives
\begin{equation}
 Y^i=\frac1r\left.\frac{\partial}{\partial z_i}
               [\mathcal Q_r(z,P)u_\Lambda]\right|_{z=\lambda}
     =\frac{m_\Lambda}{r}\gamma^i B_r(P_z,-1)u_\Lambda ,
 \qquad i=1,2 .
 \label{eq:tensorTransDerivative}
\end{equation}
For the spin projection in Eq.~\eqref{eq:tensorTransverse}, the condition $\gamma_\mu Y^\mu=0$ in Eq.~\eqref{eq:tensorY} and $n=(v_t+\lambda)/\sqrt2$ give
\begin{equation}
 \slashed n V(v_t)=\frac{\slashed nY(n)}{\sqrt2}
                  =\frac12\slashed n\big[Y(v_t)+Y(\lambda)\big] .
 \label{eq:tensorAverage}
\end{equation}
After summing over the baryon spin and dividing by $4pP_z^r$, $Y(v_t)$ gives the temporal factor $T_r(c)$ and $Y(\lambda)$ the spatial factor $F_r(c)$ of Eq.~\eqref{eq:tmRatios}. Inserting Eq.~\eqref{eq:tensorAverage} into the twist-3 term of Eq.~\eqref{eq:tensorTransverse} thus yields
\begin{equation}
 \widetilde{\mathcal M}^{T,(3,0)}_{\perp,t;n_1,n_2}
 =G_r(c)\mathcal M^T_{n_1,n_2},\qquad
 G_r(c)=\frac{T_r(c)+F_r(c)}2
       =1+\frac{r^2+1}{r+1}c+O(c^2),
 \label{eq:tensorMomentFactor}
\end{equation}
where the expansion uses Eq.~\eqref{eq:tmSmall} with $c=m_\Lambda^2/(4P_z^2)$.

\subsection{Momentum-space distribution and mass subtraction}
\label{sec:tensorSubtract}

Using $a=(1+\Delta)/2$, $b=(1-\Delta)/2$ and $\Delta=E/P_z$, take the average of the two lines in Eq.~\eqref{eq:tmRootFactors}. Since $\Delta+1=2a$, this gives
\begin{equation}
 G_r=\frac{a^{r+1}}{\Delta}
    -\frac{2ab}{\Delta^3}\frac{a^{r+1}-b^{r+1}}{r+1}
    +\frac{2ab^2}{\Delta^3}\frac{a^r-b^r}{r}.
 \label{eq:tensorRoots}
\end{equation}
Extend $\phi^T$ by zero outside $\mathcal S$. The moment of $\phi^T(x_1/s,x_2/s)$ is $s^{N+2}\mathcal M^T_{n_1,n_2}$ for either sign of $s$, as in Eq.~\eqref{eq:tmDilation}. With $N+2=r+1$, the two differences of powers in Eq.~\eqref{eq:tensorRoots} give the measures $ds/s$ and $ds/s^2$ in Eq.~\eqref{eq:tmScaleWeights}. The resummed tree-level result is
\begin{align}
 \widetilde\phi_{\perp,t}^{T,(3,0)}(x_1,x_2;P_z)
 &=\frac{1}{\Delta}\phi^T\!\left(\frac{x_1}{a},\frac{x_2}{a}\right)
  -\frac{2ab}{\Delta^3}\int_b^a\frac{ds}{s}\,
           \phi^T\!\left(\frac{x_1}{s},\frac{x_2}{s}\right)\nonumber\\
 &\quad+\frac{2ab^2}{\Delta^3}\int_b^a\frac{ds}{s^2}\,
           \phi^T\!\left(\frac{x_1}{s},\frac{x_2}{s}\right).
 \label{eq:tensorDA}
\end{align}
Equation~\eqref{eq:tensorDA} uses the fixed light-cone reference factor of Eq.~\eqref{eq:tensorFourier}. It reduces to $\phi^T(x_1,x_2)$ at $c=0$.

For comparison with lattice data one must also specify the local normalization. In the isospin limit $\phi^T(x_1,x_2)=-\phi^T(x_2,x_1)$ for the $\Lambda$, so $\mathcal M^T_{0,0}=0$ and the local tensor numerator vanishes. In Ref.~\cite{LPC:2026mvw} the tensor amplitude is divided by the nonzero local axial temporal amplitude at the same $P_z$. In the convention of Eq.~\eqref{eq:tensorFourier}, the raw tensor-to-axial correlator ratio is multiplied by $f_\Lambda^{\rm LC}/(2f_\Lambda^T)$. Denoting the resulting quasi-DA by a hat, the tree-level normalizations read
\begin{align}
 \widehat{\widetilde\phi}_{\perp,t}^{T,\mathrm{full},0}(x_1,x_2;P_z)
 &=\frac{\widetilde\phi_{\perp,t}^{T,\mathrm{full},0}(x_1,x_2;P_z)}
             {\widetilde{\mathcal M}_{\gamma^t;0,0}^{\mathrm{full},0}},
 &\quad
 \widehat{\widetilde{\mathcal M}}_{\perp,t;n_1,n_2}^{T,(3,0)}
 &=\frac{G_{N+1}(c)}{T_1(c)}\mathcal M^T_{n_1,n_2}.
 \label{eq:tensorNormalizations}
\end{align}
The first expression uses the measured axial matrix element. At tree level Eq.~\eqref{eq:ht4localnorm} gives it as $\widetilde{\mathcal M}_{\gamma^t;0,0}^{\mathrm{full},0}=T_1(c)+m_\Lambda h^{(4)}_{0,0}/[P_z(E+P_z)]$. For the twist-3 moment, the local axial denominator reduces to $T_1(c)=(1+3\Delta)/4$ by Eq.~\eqref{eq:tmLocalFactors}. The moment ratio in Eq.~\eqref{eq:tensorNormalizations} then has the expansion
\begin{equation}
 \frac{G_{N+1}}{T_1}
 =1+c\left[\frac{(N+1)^2+1}{N+2}-\frac32\right]+O(c^2)
 =1+c\left[N-\frac32+\frac{2}{N+2}\right]+O(c^2).
 \label{eq:tensorNormalizedExpansion}
\end{equation}
At the distribution level, the same normalization amounts to dividing Eq.~\eqref{eq:tensorDA} by $T_1(c)$. Since $a=1+c+O(c^2)$ and $1/[\Delta T_1(c)]=1-\tfrac72c+O(c^2)$, expanding the rescaled arguments and the scale integrals as in Eq.~\eqref{eq:tmDAFirstOrder} gives
\begin{align}
 \widehat{\widetilde\phi}_{\perp,t}^{T,(3,0)}(x_1,x_2;P_z)
 &=\frac{\widetilde\phi_{\perp,t}^{T,(3,0)}(x_1,x_2;P_z)}{T_1(c)}\nonumber\\
 &=\phi^T(x_1,x_2)
   -c\left(\tfrac72+x_1\partial_{x_1}+x_2\partial_{x_2}\right)\phi^T(x_1,x_2)\nonumber\\
 &\quad+2c\int_0^1\frac{ds}{s}\,\phi^T\!\left(\frac{x_1}{s},\frac{x_2}{s}\right)+O(c^2).
 \label{eq:tensorForwardLinear}
\end{align}
To invert Eq.~\eqref{eq:tensorForwardLinear} through first order in $c$, we reverse the signs of the two $O(c)$ terms and evaluate them on the normalized quasi-DA. Applied to lattice data with the measured axial denominator of Eq.~\eqref{eq:tensorNormalizations}, this gives
\begin{align}
 \widehat{\widetilde\phi}^{T,[1]}_{\perp,t,\mathrm{sub}}(x_1,x_2;P_z,\mu_R)
 ={}&\widehat{\widetilde\phi}^{T,\mathrm{full}}_{\perp,t}(x_1,x_2;P_z,\mu_R)\nonumber\\
 &+c\left(\tfrac72+x_1\partial_{x_1}+x_2\partial_{x_2}\right)
      \widehat{\widetilde\phi}^{T,\mathrm{full}}_{\perp,t}(x_1,x_2;P_z,\mu_R)\nonumber\\
 &-2c\int_0^1\frac{ds}{s}\,
      \widehat{\widetilde\phi}^{T,\mathrm{full}}_{\perp,t}
                 \!\left(\frac{x_1}{s},\frac{x_2}{s};P_z,\mu_R\right).
 \label{eq:tensorDataSub}
\end{align}
This relation removes the tree-level twist-3 mass term through order $c$ and preserves the antisymmetry under $x_1\leftrightarrow x_2$. Independent power corrections to the full data remain. 

\section{Numerical analysis}\label{sec:numerical}

\begin{figure}[!tp]
 \centering
 \includegraphics[width=\linewidth]{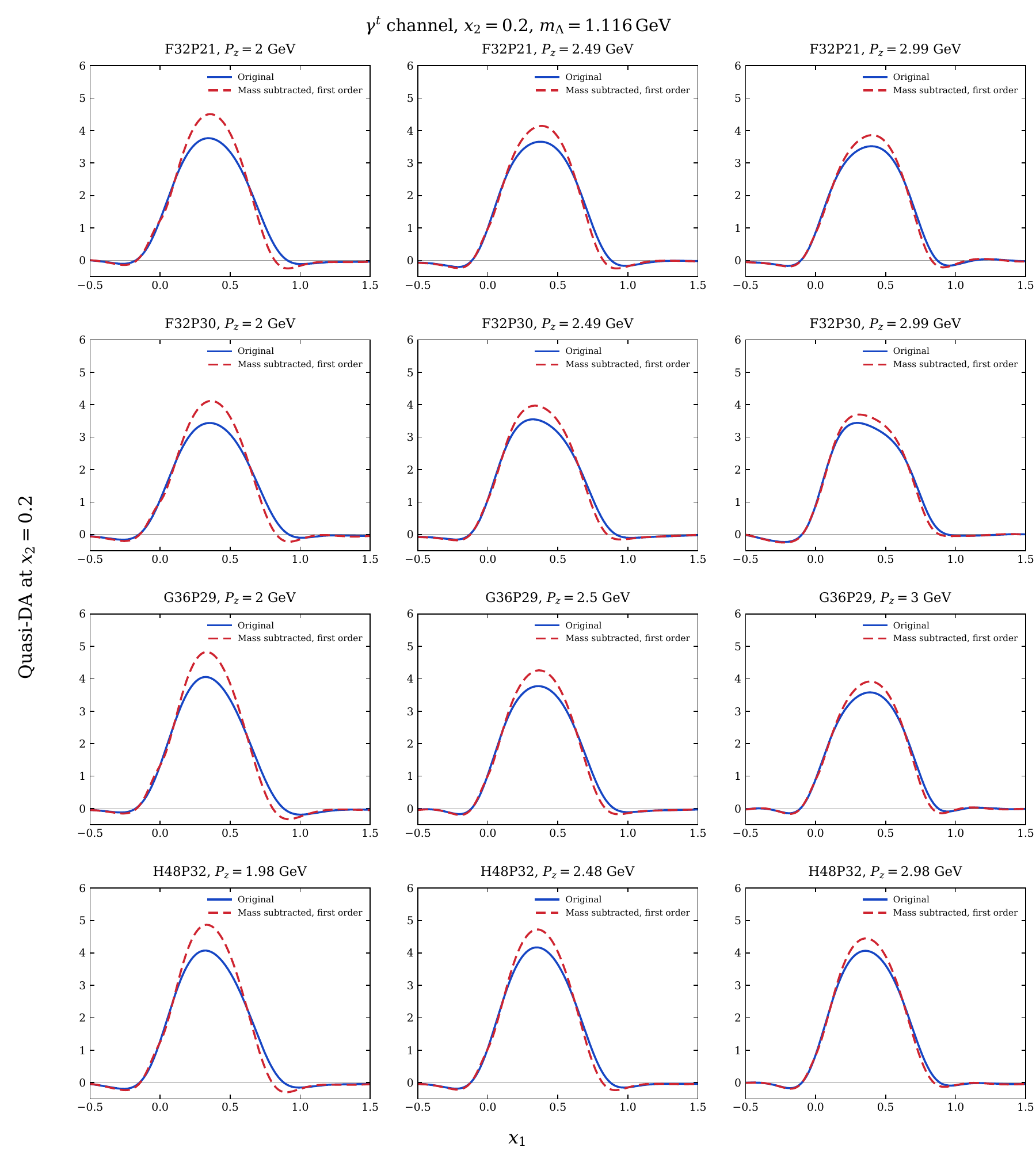}
 \caption{Locally normalized axial temporal $\Lambda$ quasi-DA at $x_2=0.2$ before (blue solid) and after (red dashed) the first-order target-mass subtraction in Eq.~\eqref{eq:match5datasub}. Each row shows one lattice ensemble from Ref.~\cite{LPC:2026mvw}. The baryon momentum $P_z$ is indicated in each panel. Both curves are obtained from the ensemble mean.}
 \label{fig:mass_correction}
\end{figure}

We apply the first-order target-mass subtraction in Eq.~\eqref{eq:match5datasub} to the two-dimensional axial temporal quasi-DA of $\Lambda$ baryon from Ref.~\cite{LPC:2026mvw}. We use the F32P21, F32P30, G36P29 and H48P32 ensembles, each at $P_z$ near $2.0$, $2.5$ and $3.0$~GeV. Ref.~\cite{LPC:2026mvw} uses $C\gamma_5\gamma^t$ in the axial diquark operator and divides the nonlocal matrix element by its local value at the same $P_z$. We therefore use the locally normalized temporal channel with $\kappa_{\gamma^t}=3$ in Eq.~\eqref{eq:match5datasub}. For this comparison we use a common mass $m_\Lambda=1.116$~GeV for all four ensembles. With the momentum of each data set, the expansion parameter $c=m_\Lambda^2/(4P_z^2)$ ranges from $0.035$ to $0.079$.

Figure~\ref{fig:mass_correction} shows the slice at $x_2=0.2$. Although only $x_1$ varies along this slice, the subtraction requires the quasi-DA away from it. For each momentum we interpolate the mean values on the full $201\times201$ grid, $-0.5\leq x_1,x_2\leq1.5$, with a bicubic spline. We take both partial derivatives in Eq.~\eqref{eq:match5datasub} on this interpolant before setting $x_2=0.2$. For the integral in the same equation, we set $t=1/s$ so that the integration in Eq.~\eqref{eq:match5datasub} follows the ray $(t x_1,0.2t)$ from $t=1$ outward. We set the unmeasured contribution beyond the region $-0.5\leq x_1,x_2\leq1.5$ to zero, so the integral used in the figure is
\begin{equation}
 \int_1^{t_{\max}}\frac{dt}{t}\,
 \widehat{\widetilde\phi}^{\rm full}_{\gamma^t}
       (t x_1,0.2t;P_z,\mu_R),\qquad
 t_{\max}=\max\{t\geq 1\mid(t x_1,0.2t)\in[-0.5,1.5]^2\}.
 \label{eq:num6ray}
\end{equation}
The contribution of the tails beyond this square is a systematic uncertainty of the subtraction.

The subtraction raises the quasi-DA near its peak on all four ensembles. To quantify the change, we compare the two curves at the $x_1$ where the original curve reaches its maximum. On F32P21 at $P_z=2.00$~GeV, for example, the quasi-DA at $x_1=0.345$ increases from $3.77$ to $4.50$, a $19.6\%$ increase. Across the four ensembles the increase ranges from $19.0\%$ to $19.7\%$ at $P_z\simeq2.0$~GeV, from $11.8\%$ to $13.3\%$ at $2.5$~GeV, and from $7.4\%$ to $9.7\%$ at $3.0$~GeV. This decrease is consistent with the explicit $1/P_z^2$ factor in the first-order subtraction. The maximum of each subtracted mean curve moves toward larger $x_1$ by $0.0025$ to $0.015$. These shifts are comparable to or smaller than the input grid spacing of $0.01$, and their statistical significance cannot be determined from the mean values alone.

It should be mentioned that the red curves in Figure~\ref{fig:mass_correction} are finite-momentum quasi-DAs. As Eq.~\eqref{eq:match5residual} shows, they still contain the leading-power one-loop matching term, higher-twist contributions and residual mass corrections of order $c^2$ and $a_s c$. Extracting the LCDA would also require inverse matching as in Eq.~\eqref{eq:match5inverse} and control of the remaining momentum dependence.

\section{Conclusion}\label{sec:conclusion}

We have derived the tree-level target-mass corrections to the vector, axial and transverse tensor quasi-DAs of the $\Lambda$ baryon. For a moment with $N$ derivatives, let $r=N+1$ and $c=m_\Lambda^2/(4P_z^2)$. The vector and axial operators have the same twist-3 tensor-spinor structure. Its spatial and temporal components multiply the moments of the corresponding LCDA by $F_r(c)$ and $T_r(c)$, respectively. For the transverse tensor operator the lattice spin projection gives $G_r(c)=(F_r(c)+T_r(c))/2$. Equations~\eqref{eq:tmSpatialDA}, \eqref{eq:tmTemporalDA} and \eqref{eq:tensorDA} resum these factors to all orders in $c$ at fixed light-cone normalization.

The local reference used for the lattice ratio determines the mass subtraction. For an axial numerator divided by its own local matrix element, the twist-3 factors are $F_r/F_1$ or $T_r/T_1$. For the $\Lambda$ vector and tensor temporal numerators, both of which vanish at zero separation, the reference is the local axial temporal matrix element. Its twist-3 part gives the factors $T_r/T_1$ and $G_r/T_1$, respectively. Equations~\eqref{eq:match5datasub} and \eqref{eq:tensorDataSub} give the resulting first-order subtractions. The full measured denominator can itself contain a power correction, as shown in Eq.~\eqref{eq:ht4localnorm}.

For the axial operator, the symmetric twist-4 and twist-5 terms contribute to fixed moments starting at order $1/P_z^2$ when their reduced matrix elements are nonzero. The temporal operator also admits a mixed-symmetry contribution at this order, as shown in Eq.~\eqref{eq:ht4leading}. The same operator decomposition applies to $V$ with its own reduced matrix elements. Section~\ref{sec:higherlattice} proposes a finite-difference construction of these matrix elements from three-quark correlators with displaced endpoints. The precision attainable with this construction remains to be determined.

Figure~\ref{fig:mass_correction} illustrates the first-order mass subtraction for the axial temporal quasi-DA on four lattice ensembles. At $x_2=0.2$ the amplitude at its original peak increases by $19.0$--$19.7\%$ at $P_z\simeq2.0$~GeV and by $7.4$--$9.7\%$ at $P_z\simeq3.0$~GeV. To extract the LCDA from these finite-momentum distributions one must also apply inverse matching and control the residual momentum dependence from higher-twist matrix elements. Equation~\eqref{eq:match5inverse} gives the leading-power inverse through one loop. Extending the subtraction through order $\alpha_s m_\Lambda^2/P_z^2$ requires the finite-mass Wilson coefficients identified in Eq.~\eqref{eq:match5oneWilson}.

\appendix

\section{Covariant translations}\label{app:bdaTransport}

Let $W(x,y)$ transport a quark from $y$ to $x$ along the straight segment. For a fixed direction $v$ and real parameter $s$, the endpoint moves along the segment, so
\begin{align}
 \frac{d}{ds}W(x,x+sv)&=-ig_s W(x,x+sv)\,v\cdot A(x+sv),\nonumber\\
 \frac{d}{ds}\big[W(x,x+sv)q(x+sv)\big]
 &=W(x,x+sv)\,(v\cdot Dq)(x+sv).
 \label{eq:Psider1}
\end{align}
The second line uses $D_\mu=\partial_\mu-ig_s A_\mu$. Replacing $q$ successively by $(v\cdot D)^nq$ gives the $n$th derivative at $s=0$. The corresponding covariant translation has the formal local expansion
\begin{equation}
 W(x,x+sv)q(x+sv)
 =\sum_{n=0}^{\infty}\frac{s^n}{n!}\,[(v\cdot D)^nq](x)
 =[e^{s v\cdot D}q](x).
 \label{eq:expoDeriv}
\end{equation}

For the one- and two-derivative insertions in Appendix~\ref{app:ht4convolutions}, apply Eq.~\eqref{eq:expoDeriv} to the quark field and then to the entire expression to the right of each insertion. Taking the base point to be the origin and denoting signed distances along $v$ by $a,b,c$ gives
\begin{align}
 [e^{a v\cdot D}D_\nu e^{b v\cdot D}q](0)
 &=W(0,av)D_\nu(av)W(av,(a+b)v)q((a+b)v),
 \label{eq:formula1}\\
 [e^{c v\cdot D}D_\mu e^{a v\cdot D}D_\nu e^{b v\cdot D}q](0)
 &=W(0,cv)D_\mu(cv)W(cv,(c+a)v)\nonumber\\
 &\quad{}\times D_\nu((c+a)v)W((c+a)v,(c+a+b)v)q((c+a+b)v).
 \label{eq:formula2}
\end{align}
Each $D_\mu(x)$ acts on the transported field to its right at the base point $x$, with the subsequent displacement vectors held fixed. For instance, the derivative in Eq.~\eqref{eq:formula1} acts on $W(x,x+bv)q(x+bv)$ before $x$ is set to $av$.

\section{Tensor-spinor projection}
\label{app:bdaProjector}

\subsection{Trace subtraction}
\label{app:bdaConstruction}

Fix the derivative counts $(n_1,n_2)$ in Eq.~\eqref{eq:bdaLocal}, so that $r=n_1+n_2+1$.
For a symmetric tensor-spinor of this rank, write
\[
 X_r(z)=z_{\mu_1}\cdots z_{\mu_r}X^{(\mu_1\cdots\mu_r)} .
\]
Each vector index makes the same contribution when we differentiate with respect to $z$. The two traces therefore take the form
\begin{align}
 \Box_z X_r&=r(r-1)z_{\mu_3}\cdots z_{\mu_r}
      g_{\mu_1\mu_2}X^{(\mu_1\cdots\mu_r)},\qquad r\geq2,\nonumber\\
 \slashed{\partial}_z X_r&=r z_{\mu_2}\cdots z_{\mu_r}
      \gamma_{\mu_1}X^{(\mu_1\cdots\mu_r)},\qquad r\geq1 .
 \label{eq:bdaAppTraces}
\end{align}
Thus the required polynomial lies in the kernel of $\slashed{\partial}_z$. We first construct its metric-traceless part and then remove the gamma trace that remains within that subspace.

In Eq.~\eqref{eq:bdaAppTraces}, each $\Box_z$ takes one trace and lowers the degree by two. Multiplication by $z^2$ restores the two vector indices as a metric tensor. We therefore seek the harmonic part in the form
\begin{equation}
 \mathcal H_r X_r=\sum_{k=0}^{\lfloor r/2\rfloor}h_{rk}(z^2)^k\Box_z^k X_r,
 \qquad h_{r0}=1,\qquad \Box_z\mathcal H_r X_r=0 .
 \label{eq:bdaHAnsatz}
\end{equation}
To determine $h_{rk}$, let $f_\ell(z)$ be homogeneous of degree $\ell$, so that $z\cdot\partial_z f_\ell=\ell f_\ell$. In $d$ spacetime dimensions, $\partial_{z^\mu}(z^2)^k=2kz_\mu(z^2)^{k-1}$ and $\Box_z(z^2)^k=2k(d+2k-2)(z^2)^{k-1}$ for $k\geq1$. The product rule then gives
\begin{align}
 \Box_z[(z^2)^k f_\ell]
 &=(z^2)^k\Box_z f_\ell
   +2k(z^2)^{k-1}\big[(d+2k-2)f_\ell+2z\cdot\partial_z f_\ell\big]\nonumber\\
 &=(z^2)^k\Box_z f_\ell+2k(d+2\ell+2k-2)(z^2)^{k-1}f_\ell .
 \label{eq:bdaHStep}
\end{align}
Set $d=4$ and $f_\ell=\Box_z^k X_r$ in Eq.~\eqref{eq:bdaHStep}. Its degree is $\ell=r-2k$, so the second coefficient reduces to $4k(r-k+1)$. Applying $\Box_z$ to Eq.~\eqref{eq:bdaHAnsatz} and collecting terms with the same number of traces yields
\begin{equation}
 \Box_z\mathcal H_r X_r
 =\sum_{k=1}^{\lfloor r/2\rfloor}
   [h_{r,k-1}+4k(r-k+1)h_{rk}](z^2)^{k-1}\Box_z^k X_r .
 \label{eq:bdaHCollect}
\end{equation}
The term beyond the upper limit vanishes because $\Box_z^{\lfloor r/2\rfloor+1}X_r=0$. Setting each bracket in Eq.~\eqref{eq:bdaHCollect} to zero and iterating from $h_{r0}=1$ gives
\begin{equation}
 h_{rk}=-\frac{h_{r,k-1}}{4k(r-k+1)}
 =\prod_{j=1}^k\frac{-1}{4j(r-j+1)}
 =\frac{(-1)^k(r-k)!}{4^k k!\,r!}.
 \label{eq:bdaHCoef}
\end{equation}
This gives the harmonic projector in Eq.~\eqref{eq:bdaHarmonic}. If $X_r$ is already harmonic, only the $k=0$ term of Eq.~\eqref{eq:bdaHAnsatz} survives. Thus $\mathcal H_r$ is the identity on its image and $\mathcal H_r^2=\mathcal H_r$.

To remove the remaining gamma trace, write $Y_r=\mathcal H_rX_r$. Its gamma trace $\slashed{\partial}_zY_r$ has degree $r-1$ and is itself gamma-traceless, since $\slashed{\partial}_z^{\,2}Y_r=\Box_zY_r=0$. To form a degree-$r$ subtraction from this trace, we restore one vector index by multiplying by $\slashed z$. For an arbitrary spinor polynomial $F$, differentiating $\slashed z F$ gives
\begin{align}
 \slashed{\partial}_z(\slashed z F)
 &=\gamma^\mu\gamma_\mu F+\gamma^\mu\slashed z\,\partial_{z^\mu}F\nonumber\\
 &=4F+2z\cdot\partial_z F-\slashed z\slashed{\partial}_zF .
 \label{eq:bdaClifford}
\end{align}
Taking $F=\slashed{\partial}_zY_r$ in Eq.~\eqref{eq:bdaClifford} gives
\begin{align}
 \slashed{\partial}_z(\slashed z\slashed{\partial}_zY_r)
 &=[4+2(r-1)]\slashed{\partial}_zY_r-\slashed z\Box_zY_r\nonumber\\
 &=2(r+1)\slashed{\partial}_zY_r .
 \label{eq:bdaAppGamma}
\end{align}
In fact, for any gamma-traceless $F$ of degree $r-1$, Eq.~\eqref{eq:bdaClifford} gives $\slashed{\partial}_z[\slashed z F/(2(r+1))]=F$. In particular, $\slashed z\slashed{\partial}_zY_r/[2(r+1)]$ has the same gamma trace as $Y_r$. Subtracting it gives the projector in Eq.~\eqref{eq:bdaHarmonic}, with
\begin{align}
 \slashed{\partial}_z\mathcal P_r X_r
 &=\slashed{\partial}_zY_r
   -\frac{\slashed{\partial}_z(\slashed z\slashed{\partial}_zY_r)}{2(r+1)}=0,\nonumber\\
 \Box_z\mathcal P_r X_r
 &=\slashed{\partial}_z\big(\slashed{\partial}_z\mathcal P_r X_r\big)=0 .
 \label{eq:bdaAppProjected}
\end{align}
If $\slashed{\partial}_zX_r=0$ to begin with, then $\Box_zX_r=0$, $\mathcal H_rX_r=X_r$, and the gamma-trace subtraction also vanishes. It follows that $\mathcal P_r^2=\mathcal P_r$.

\subsection{Covariant matrix element}
\label{app:bdaCovariant}

To compute $A_r$ and $B_r$ in Eq.~\eqref{eq:bdaGeneralME}, we apply $\mathcal P_r$ to $\chi^r u_\Lambda$. The first step in Eq.~\eqref{eq:bdaHarmonic} is the harmonic projection of $\chi^r$. Since $\partial_{z^\mu}\chi=P_\mu$ and $P^2=m_\Lambda^2$, each application of $\Box_z$ reduces the power of $\chi$ by two. For $0\leq k\leq\lfloor r/2\rfloor$ this gives $\Box_z^k\chi^r=m_\Lambda^{2k}r!\,\chi^{r-2k}/(r-2k)!$. Denote the resulting scalar polynomial by $h_r$. Substitution into the harmonic projector gives
\begin{equation}
 h_r(\chi,\rho)=\mathcal H_r\chi^r
 =\sum_{k=0}^{\lfloor r/2\rfloor}
 \frac{(-1)^k(r-k)!}{4^k k!(r-2k)!}
       (m_\Lambda^2\rho)^k\chi^{r-2k},
 \qquad \chi=P\cdot z,\quad \rho=z^2 .
 \label{eq:bdaAppScalar}
\end{equation}
For the gamma-trace subtraction, differentiate with respect to both arguments of $h_r$. With $\partial_{z^\mu}\rho=2z_\mu$, $(\slashed z)^2=\rho$ and $\slashed P u_\Lambda=m_\Lambda u_\Lambda$, one finds
\begin{align}
 \slashed z\slashed{\partial}_z(h_r u_\Lambda)
 &=\slashed z\big[\slashed P\,\partial_\chi h_r
                 +2\slashed z\,\partial_\rho h_r\big]u_\Lambda\nonumber\\
 &=\big[m_\Lambda\slashed z\,\partial_\chi h_r
                 +2\rho\,\partial_\rho h_r\big]u_\Lambda .
 \label{eq:bdaAppHDerivative}
\end{align}
Substituting Eq.~\eqref{eq:bdaAppHDerivative} into Eq.~\eqref{eq:bdaHarmonic} gives
\begin{equation}
 \mathcal Q_r(z,P)u_\Lambda
 =\left[\underbrace{h_r-\frac{\rho}{r+1}\partial_\rho h_r}_{A_r(\chi,\rho)}
       +m_\Lambda\slashed z\,
          \underbrace{\left(-\frac{1}{2(r+1)}\partial_\chi h_r\right)}_{B_r(\chi,\rho)}
   \right]u_\Lambda .
 \label{eq:bdaExplicitQ}
\end{equation}

To see why the scalar and spinor terms share a single normalization, take an arbitrary gamma-traceless polynomial $[A(\chi,\rho)+m_\Lambda\slashed z B(\chi,\rho)]u_\Lambda$ of degree $r$. Here $A$ has degree $r$ in $z$ and $B$ has degree $r-1$. We leave their normalization free. Subscripts $\chi$ and $\rho$ denote partial derivatives. Applying the chain rule and using $\slashed P\slashed z+\slashed z\slashed P=2\chi$ gives
\begin{align}
 \slashed{\partial}_z(Au_\Lambda)
   &=(m_\Lambda A_\chi+2\slashed z A_\rho)u_\Lambda,\nonumber\\
 \slashed{\partial}_z(\slashed z B u_\Lambda)
   &=\big[4B+2\chi B_\chi+2\rho B_\rho
                     -m_\Lambda\slashed z B_\chi\big]u_\Lambda .
 \label{eq:bdaAppDirac}
\end{align}
Homogeneity gives $\chi B_\chi+2\rho B_\rho=(r-1)B$. The coefficients of $u_\Lambda$ and $\slashed z u_\Lambda$ in the gamma trace must vanish separately for generic $z$. Using Eq.~\eqref{eq:bdaAppDirac} and $m_\Lambda>0$, these two conditions reduce to
\begin{equation}
 A_\chi+2(r+1)B-2\rho B_\rho=0,\qquad
 2A_\rho=m_\Lambda^2 B_\chi .
 \label{eq:bdaAppAB}
\end{equation}
Write the two polynomials as
\begin{equation}
 A=\sum_{k=0}^{\lfloor r/2\rfloor}
       a_k(m_\Lambda^2\rho)^k\chi^{r-2k},\qquad
 B=\sum_{k=0}^{\lfloor(r-1)/2\rfloor}
       b_k(m_\Lambda^2\rho)^k\chi^{r-1-2k}.
 \label{eq:bdaAppSeries}
\end{equation}
At order $\rho^k$, the first condition in Eq.~\eqref{eq:bdaAppAB} relates $a_k$ and $b_k$. In the second condition the derivative $A_\rho$ lowers the power of $\rho$ by one, so it relates $b_k$ and $a_{k+1}$. Equating the coefficients gives
\begin{align}
 b_k&=-\frac{r-2k}{2(r+1-k)}a_k,
 &0\leq k\leq\lfloor(r-1)/2\rfloor,\nonumber\\
 a_{k+1}&=\frac{r-1-2k}{2(k+1)}b_k,
 &0\leq k<\lfloor r/2\rfloor .
 \label{eq:bdaAppRecurrence}
\end{align}
Starting from $a_0$, the two lines of Eq.~\eqref{eq:bdaAppRecurrence} determine the successive coefficients up to the limits in Eq.~\eqref{eq:bdaAppSeries}. Thus the only free coefficient is the multiplier of $\chi^r$ in $A$. The kinematic polynomial in Eq.~\eqref{eq:bdaGeneralME} uses $a_0=1$. Multiplying it by $f_\Lambda^{\rm LC}C_{n_1,n_2}$ restores the normalization of the three-quark matrix element, including the case $C_{n_1,n_2}=0$.

\section{Moment factors and scale integrals}
\label{app:tmRoots}

\subsection{Root representation}

To obtain the first line of Eq.~\eqref{eq:tmRootFactors}, differentiate $K_r=(a^{r+1}-b^{r+1})/\Delta$. From Eq.~\eqref{eq:tmRoots}, $da/dc=1/\Delta$, $db/dc=-1/\Delta$ and $d\Delta/dc=2/\Delta$. Hence
\begin{equation}
 K_r'=\frac{(r+1)(a^r+b^r)}{\Delta^2}
       -\frac{2(a^{r+1}-b^{r+1})}{\Delta^3}.
 \label{eq:tmKRootDer}
\end{equation}
Inserting Eq.~\eqref{eq:tmKRootDer} into Eq.~\eqref{eq:tmF} and collecting the terms proportional to $1/(r+1)$ gives
\begin{align}
 F_r&=\frac{a(a^{r+1}-b^{r+1})-c(a^r+b^r)}{\Delta}
      -\frac{b(a^{r+1}-b^{r+1})}{(r+1)\Delta^2}\nonumber\\
    &=\frac{a^{r+1}}{\Delta}
      -\frac{b(a^{r+1}-b^{r+1})}{(r+1)\Delta^2}.
 \label{eq:tmAppF}
\end{align}
The second line uses $ab=-c$ and $a^2-c=a$.

For the temporal factor, it is convenient to differentiate the two roots separately. Define the homogeneous function of degree $r$
\begin{equation}
 L_r(a,b)=\frac{a^{r+1}}{a-b}
   -\frac{b}{(a-b)^2}\frac{a^{r+1}-b^{r+1}}{r+1}.
 \label{eq:tmHomogeneous}
\end{equation}
On $a+b=1$ it equals $F_r(c)$. Treating $a,b$ as independent variables when taking partial derivatives gives $rL_r=(a\partial_a+b\partial_b)L_r$. Along $a+b=1$ we have $dF_r/d\Delta=(\partial_a-\partial_b)L_r/2$. Equation~\eqref{eq:tmT} therefore becomes
\begin{align}
 rT_r&=r\Delta F_r-(\Delta^2-1)\frac{dF_r}{d\Delta}
     =(a\partial_a-b\partial_b)L_r
     =rF_r-2b\,\partial_b L_r ,
 \label{eq:tmRootTimeDer}
\end{align}
where the middle equality uses $\Delta=a-b$ and $\Delta^2-1=-4ab$. Differentiating Eq.~\eqref{eq:tmHomogeneous} gives
\begin{align}
 (a-b)^3\partial_b L_r
 &=(a-b)(a^{r+1}+b^{r+1})
      -(a+b)\frac{a^{r+1}-b^{r+1}}{r+1}\nonumber\\
 &=r(a+b)\frac{a^{r+1}-b^{r+1}}{r+1}-2ab(a^r-b^r).
 \label{eq:tmRootPartial}
\end{align}
Substitution into Eq.~\eqref{eq:tmRootTimeDer} yields the second line of Eq.~\eqref{eq:tmRootFactors}.

\subsection{Integration over both branches}
\label{app:tmBranches}
For $c>0$, the roots in Eq.~\eqref{eq:tmRoots} satisfy $b<0<a$. Split the scale integral at zero and set $s=au$ on $0<s<a$ and $s=bu$ on $b<s<0$. The positive branch runs from $u=0$ to $1$, and the negative branch runs from $u=1$ to $0$. For the weight $ds/s$, the two substitutions give
\begin{align}
 \int_0^a\frac{ds}{s}\,\phi(x_1/s,x_2/s)
 &=\int_0^1\frac{du}{u}\,\phi\!\left(\frac{x_1}{au},\frac{x_2}{au}\right),\nonumber\\
 \int_b^0\frac{ds}{s}\,\phi(x_1/s,x_2/s)
 &=\int_1^0\frac{du}{u}\,\phi\!\left(\frac{x_1}{bu},\frac{x_2}{bu}\right)
 =-\int_0^1\frac{du}{u}\,\phi\!\left(\frac{x_1}{bu},\frac{x_2}{bu}\right).
 \label{eq:tmSplitOne}
\end{align}
With $a=f_+/2$ and $b=-f_-/2$, the second line has LCDA arguments $-2x_i/(u f_-)$ for $i=1,2$. Substituting Eq.~\eqref{eq:tmSplitOne} into Eq.~\eqref{eq:tmSpatialDA} and using $-b/\Delta^2=f_-/(2\Delta^2)$ gives Eq.~\eqref{eq:tmSpatialDAu}.

For the $ds/s^2$ term in Eq.~\eqref{eq:tmTemporalDA}, the measure supplies an additional factor of $1/a$ or $1/b$, giving
\begin{align}
 \int_0^a\frac{ds}{s^2}\,\phi(x_1/s,x_2/s)
 &=\frac{1}{a}\int_0^1\frac{du}{u^2}\,\phi\!\left(\frac{x_1}{au},\frac{x_2}{au}\right),\nonumber\\
 \int_b^0\frac{ds}{s^2}\,\phi(x_1/s,x_2/s)
 &=\frac{1}{b}\int_1^0\frac{du}{u^2}\,\phi\!\left(\frac{x_1}{bu},\frac{x_2}{bu}\right)
 =-\frac{1}{b}\int_0^1\frac{du}{u^2}\,\phi\!\left(\frac{x_1}{bu},\frac{x_2}{bu}\right).
 \label{eq:tmSplitTwo}
\end{align}
Since $b<0$, both weights in Eq.~\eqref{eq:tmSplitTwo} are positive. Multiplying them by $4ab^2/\Delta^3$ gives $4b^2/\Delta^3=f_-^2/\Delta^3$ and $-4ab/\Delta^3=f_-f_+/\Delta^3$, respectively. The $du/u$ difference in Eq.~\eqref{eq:tmSplitOne} carries the coefficient $-b(\Delta+2)/\Delta^3=f_-(\Delta+2)/(2\Delta^3)$. Inserting these three coefficients into Eq.~\eqref{eq:tmTemporalDA} gives Eq.~\eqref{eq:tmTemporalDAu}.

To check convergence at $u=0$, assume $\|\phi\|_1=\int_{\mathcal S}dy_1\,dy_2\,|\phi(y_1,y_2)|<\infty$ and extend $\phi$ by zero outside $\mathcal S$. For either branch, write $s=s_0 u$ with $s_0=a$ or $b$. At fixed $u>0$, the change of variables $x_i=s_0 u y_i$ gives
\begin{equation}
 \int_{\mathbb R^2}dx_1\,dx_2\,
 \left|\phi\!\left(\frac{x_1}{s_0 u},\frac{x_2}{s_0 u}\right)\right|
 =|s_0|^2 u^2\|\phi\|_1 .
 \label{eq:tmAbsolute}
\end{equation}
Thus for $0<\eta\le1$ and either weight $du/u^j$ appearing in Eqs.~\eqref{eq:tmSpatialDAu} and \eqref{eq:tmTemporalDAu}, the absolute integral from the endpoint to $\eta$ is
\begin{equation}
 \int_0^\eta\frac{du}{u^j}\int_{\mathbb R^2}dx_1\,dx_2\,
 \left|\phi\!\left(\frac{x_1}{s_0 u},\frac{x_2}{s_0 u}\right)\right|
 =|s_0|^2\|\phi\|_1\int_0^\eta u^{2-j}\,du
 =|s_0|^2\|\phi\|_1
 \begin{cases}
  \eta^2/2, & j=1,\\
  \eta, & j=2 .
 \end{cases}
 \label{eq:tmEndpointU}
\end{equation}
The two bounds vanish as $\eta\to0$. Thus the scale integral on each branch is absolutely convergent for almost every $(x_1,x_2)$ and needs no principal-value prescription. 

\section{Higher-twist projections}
\label{app:ht4details}

\subsection{Symmetric operators}
\label{app:ht4project}

Fix a local moment of rank $r=n_1+n_2+1$ and write $\rho=z^2$. In Eq.~\eqref{eq:ht4harmonicexp} the coefficient $X_{r-2a}$ of $\rho^a$ is harmonic and homogeneous of degree $r-2a$. To extract a specified coefficient $X_{r-2k}$, we first take $k$ Laplacians and then remove the traces that remain.

For any harmonic $X_j$ of degree $j$, the product rule in four dimensions gives
\begin{align}
 \Box_z(\rho^a X_j)
 &=\rho^a\Box_z X_j+4a\rho^{a-1}z\cdot\partial_zX_j
       +4a(a+1)\rho^{a-1}X_j \nonumber\\
 &=4a(j+a+1)\rho^{a-1}X_j \qquad(a\ge1).
 \label{eq:ht4traceLayer}
\end{align}
In the second line we used $z\cdot\partial_zX_j=jX_j$. Each further Laplacian lowers $a$ by one and leaves $j$ fixed. For $0\le k\le a$, Eq.~\eqref{eq:ht4traceLayer} therefore gives
\begin{align}
 \Box_z^k(\rho^a X_j)
 &=4^k\prod_{b=0}^{k-1}(a-b)(j+a+1-b)\,\rho^{a-k}X_j\nonumber\\
 &=4^k\frac{a!}{(a-k)!}\frac{(j+a+1)!}{(j+a+1-k)!}\,\rho^{a-k}X_j .
 \label{eq:ht4iteratedtrace}
\end{align}
At $k=a$ only a constant multiple of $X_j$ remains. One more Laplacian gives zero. Thus all terms with $a<k$ in Eq.~\eqref{eq:ht4harmonicexp} disappear under $\Box_z^k$.

For $a>k$ the surviving polynomial has a factor of $\rho$. To see why the harmonic projector removes it, take any homogeneous degree-$l$ polynomial of the form $\rho Y$. Every term of $\mathcal H_l(\rho Y)$ in Eq.~\eqref{eq:bdaHarmonic} is divisible by $\rho$, including its first term $\rho Y$. The result is also harmonic. If it were nonzero, write it as $\rho^b U$, taking $b\ge1$ as large as possible. Then $U$ has degree $l-2b$ and Eq.~\eqref{eq:bdaHStep} implies
\begin{equation}
 0=\Box_z(\rho^bU)
 =\rho^{b-1}\big[4b(l-b+1)U+\rho\Box_z U\big].
 \label{eq:ht4divisible}
\end{equation}
The coefficient $4b(l-b+1)$ is nonzero. Equation~\eqref{eq:ht4divisible} would make $U$ divisible by $\rho$, contrary to the choice of $b$. Hence $\mathcal H_l(\rho Y)=0$.

Now apply $\mathcal H_{r-2k}\Box_z^k$ to Eq.~\eqref{eq:ht4harmonicexp}. After the two eliminations just proved, the sole surviving term has $a=k$ and $j=r-2k$. Since $\mathcal H_jX_j=X_j$, Eq.~\eqref{eq:ht4iteratedtrace} yields
\begin{equation}
 \mathcal H_j\Box_z^k\mathcal O_r
 =\mathcal H_j\Box_z^k(\rho^kX_j)
 =4^kk!\frac{(j+k+1)!}{(j+1)!}\,X_j ,
 \qquad j=r-2k .
 \label{eq:ht4layerderiv}
\end{equation}
Solving Eq.~\eqref{eq:ht4layerderiv} for $X_j$ and using $j+k+1=r-k+1$ gives Eq.~\eqref{eq:ht4layer} in the main text.

There is still an open spinor index on $X_j$. For $j\ge1$, its gamma trace $\slashed{\partial}_zX_j$ is gamma-traceless, since $\slashed{\partial}_z^{\,2}X_j=\Box_zX_j=0$. It has degree $j-1$. The Clifford identity in Eq.~\eqref{eq:bdaClifford} consequently gives
\begin{align}
 \slashed{\partial}_z(\slashed z\slashed{\partial}_z X_j)
 &=[4+2(j-1)]\slashed{\partial}_zX_j-\slashed z\Box_zX_j
 =2(j+1)\slashed{\partial}_zX_j,\nonumber\\
 X_j&=\mathcal P_jX_j+\frac{\slashed z\slashed{\partial}_z}{2(j+1)}X_j,
 \qquad \slashed{\partial}_z\mathcal P_jX_j=0 .
 \label{eq:ht4layergamma}
\end{align}
The first term transforms as a gamma-traceless tensor-spinor of rank $j$ and has spin $j+1/2$. The second is determined by the gamma-traceless tensor-spinor $\slashed{\partial}_zX_j$ of rank $j-1$, hence has spin $j-1/2$. Multiplying back the $k$ metrics preserves each spin, while the canonical dimension of the original operator is $r+7/2$. Subtracting the two spins from this dimension gives $3+2k$ and $4+2k$, respectively. Equations~\eqref{eq:ht4layer} and \eqref{eq:ht4layergamma} then give
\begin{align}
 \mathcal O_r^{(3+2k)}
 &=\rho^k\mathcal P_jX_j
 =\frac{\rho^k(j+1)!}{4^k k!(r-k+1)!}\mathcal P_j\Box_z^k\mathcal O_r,
 \nonumber\\
 \mathcal O_r^{(4+2k)}
 &=\rho^k\frac{\slashed z\slashed{\partial}_z}{2(j+1)}X_j
 =\frac{\rho^k j!}{2\cdot4^k k!(r-k+1)!}
    \slashed z\slashed{\partial}_z\mathcal H_j\Box_z^k\mathcal O_r .
 \label{eq:ht4families}
\end{align}
Here $\mathcal P_j\mathcal H_j=\mathcal P_j$. The first line runs from $k=0$ to $\lfloor r/2\rfloor$. The second runs to $\lfloor(r-1)/2\rfloor$, since a degree-zero $X_0$ has vanishing gamma trace.

\subsection{The mixed temporal operator}
\label{app:ht4mixed}

Fix $(n_1,n_2)$ and let $r=n_1+n_2+1$. The curl in Eq.~\eqref{eq:ht4curl} vanishes for $r=1$. For $r\ge2$ write its coefficients as
\begin{equation}
 B_{\mu\nu}(z)=\partial_\nu V_\mu-\partial_\mu V_\nu
 =z^{\rho_1}\cdots z^{\rho_{r-2}}B_{\mu\nu;\rho_1\cdots\rho_{r-2}} .
 \label{eq:ht4mixcoeff}
\end{equation}
The pair $\mu,\nu$ is antisymmetric, the $\rho_i$ are symmetric, and the Dirac index is open. An antisymmetric pair has Lorentz representation $(1,0)\oplus(0,1)$. Each remaining vector index carries $(1/2,1/2)$ and the open spinor carries $(1/2,0)\oplus(0,1/2)$. Thus the largest possible $s=j_L+j_R$ and its geometric twist are
\begin{equation}
 s_\star=1+(r-2)+\tfrac12=r-\tfrac12,\qquad
 \tau_\star=(r+\tfrac72)-s_\star=4 .
 \label{eq:ht4mixspin}
\end{equation}
We construct the projector $\Pi_r^{(4)}$ onto all components with $s=s_\star$ before taking the temporal contraction in Eq.~\eqref{eq:ht4channelbalance}. A single gamma contraction of $B$ leaves $r-1$ vector indices and a spinor, which can still carry $s_\star$. The symmetric-tensor prescription in Eq.~\eqref{eq:bdaHarmonic} therefore does not isolate the mixed twist-4 operator.

To express the projector in two-component spinor notation, take $\sigma^\mu=(\mathbf 1_2,\boldsymbol{\sigma})$ and $\bar\sigma^\mu=(\mathbf 1_2,-\boldsymbol{\sigma})$, where $\boldsymbol{\sigma}$ are the Pauli matrices. An undotted index $A$ transforms as $(1/2,0)$ and a dotted index $\dot A$ as $(0,1/2)$. A vector index is converted into one of each by the invertible map~\cite{Dreiner:2008tw}
\begin{equation}
 w_{A\dot A}=\sigma^\mu_{A\dot A}w_\mu,\qquad
 w_\mu=\tfrac12\bar\sigma_\mu^{\dot A A}w_{A\dot A},\qquad
 \tfrac12\operatorname{tr}(\bar\sigma_\mu\sigma^\nu)=\delta_\mu^{\ \nu}.
 \label{eq:ht4mixsigma}
\end{equation}
Apply Eq.~\eqref{eq:ht4mixsigma} to each of the $r$ vector indices in Eq.~\eqref{eq:ht4mixcoeff}. Hold the last $r-2$ index pairs and the open Weyl index fixed. Interchanging the first two pairs $A\dot A,B\dot B$ changes the sign of the tensor. Its part symmetric in $A,B$ must then be antisymmetric in $\dot A,\dot B$, and conversely. A two-index antisymmetric spinor is proportional to $\epsilon$. With $\epsilon_{12}=\epsilon_{\dot1\dot2}=1$ and $\epsilon^{12}=\epsilon^{\dot1\dot2}=-1$, we obtain
\begin{align}
 B_{A\dot A,B\dot B;\ldots}
 &=\epsilon_{\dot A\dot B} B_{AB;\ldots}
    +\epsilon_{AB}\bar B_{\dot A\dot B;\ldots},\nonumber\\
 B_{AB;\ldots}&=\tfrac12\epsilon^{\dot B\dot A}B_{A\dot A,B\dot B;\ldots},
 \qquad
 \bar B_{\dot A\dot B;\ldots}=\tfrac12\epsilon^{BA}B_{A\dot A,B\dot B;\ldots}.
 \label{eq:ht4mixblocks}
\end{align}
The two coefficients are symmetric in their displayed pair. The bar labels the dotted block. For each coefficient, take the left- and right-handed parts of the open Dirac index in turn, lowering the right-handed index with $\epsilon_{\dot C\dot D}$.

In each of the resulting four blocks, symmetrize all undotted indices together and all dotted indices together, including the open Weyl index in its own group. Each symmetrization has unit weight. A symmetric string of $p$ two-component indices has spin $p/2$. An antisymmetric pair factors out an $\epsilon$ and lowers this spin by one. Thus the symmetrizations retain precisely the components counted in Eq.~\eqref{eq:ht4mixspin}. The strings to be symmetrized are listed explicitly below. Each $A_i,\dot A_i$ comes from the vector index $\rho_i$ in Eq.~\eqref{eq:ht4mixcoeff}.
\begin{equation}
\begin{array}{c|l|l}
 \text{coefficient and open index} &\text{undotted indices}&\text{dotted indices}\\ \hline
 B_{AB;\ldots;C} & A,B,A_1,\ldots,A_{r-2},C & \dot A_1,\ldots,\dot A_{r-2}\\
 B_{AB;\ldots;\dot C} & A,B,A_1,\ldots,A_{r-2} & \dot A_1,\ldots,\dot A_{r-2},\dot C\\
 \bar B_{\dot A\dot B;\ldots;C} & A_1,\ldots,A_{r-2},C & \dot A,\dot B,\dot A_1,\ldots,\dot A_{r-2}\\
 \bar B_{\dot A\dot B;\ldots;\dot C} & A_1,\ldots,A_{r-2}
        & \dot A,\dot B,\dot A_1,\ldots,\dot A_{r-2},\dot C
\end{array}
\label{eq:ht4mixstrings}
\end{equation}
Counting the two strings in each row gives the retained representations,
\begin{equation}
 \left(\tfrac{r+1}{2},\tfrac{r-2}{2}\right),\qquad
 \left(\tfrac r2,\tfrac{r-1}{2}\right)
 \qquad\text{and their left--right conjugates}.
 \label{eq:ht4top}
\end{equation}
For each chirality of the open spinor, substitute the two symmetrized coefficients back into the first line of Eq.~\eqref{eq:ht4mixblocks}, keeping the displayed $\epsilon$ tensors outside the symmetrizations. Apply $\bar\sigma_\mu/2$ from Eq.~\eqref{eq:ht4mixsigma} to each of the $r$ index pairs and restore the two Weyl components of the open Dirac index. This gives $(\Pi_r^{(4)}B)_{\mu\nu;\rho_1\cdots\rho_{r-2}}$ in the original tensor notation.

To recover the mixed part of $V_\mu$ after projection, use its homogeneity $z\cdot\partial_z V_\mu=(r-1)V_\mu$ in Eq.~\eqref{eq:ht4mixcoeff}, which gives
\begin{equation}
 z^\nu B_{\mu\nu}
 =(r-1)V_\mu-\big[\partial_\mu(z^\nu V_\nu)-V_\mu\big]
 =r V_\mu-\partial_\mu\mathcal O_r=rV_\mu^{\rm M}.
 \label{eq:ht4mixrecover}
\end{equation}
The factor $1/r$ in the temporal channel follows from Eq.~\eqref{eq:ht4mixrecover}. Use the twist-4 projection in Eq.~\eqref{eq:ht4mixstrings} and its complement $\Pi_r^{(\ge5)}=1-\Pi_r^{(4)}$. In either case $v_t^0=1$ selects $\mu=0$, and each of the $r-1$ factors $\lambda^3=-1$ selects a lower index $3$. At fixed $(n_1,n_2)$ this gives
\begin{equation}
 \mathcal O_{\gamma^t,{\rm M}}^{(\tau)}
 =\frac{v_t^\mu\lambda^\nu}{r}(\Pi_r^{(\tau)}B)_{\mu\nu}(\lambda)
 =\frac{(-1)^{r-1}}{r}(\Pi_r^{(\tau)}B)_{03;\,\underbrace{\scriptstyle 3\cdots3}_{r-2}},
 \qquad \tau=4,\ \ge5 .
 \label{eq:ht4mixedproject}
\end{equation}
Their sum is the mixed term in Eq.~\eqref{eq:ht4channelbalance}. Appendix~\ref{app:ht4mixedpower} derives the $O(P_z^{-2})$ bound for each projected moment. A measurement of the unprojected $03;3\cdots3$ component alone gives only their sum.

Finally restore the derivative counts on $B_{\mu\nu,n_1,n_2}$ and use the distance weights in Eq.~\eqref{eq:ht4gen}. At each rank $r=n_1+n_2+1$, integration of $\beta^{n_1+n_2}$ supplies $1/r$. Hence for either value of $\tau$ we obtain
\begin{align}
 \widetilde{\mathcal O}_{t,{\rm M}}^{(\tau)}(\ell_1,\ell_2)
 &=v_t^\mu\lambda^\nu
 \sum_{n_1+n_2\ge1}\frac{(-i\ell_1)^{n_1}(-i\ell_2)^{n_2}}{n_1!n_2!}
 \frac{(\Pi_r^{(\tau)}B_{n_1,n_2})_{\mu\nu}(\lambda)}{r}\nonumber\\
 &=v_t^\mu\lambda^\nu\int_0^1d\beta\!
 \sum_{n_1+n_2\ge1}\frac{(-i\beta\ell_1)^{n_1}(-i\beta\ell_2)^{n_2}}{n_1!n_2!}
  (\Pi_r^{(\tau)}B_{n_1,n_2})_{\mu\nu}(\lambda).
 \label{eq:ht4mixdistance}
\end{align}
In Eq.~\eqref{eq:ht4mixdistance}, $n_1,n_2$ are nonnegative and the projector acts at fixed rank before contraction with $v_t^\mu\lambda^\nu$. Summing the two twist sectors restores the full mixed term in Eq.~\eqref{eq:ht4genchannels}.

\subsection{Wilson-line representation}
\label{app:ht4convolutions}

The three-quark operator to be differentiated in Eq.~\eqref{eq:ht4gensym} is
\begin{equation}
 \mathcal O(L_1,L_2;z)=\epsilon^{abc}
 [W(0,L_1z)u(L_1z)]_a^T C\gamma_5\slashed z\,
 [W(0,L_2z)d(L_2z)]_b\,s_c(0).
 \label{eq:wcuds}
\end{equation}
We evaluate $\tfrac12\Box_z\mathcal O(L_1,L_2;z)$ at arbitrary $z$, holding the signed distances $L_1,L_2$ fixed. The derivatives act on both transported quark fields and on the diquark matrix in Eq.~\eqref{eq:wcuds}.

Equation~\eqref{eq:expoDeriv} identifies the transported field with $[e^{L_i z\cdot D}q](0)$. The exponent $B(z)=L_i z\cdot D$ is linear in $z$, so $\partial B/\partial z^\mu=L_iD_\mu$. Since the covariant derivatives need not commute, we use the operator identity
\begin{equation}
 \frac{\partial}{\partial z^\mu}e^{B(z)}
 =\int_0^1 du\, e^{uB(z)}
     \frac{\partial B(z)}{\partial z^\mu}\, e^{(1-u)B(z)} .
 \label{eq:expformula}
\end{equation}
For the $u$ line, setting $t=u L_1$ in Eq.~\eqref{eq:expformula} and using Eq.~\eqref{eq:formula1} yields
\begin{align}
 \partial_\mu[W(0,L_1z)u(L_1z)]
 &=L_1\int_0^1du\,
    [e^{uL_1z\cdot D}D_\mu e^{(1-u)L_1z\cdot D}u](0)\nonumber\\
 &=\int_0^{L_1}dt\,
    W(0,tz)D_\mu(tz)W(tz,L_1z)u(L_1z).
 \label{eq:wconeline}
\end{align}
Here and below $\partial_\mu=\partial/\partial z^\mu$. The insertion acts on the whole transported field to its right. In the second line of Eq.~\eqref{eq:wconeline} this means
\begin{align}
 D_\mu(tz)W(tz,L_1z)u(L_1z)
 =\left.D_\mu^{(x)}
     [W(x,x+(L_1-t)z)u(x+(L_1-t)z)]\right|_{x=tz},
\end{align}
where the remaining displacement $(L_1-t)z$ is held fixed in $D_\mu^{(x)}$. The same formula with $L_2,d$ gives the derivative on the $d$ line.

Apply Eq.~\eqref{eq:expformula} once more to the first line of Eq.~\eqref{eq:wconeline}. With $B=L_1z\cdot D$, the two exponentials give separate terms,
\begin{align}
 \partial_\nu\partial_\mu e^B
 &=L_1^2\int_0^1du\int_0^1dv\,\Big[
 u\,e^{uvB}D_\nu e^{u(1-v)B}D_\mu e^{(1-u)B}\nonumber\\
 &\hspace{34mm}{}+(1-u)\,e^{uB}D_\mu e^{(1-u)vB}D_\nu
                 e^{(1-u)(1-v)B}\Big].
 \label{eq:wcsecondraw}
\end{align}
The insertion fractions in the first term are $(uv,u)$ and those in the second are $(u,u+(1-u)v)$. Both pairs run over $0\le s_1\le s_2\le1$, with Jacobians $u$ and $1-u$, respectively. With $t_j=L_1s_j$, Eq.~\eqref{eq:formula2} therefore gives
\begin{align}
 &\partial_\nu\partial_\mu[W(0,L_1z)u(L_1z)]
 =\int_0^{L_1}dt_2\int_0^{t_2}dt_1\,
 W(0,t_1z)\nonumber\\
 &\qquad{}\times
 \big[D_\nu(t_1z)W(t_1z,t_2z)D_\mu(t_2z)
    +D_\mu(t_1z)W(t_1z,t_2z)D_\nu(t_2z)\big]\,
 W(t_2z,L_1z)u(L_1z).
 \label{eq:wcsecondline}
\end{align}
 Contracting with $g^{\mu\nu}/2$ makes the two terms identical and leaves one ordered double integral with coefficient one.

Before contracting a second derivative, differentiate Eq.~\eqref{eq:wcuds} once by the product rule,
\begin{align}
 \partial_\mu\mathcal O(L_1,L_2;z)=\epsilon^{abc}\Big\{&
 \big[\partial_\mu(W(0,L_1z)u(L_1z))\big]_a^T
 C\gamma_5\slashed z[W(0,L_2z)d(L_2z)]_b \nonumber\\
 &+[W(0,L_1z)u(L_1z)]_a^T
 C\gamma_5\gamma_\mu[W(0,L_2z)d(L_2z)]_b \nonumber\\
 &+[W(0,L_1z)u(L_1z)]_a^T C\gamma_5\slashed z
 \big[\partial_\mu(W(0,L_2z)d(L_2z))\big]_b\Big\}s_c(0).
 \label{eq:wcudsfirst}
\end{align}
Applying $\tfrac12\partial^\mu$ to Eq.~\eqref{eq:wcudsfirst} gives five terms. Two insertions on the same line are ordered as in Eq.~\eqref{eq:wcsecondline}, whereas insertions on different lines have independent positions. With the single insertions from Eq.~\eqref{eq:wconeline}, the result at arbitrary $z$ is
\begin{align}
 &\tfrac12\Box_z\mathcal O(L_1,L_2;z)
 =\epsilon^{abc}\Bigg\{\nonumber\\
 &\quad \int_0^{L_1}dt\,
 [W(0,tz)D_\mu(tz)W(tz,L_1z)u(L_1z)]_a^T
 \nonumber\\[-1mm]
 &\hspace{34mm}{}\times C\gamma_5\gamma^\mu
 [W(0,L_2z)d(L_2z)]_b\,s_c(0)\nonumber\\
 &\quad +\int_0^{L_2}dw\,
 [W(0,L_1z)u(L_1z)]_a^T C\gamma_5\gamma^\mu
 \nonumber\\[-1mm]
 &\hspace{34mm}{}\times
 [W(0,wz)D_\mu(wz)W(wz,L_2z)d(L_2z)]_b\,s_c(0)
 \nonumber\\
 &\quad +\int_0^{L_1}dt_2\int_0^{t_2}dt_1\,
 [W(0,t_1z)D_\mu(t_1z)W(t_1z,t_2z)
 \nonumber\\[-1mm]
 &\hspace{12mm}{}\times D^\mu(t_2z)W(t_2z,L_1z)u(L_1z)]_a^T
 C\gamma_5\slashed z [W(0,L_2z)d(L_2z)]_b\,s_c(0)
 \nonumber\\
 &\quad +\int_0^{L_2}dw_2\int_0^{w_2}dw_1\,
 [W(0,L_1z)u(L_1z)]_a^T C\gamma_5\slashed z
 \nonumber\\[-1mm]
 &\hspace{12mm}{}\times [W(0,w_1z)D_\mu(w_1z)W(w_1z,w_2z)
 D^\mu(w_2z)W(w_2z,L_2z)d(L_2z)]_b\,s_c(0)
 \nonumber\\
 &\quad +\int_0^{L_1}dt\int_0^{L_2}dw\,
 [W(0,tz)D_\mu(tz)W(tz,L_1z)u(L_1z)]_a^T
 \nonumber\\[-1mm]
 &\hspace{12mm}{}\times C\gamma_5\slashed z
 [W(0,wz)D^\mu(wz)W(wz,L_2z)d(L_2z)]_b\,s_c(0)
 \Bigg\}.
 \label{eq:wcfiveterms}
\end{align}

\subsection{Power counting for the mixed moment}
\label{app:ht4mixedpower}

Fix $(n_1,n_2)$, $m_\Lambda>0$ and the renormalization scale. For $r=n_1+n_2+1=1$ the curl $B_{\mu\nu}$ in Eq.~\eqref{eq:ht4curl} vanishes. For $r\ge2$, Eq.~\eqref{eq:ht4mixcoeff} leaves two antisymmetric indices $\mu,\nu$ and $r-2$ symmetric indices contracted with the argument of $B_{\mu\nu}$. Using $v_t=(n+\bar n)/\sqrt2$ and $\lambda=(n-\bar n)/\sqrt2$, we first contract the antisymmetric pair in Eq.~\eqref{eq:ht4raw},
\begin{align}
 v_t^\mu\lambda^\nu B_{\mu\nu}(\lambda)
 &=\tfrac12(n^\mu n^\nu-n^\mu\bar n^\nu+\bar n^\mu n^\nu
                      -\bar n^\mu\bar n^\nu)B_{\mu\nu}(\lambda)\nonumber\\
 &=-n^\mu\bar n^\nu B_{\mu\nu}(\lambda).
 \label{eq:ht4mixedpair}
\end{align}
The terms with two identical vectors vanish and exchanging $\mu,\nu$ in the third term gives the second term once more. With the normalization in Eq.~\eqref{eq:bdaQuasiMoment}, the mixed temporal moment is then
\begin{equation}
 \widetilde{\mathcal M}^{\rm M,0}_{\gamma^t;n_1,n_2}
 =-\frac{n^\mu\bar n^\nu}{4f_\Lambda^{\rm LC}r\,pP_z^r}
    \sum_\sigma\bar u_\Lambda(P,\sigma)\slashed n
       \langle0|B_{\mu\nu}(\lambda)|\Lambda(P,\sigma)\rangle .
 \label{eq:ht4mixedmoment}
\end{equation}

To count the remaining powers of $P_z$, write Eq.~\eqref{eq:ht4mixedmoment} in the baryon rest frame. Let $P_\star=(m_\Lambda,0,0,0)$ and let $L$ be the boost along the third axis that takes $P_\star$ to $P$. For any vector $a$, its inverse acts on the longitudinal components as
\begin{equation}
 \begin{pmatrix}(L^{-1}a)^0\\(L^{-1}a)^3\end{pmatrix}
 =\frac1{m_\Lambda}\begin{pmatrix} E&-P_z\\-P_z&E\end{pmatrix}
                      \begin{pmatrix}a^0\\a^3\end{pmatrix}.
 \label{eq:ht4mixedboostmatrix}
\end{equation}
Applying Eq.~\eqref{eq:ht4mixedboostmatrix} to the vectors in Eq.~\eqref{eq:bdaDirections} gives
\begin{align}
 L^{-1}n&=\frac{E+P_z}{m_\Lambda}\,n,\qquad
 L^{-1}\bar n=\frac{E-P_z}{m_\Lambda}\,\bar n,\nonumber\\
 (L^{-1}n)^\mu(L^{-1}\bar n)^\nu
 &=\frac{E^2-P_z^2}{m_\Lambda^2}\,n^\mu\bar n^\nu=n^\mu\bar n^\nu .
 \label{eq:ht4mixedboostpair}
\end{align}
There is one additional $n$ in the spin projection. Under the Dirac boost $S(L)$, both $u_\Lambda$ and the open spinor index of the matrix element acquire $S(L)$, whereas $\bar u_\Lambda$ acquires $S(L)^{-1}$ on the right. Since $p=P\cdot n=P_\star\cdot(L^{-1}n)$, the normalized spin insertion becomes
\begin{equation}
 \frac{S(L)^{-1}\slashed n S(L)}{p}
 =\frac{\gamma_\mu(L^{-1}n)^\mu}{P_\star\cdot(L^{-1}n)}
 =\frac{\slashed n}{P_\star\cdot n},\qquad P_\star\cdot n=\frac{m_\Lambda}{\sqrt2}.
 \label{eq:ht4mixedboostspin}
\end{equation}
Using Eqs.~\eqref{eq:ht4mixedboostpair} and \eqref{eq:ht4mixedboostspin} in Eq.~\eqref{eq:ht4mixedmoment}, we obtain
\begin{equation}
 \widetilde{\mathcal M}^{\rm M,0}_{\gamma^t;n_1,n_2}
 =-\frac{n^\mu\bar n^\nu}{4f_\Lambda^{\rm LC}r(P_\star\cdot n)P_z^r}
   \sum_\sigma\bar u_\Lambda(P_\star,\sigma)\slashed n
   \langle0|B_{\mu\nu}(L^{-1}\lambda)|\Lambda(P_\star,\sigma)\rangle .
 \label{eq:ht4mixedrestmoment}
\end{equation}

The numerator of Eq.~\eqref{eq:ht4mixedrestmoment} now depends on $P_z$ only through $L^{-1}\lambda$. For $\lambda=(0,0,0,-1)$, Eq.~\eqref{eq:ht4mixedboostmatrix} gives
\begin{equation}
 L^{-1}\lambda=\frac{(P_z,0,0,-E)}{m_\Lambda}
 =\frac{P_z}{m_\Lambda}\big[\sqrt2\,n+(\Delta-1)\lambda\big],
 \qquad \Delta=\frac{E}{P_z}.
 \label{eq:ht4boost}
\end{equation}
By Eq.~\eqref{eq:ht4mixcoeff}, $B_{\mu\nu}(z)$ is homogeneous of degree $r-2$ in $z$. Factoring out the common multiplier in Eq.~\eqref{eq:ht4boost} and expanding the remaining polynomial gives the finite sum
\begin{align}
 \frac{B_{\mu\nu}(L^{-1}\lambda)}{P_z^r}
 &=\frac{B_{\mu\nu}\big(\sqrt2\,n+(\Delta-1)\lambda\big)}
         {m_\Lambda^{r-2}P_z^2}\nonumber\\
 &=\frac{1}{m_\Lambda^{r-2}P_z^2}
   \sum_{k=0}^{r-2}\frac{(\Delta-1)^k}{k!}
      \left.(\lambda\cdot\partial_z)^k B_{\mu\nu}(z)\right|_{z=\sqrt2\,n}.
 \label{eq:ht4mixedrestseries}
\end{align}
The $k=0$ term contains $B_{\mu\nu}(\sqrt2\,n)=(\sqrt2)^{r-2}B_{\mu\nu}(n)$. At fixed $r$ the other coefficients in Eq.~\eqref{eq:ht4mixedrestseries} also have momentum-independent rest-frame matrix elements. Using the mass-shell condition once more gives the power of each term,
\begin{equation}
 \Delta-1=\frac{E-P_z}{P_z}
         =\frac{m_\Lambda^2}{P_z(E+P_z)},\qquad
 \frac{(\Delta-1)^k}{P_z^2}
 =\frac{(m_\Lambda^2/2)^k}{P_z^{2+2k}}
     \left[1+O\!\left(\frac{m_\Lambda^2}{P_z^2}\right)\right].
 \label{eq:ht4mixedorders}
\end{equation}
Inserting the $k=0$ term of Eq.~\eqref{eq:ht4mixedrestseries} into Eq.~\eqref{eq:ht4mixedrestmoment} gives the coefficient of $1/P_z^2$ in Eq.~\eqref{eq:ht4mixedpower},
\begin{align}
 H^{\rm M}_{n_1,n_2}
 ={}&-\frac{n^\mu\bar n^\nu}{4f_\Lambda^{\rm LC}r(P_\star\cdot n)}
       \left(\frac{\sqrt2}{m_\Lambda}\right)^{r-2}\nonumber\\[-2pt]
 &{}\times\sum_\sigma\bar u_\Lambda(P_\star,\sigma)\slashed n
       \langle0|B_{\mu\nu}(n)|\Lambda(P_\star,\sigma)\rangle .
 \label{eq:ht4mixedrestcoef}
\end{align}
All terms with $k\ge1$ contribute at $O(P_z^{-4})$ or beyond by Eq.~\eqref{eq:ht4mixedorders}. Replacing $B_{\mu\nu}$ by either of its Lorentz projections in Eq.~\eqref{eq:ht4mixedproject} preserves the antisymmetric pair and the degree $r-2$. The same power bound therefore holds separately for each projected moment.

\section{Moments and mass subtraction}
\label{app:match5details}

\subsection{Light-cone moment weights}
\label{sec:match5weights}

Equation~\eqref{eq:bdaLC} defines the LCDA with the strange quark at the origin. To differentiate with respect to its position, take all three coordinates $\xi_j n$ to be independent. Translating the anchored matrix element by $\xi_3 n$ gives a phase $e^{-ip\xi_3}$.  With $y_3=1-y_1-y_2$, the translated matrix element becomes
\begin{align}
 &\langle0|\mathcal O_{\mathrm{LC},+,R}(\xi_1,\xi_2,\xi_3;\mu)|\Lambda(P,\sigma)\rangle
 \nonumber\\
 &\quad=e^{-ip\xi_3}
       \langle0|\mathcal O_{\mathrm{LC},+,R}(\xi_1-\xi_3,\xi_2-\xi_3,0;\mu)
                        |\Lambda(P,\sigma)\rangle \nonumber\\
 &\quad=f_\Lambda^{\rm LC}(\mu)p u_{\Lambda,+}\int_{\mathcal S}dy_1\,dy_2\,
     e^{-ip[\xi_3+y_1(\xi_1-\xi_3)+y_2(\xi_2-\xi_3)]}\phi(y_1,y_2,\mu)
     \nonumber\\
 &\quad=f_\Lambda^{\rm LC}(\mu)p u_{\Lambda,+}\int_{\mathcal S}dy_1\,dy_2\,
       e^{-ip(y_1\xi_1+y_2\xi_2+y_3\xi_3)}\phi(y_1,y_2,\mu).
 \label{eq:match5translate}
\end{align}
For a bare quark field transported along $n$, differentiating its endpoint along the ray inserts $n\cdot D$. After renormalization, the degree-$N$ coefficient of the light-ray operator defines the local moments in the same scheme as that in Ref.~\cite{Braun:1999te}. Applying Eq.~\eqref{eq:bdaLocalMoment} to the three endpoints of Eq.~\eqref{eq:match5translate} gives
\begin{align}
 &\Pi_+\langle0|\mathcal O^{(3)}_{N;(k_1,k_2,k_3),R}(n;\mu)|\Lambda(P,\sigma)\rangle
 \nonumber\\
 &\quad=\left.\prod_{j=1}^{3}(i\partial_{\xi_j})^{k_j}
   \langle0|\mathcal O_{\mathrm{LC},+,R}(\xi_1,\xi_2,\xi_3;\mu)
                      |\Lambda(P,\sigma)\rangle\right|_{\xi_1=\xi_2=\xi_3=0}
 \nonumber\\
 &\quad=f_\Lambda^{\rm LC}(\mu)p^{N+1}u_{\Lambda,+}
  \int_{\mathcal S}dy_1\,dy_2\, y_1^{k_1}y_2^{k_2}y_3^{k_3}\phi(y_1,y_2,\mu).
 \label{eq:match5momentder}
\end{align}
Each factor $i\partial_{\xi_j}$ contributes $p y_j$, whereas a total derivative $i\sum_j\partial_{\xi_j}$ supplies $p(y_1+y_2+y_3)=p$. Thus a total derivative raises $N$ by one while multiplying the fractional weight by unity. Comparing Eq.~\eqref{eq:match5momentder} with Eq.~\eqref{eq:match5lcfix} gives the moment for one derivative assignment. Expanding $y_3^{k_3}$ then yields
\begin{align}
 \mathcal M_{N;(k_1,k_2,k_3)}(\mu)
 &=\int_{\mathcal S}dy_1\,dy_2\,
        y_1^{k_1}y_2^{k_2}(1-y_1-y_2)^{k_3}\phi(y_1,y_2,\mu)\nonumber\\
 &=\sum_{i=0}^{k_3}\sum_{j=0}^{k_3-i}
   \frac{(-1)^{i+j}k_3!}{i!\,j!\,(k_3-i-j)!}\,
      \mathcal M_{k_1+i,k_2+j}(\mu).
 \label{eq:match5anchor}
\end{align}
For $k_3=0$ the original two-index moment is recovered. The lower-degree moments that occur for $k_3>0$ are a consequence of eliminating $y_3$ at fixed $N$.

For a general basis operator $I$, take the vacuum-to-baryon matrix element of Eq.~\eqref{eq:match5betabasis} and use Eq.~\eqref{eq:match5lcfix} on both sides. Equating the coefficients of $f_\Lambda^{\rm LC}(\mu)p^{N+1}u_{\Lambda,+}$ and substituting the integral in Eq.~\eqref{eq:match5anchor} gives
\begin{align}
 \mathcal M_{N;I}(\mu)
 &=\sum_{k_1+k_2+k_3=N}\beta_{I;k_1k_2k_3}\,
              \mathcal M_{N;(k_1,k_2,k_3)}(\mu)\nonumber\\
 &=\sum_{k_1+k_2+k_3=N}\beta_{I;k_1k_2k_3}
   \int_{\mathcal S}dy_1\,dy_2\,
   y_1^{k_1}y_2^{k_2}(1-y_1-y_2)^{k_3}\phi(y_1,y_2,\mu)\nonumber\\
 &=\int_{\mathcal S}dy_1\,dy_2\,
   \underbrace{\left[\sum_{k_1+k_2+k_3=N}\beta_{I;k_1k_2k_3}
   y_1^{k_1}y_2^{k_2}(1-y_1-y_2)^{k_3}\right]}_{\displaystyle w_{NI}(y_1,y_2)}
   \phi(y_1,y_2,\mu).
 \label{eq:match5appBetaWeight}
\end{align}
Substituting the last line of Eq.~\eqref{eq:match5anchor} into the first line of Eq.~\eqref{eq:match5appBetaWeight} gives the same matrix element in the two-index moment basis of Eq.~\eqref{eq:bdaMoment},
\begin{equation}
 \mathcal M_{N;I}(\mu)
 =\sum_{k_1+k_2+k_3=N}\beta_{I;k_1k_2k_3}
   \sum_{i=0}^{k_3}\sum_{j=0}^{k_3-i}
   \frac{(-1)^{i+j}k_3!}{i!\,j!\,(k_3-i-j)!}\,
   \mathcal M_{k_1+i,k_2+j}(\mu).
 \label{eq:match5appAnchoredWeight}
\end{equation}

\subsection{Inverse target-mass map}
\label{app:match5inverse}

At fixed $P_z,\mu_R,\mu$, the inverse of the full tree map acts on the measured quasi-DA in Eq.~\eqref{eq:match5fullforward} as
\begin{align}
 \widehat{\widetilde\phi}_{\Gamma,\rm sub}^{\rm inv}
 &\equiv(\widehat{\mathcal T}_{\Gamma,c}^{(0)})^{-1}
                   \widehat{\widetilde\phi}^{\rm full}_\Gamma\nonumber\\
 &=\phi+a_s(\widehat{\mathcal T}_{\Gamma,c}^{(0)})^{-1}
                  [C_{\Gamma,\mathrm{LP}}^{(1)}\otimes\phi]
  +(\widehat{\mathcal T}_{\Gamma,c}^{(0)})^{-1}
                  [\mathcal R_{\Gamma,\mathrm{mix}}+\mathcal R_{\Gamma,\mathrm{HT}}]
  +O(a_s^2).
 \label{eq:match5exactsub}
\end{align}
The inverse acts on both momentum fractions in each term. We give its integral representation below.

Let $g(x_1,x_2)=[\widehat{\mathcal T}_{\Gamma,c}^{(0)}\phi](x_1,x_2)$ at fixed mass and momentum. We use the roots $a,b$ of Eq.~\eqref{eq:tmRoots} and the local factors $d_{\gamma^z}^{(0)}=F_1(c)$ and $d_{\gamma^t}^{(0)}=T_1(c)$. Evaluate Eqs.~\eqref{eq:tmSpatialDA} and \eqref{eq:tmTemporalDA}, divided by the respective local factors at $(ax_1,ax_2)$. Changing variables from $s$ to $u=s/a$ sends the endpoints $b,a$ to $b/a,1$. The measures $ds/s$ and $ds/s^2$ become $du/u$ and $du/(a u^2)$, respectively. Collecting the terms gives
\begin{align}
 d_\Gamma^{(0)}\Delta\,g(ax_1,ax_2)
 &=\phi(x_1,x_2)-[\mathcal L_\Gamma\phi](x_1,x_2),\nonumber\\
 [\mathcal L_{\gamma^z}\phi](x_1,x_2)
 &=\int_{b/a}^{1}du\,\frac{b}{\Delta u}\,
      \phi\!\left(\frac{x_1}{u},\frac{x_2}{u}\right),\nonumber\\
 [\mathcal L_{\gamma^t}\phi](x_1,x_2)
 &=\int_{b/a}^{1}du\left[\frac{b(\Delta+2)}{\Delta^2 u}
                    -\frac{4b^2}{\Delta^2 u^2}\right]
      \phi\!\left(\frac{x_1}{u},\frac{x_2}{u}\right).
 \label{eq:match5rescale}
\end{align}
Solving the first line of Eq.~\eqref{eq:match5rescale} for $\phi$ gives $\phi(x_1,x_2)=d_\Gamma^{(0)}\Delta\,g(ax_1,ax_2)+[\mathcal L_\Gamma\phi](x_1,x_2)$. Substitute this expression repeatedly in the integral term. The common dilation by $a$ commutes with $\mathcal L_\Gamma$. After $K+1$ substitutions, the remainder is $\mathcal L_\Gamma^{K+1}\phi$. When this remainder tends to zero, the inverse used in Eq.~\eqref{eq:match5exactsub} is
\begin{equation}
 [(\widehat{\mathcal T}_{\Gamma,c}^{(0)})^{-1}g](x_1,x_2)
   =d_\Gamma^{(0)}\Delta\sum_{k=0}^{\infty}
                [\mathcal L_\Gamma^k g](ax_1,ax_2),\qquad \mathcal L_\Gamma^0=1 .
 \label{eq:match5series}
\end{equation}

We test convergence of Eq.~\eqref{eq:match5series} in $L^1(\mathbb R^2)$, the space of absolutely integrable functions of the two momentum fractions. For $g$ in this space we use the norm
\begin{equation}
 \|g\|_1\equiv\int_{\mathbb R^2}dx_1\,dx_2\,|g(x_1,x_2)|.
\end{equation}
The subscript $1$ denotes the first power of $|g|$. To see how much one application of $\mathcal L_\Gamma$ can change this integral, seek a constant $C\ge0$ such that
\begin{equation}
 \int_{\mathbb R^2}dx_1\,dx_2\,\big|[\mathcal L_\Gamma g](x_1,x_2)\big|
 \le C\int_{\mathbb R^2}dx_1\,dx_2\,|g(x_1,x_2)|
 \qquad\text{for every }g\in L^1(\mathbb R^2).
\end{equation}
The smallest such $C$ is the operator norm $\|\mathcal L_\Gamma\|_{1\to1}$. The notation $1\to1$ means that both the input and the output are measured by the integral of their absolute value. To obtain a bound for each channel in Eq.~\eqref{eq:match5rescale}, set $x_i=u y_i$ for $u\ne0$. The Jacobian gives $\|g(\,\cdot\,/u,\,\cdot\,/u)\|_1=|u|^2\|g\|_1$. The triangle inequality then yields
\begin{align}
 \|\mathcal L_{\gamma^z}\|_{1\to1}
 &\le\frac{|b|}{\Delta}\int_{b/a}^1du\,|u|
 =\frac{|b|}{2\Delta}\big[1+(b/a)^2\big]<1,\nonumber\\
 \|\mathcal L_{\gamma^t}\|_{1\to1}
 &\le\frac{1}{\Delta^2}\int_{b/a}^1du\,|b(\Delta+2)u-4b^2|
 =\frac{(\Delta-1)(2\Delta+1)}{(\Delta+1)^2}.
 \label{eq:match5normbound}
\end{align}
In the temporal integral the expression inside the absolute value is nonpositive throughout the interval. Its maximum occurs at $u=b/a$, where it equals $-2\Delta b^2/(\Delta+1)$. Whenever $\|\mathcal L_\Gamma\|_{1\to1}<1$, the remainder obeys
\[
 \|\mathcal L_\Gamma^{K+1}\phi\|_1
 \le \|\mathcal L_\Gamma\|_{1\to1}^{K+1}\|\phi\|_1
 \longrightarrow 0 \qquad (K\to\infty).
\]
Thus Eq.~\eqref{eq:match5normbound} proves convergence in $L^1$ for every finite $c\ge0$ in the spatial channel and for $0\le c<(11+3\sqrt{17})/8$ in the temporal channel. The latter is a sufficient range that includes the large-momentum regime. At $c=0$ the two integral operators vanish and Eq.~\eqref{eq:match5series} reduces to the identity.

\acknowledgments
We thank Shuai Zhao for valuable discussions. We also thank Jun Hua and Mu-Hua Zhang for providing the quasi-DA data. This work is supported in part by Natural Science Foundation of China under Grants No.~12305103, 12205180, 12565014, 12635004. The work of Jun Zeng is also supported by the Talent Research Startup Foundation of Hainan Normal University: HSZK-KYQD-202523, the Opening Foundation of Shanghai Key Laboratory of Particle Physics and Cosmology under Grant No. 22DZ2229013-5, and the Hainan Provincial Natural Science Foundation of China under Grant No. 126MS0134.





\begin{thebibliography}{99}

\bibitem{Lepage:1980fj}
G.~P.~Lepage and S.~J.~Brodsky,
``Exclusive Processes in Perturbative Quantum Chromodynamics,''
Phys. Rev. D \textbf{22}, 2157 (1980)
doi:10.1103/PhysRevD.22.2157

\bibitem{Efremov:1978rn}
A.~V.~Efremov and A.~V.~Radyushkin,
``Asymptotical Behavior of Pion Electromagnetic Form-Factor in QCD,''
Theor. Math. Phys. \textbf{42}, 97-110 (1980)
doi:10.1007/BF01032111

\bibitem{LHCb:2025ray}
R.~Aaij \textit{et al.} [LHCb],
``Observation of charge{\textendash}parity symmetry breaking in baryon decays,''
Nature \textbf{643}, no.8074, 1223-1228 (2025)
doi:10.1038/s41586-025-09119-3
[arXiv:2503.16954 [hep-ex]].

\bibitem{Shih:1998pb}
H.~H.~Shih, S.~C.~Lee and H.~n.~Li,
``The Lambda(b) ---{\ensuremath{>}} p lepton anti-neutrino decay in perturbative QCD,''
Phys. Rev. D \textbf{59}, 094014 (1999)
doi:10.1103/PhysRevD.59.094014
[arXiv:hep-ph/9810515 [hep-ph]].

\bibitem{Keum:2000wi}
Y.~Y.~Keum, H.~N.~Li and A.~I.~Sanda,
``Penguin enhancement and $B \to K \pi$ decays in perturbative QCD,''
Phys. Rev. D \textbf{63}, 054008 (2001)
doi:10.1103/PhysRevD.63.054008
[arXiv:hep-ph/0004173 [hep-ph]].

\bibitem{Lu:2000em}
C.~D.~Lu, K.~Ukai and M.~Z.~Yang,
``Branching ratio and CP violation of B ---{\ensuremath{>}} pi pi decays in perturbative QCD approach,''
Phys. Rev. D \textbf{63}, 074009 (2001)
doi:10.1103/PhysRevD.63.074009
[arXiv:hep-ph/0004213 [hep-ph]].

\bibitem{Keum:2000ph}
Y.~Y.~Keum, H.~n.~Li and A.~I.~Sanda,
``Fat penguins and imaginary penguins in perturbative QCD,''
Phys. Lett. B \textbf{504}, 6-14 (2001)
doi:10.1016/S0370-2693(01)00247-7
[arXiv:hep-ph/0004004 [hep-ph]].

\bibitem{Guo:2005qa}
P.~Guo, H.~W.~Ke, X.~Q.~Li, C.~D.~Lu and Y.~M.~Wang,
``Diquarks and the semi-leptonic decay of Lambda(b) in the hyrid scheme,''
Phys. Rev. D \textbf{75}, 054017 (2007)
doi:10.1103/PhysRevD.75.054017
[arXiv:hep-ph/0501058 [hep-ph]].

\bibitem{He:2006ud}
X.~G.~He, T.~Li, X.~Q.~Li and Y.~M.~Wang,
``PQCD calculation for Lambda(b) ---{\ensuremath{>}} Lambda gamma in the standard model,''
Phys. Rev. D \textbf{74}, 034026 (2006)
doi:10.1103/PhysRevD.74.034026
[arXiv:hep-ph/0606025 [hep-ph]].

\bibitem{Lu:2009cm}
C.~D.~Lu, Y.~M.~Wang, H.~Zou, A.~Ali and G.~Kramer,
``Anatomy of the pQCD approach to the baryonic decays $\Lambda_b \to p \pi, pK$,''
Physical review / D \textbf{80}, 1-29 (2009)
doi:10.1103/PhysRevD.80.034011
[arXiv:0906.1479 [hep-ph]].

\bibitem{Han:2022srw}
J.~J.~Han, Y.~Li, H.~n.~Li, Y.~L.~Shen, Z.~J.~Xiao and F.~S.~Yu,
``$\Lambda _b\rightarrow p$ transition form factors in perturbative QCD,''
Eur. Phys. J. C \textbf{82}, no.8, 686 (2022)
doi:10.1140/epjc/s10052-022-10642-0
[arXiv:2202.04804 [hep-ph]].

\bibitem{Li:2025rsm}
Y.~Li, J.~Chen, Y.~X.~Wang and Z.~T.~Zou,
``Investigation of {\ensuremath{\Lambda}}b{\textrightarrow}{\ensuremath{\Lambda}}c{\ensuremath{\ell}}-{\ensuremath{\nu}}{\ensuremath{\ell}}{\textasciimacron} decays in the perturbative QCD approach,''
Phys. Rev. D \textbf{113}, no.1, 013003 (2026)
doi:10.1103/8ztx-56ys
[arXiv:2509.02257 [hep-ph]].

\bibitem{Yang:2025yaw}
L.~Yang, J.~J.~Han, Q.~Chang and F.~S.~Yu,
``The $\Lambda _{b} \rightarrow \Lambda $ transition form factors in perturbative QCD approach,''
Eur. Phys. J. C \textbf{86}, no.2, 103 (2026)
doi:10.1140/epjc/s10052-026-15295-x
[arXiv:2508.18069 [hep-ph]].

\bibitem{Han:2024kgz}
J.~J.~Han, J.~X.~Yu, Y.~Li, H.~n.~Li, J.~P.~Wang, Z.~J.~Xiao and F.~S.~Yu,
``Establishing CP Violation in b-Baryon Decays,''
Phys. Rev. Lett. \textbf{134}, no.22, 221801 (2025)
doi:10.1103/ynnx-f63h
[arXiv:2409.02821 [hep-ph]].

\bibitem{Han:2025tvc}
J.~J.~Han, J.~X.~Yu, Y.~Li, H.~n.~Li, J.~P.~Wang, Z.~J.~Xiao and F.~S.~Yu,
``CP violation in two-body hadronic {\ensuremath{\Lambda}}b decays in the PQCD approach,''
Phys. Rev. D \textbf{112}, no.5, 053007 (2025)
doi:10.1103/lvsn-v3xj
[arXiv:2506.07197 [hep-ph]].

\bibitem{Huang:2025bbk}
Y.~K.~Huang, Y.~Ji, B.~X.~Shi and Y.~M.~Wang,
``Two-Loop Renormalization-Group Evolution for the Nucleon Distribution Amplitude,''
[arXiv:2512.20471 [hep-ph]].

\bibitem{Chernyak:1987nu}
V.~L.~Chernyak, A.~A.~Ogloblin and I.~R.~Zhitnitsky,
``Wave Functions of Octet Baryons,''
Yad. Fiz. \textbf{48}, 1410-1422 (1988)
doi:10.1007/BF01557663

\bibitem{Ball:2008fw}
P.~Ball, V.~M.~Braun and E.~Gardi,
``Distribution Amplitudes of the Lambda(b) Baryon in QCD,''
Phys. Lett. B \textbf{665}, 197-204 (2008)
doi:10.1016/j.physletb.2008.06.004
[arXiv:0804.2424 [hep-ph]].

\bibitem{Ji:2013dva}
X.~Ji,
``Parton Physics on a Euclidean Lattice,''
Phys. Rev. Lett. \textbf{110}, 262002 (2013)
doi:10.1103/PhysRevLett.110.262002
[arXiv:1305.1539 [hep-ph]].

\bibitem{Ji:2014gla}
X.~Ji,
``Parton Physics from Large-Momentum Effective Field Theory,''
Sci. China Phys. Mech. Astron. \textbf{57}, 1407-1412 (2014)
doi:10.1007/s11433-014-5492-3
[arXiv:1404.6680 [hep-ph]].

\bibitem{Deng:2023csv}
Z.~F.~Deng, C.~Han, W.~Wang, J.~Zeng and J.~L.~Zhang,
``Light-cone distribution amplitudes of a light baryon in large-momentum effective theory,''
JHEP \textbf{07}, 191 (2023)
doi:10.1007/JHEP07(2023)191
[arXiv:2304.09004 [hep-ph]].

\bibitem{Han:2023xbl}
C.~Han, Y.~Su, W.~Wang and J.~L.~Zhang,
``Hybrid renormalization for quasi distribution amplitudes of a light baryon,''
JHEP \textbf{12}, 044 (2023)
doi:10.1007/JHEP12(2023)044
[arXiv:2308.16793 [hep-ph]].

\bibitem{Han:2024ucv}
C.~Han, W.~Wang, J.~Zeng and J.~L.~Zhang,
``Lightcone and quasi distribution amplitudes for light octet and decuplet baryons,''
JHEP \textbf{07}, 019 (2024)
doi:10.1007/JHEP07(2024)019
[arXiv:2404.04855 [hep-ph]].

\bibitem{Shi:2026mjb}
Y.~J.~Shi and J.~Zeng,
``Factorization formula connecting the {\ensuremath{\Lambda}}$_{b}$ LCDA in QCD and boosted HQET,''
JHEP \textbf{05}, 212 (2026)
doi:10.1007/JHEP05(2026)212
[arXiv:2602.14187 [hep-ph]].

\bibitem{Shi:2026fel}
Y.~J.~Shi, J.~Xu and S.~Zhao,
``Heavy quark mass dependence of the $\Lambda_Q$ light-cone distribution amplitude in QCD,''
[arXiv:2607.17471 [hep-ph]].

\bibitem{LatticeParton:2024vck}
M.~H.~Chu \textit{et al.} [Lattice Parton],
``Light cone distribution amplitude for the {\ensuremath{\Lambda}} baryon from lattice QCD,''
Phys. Rev. D \textbf{111}, no.3, 034510 (2025)
doi:10.1103/PhysRevD.111.034510
[arXiv:2411.12554 [hep-lat]].

\bibitem{LatticePartonCollaborationLPC:2025vhd}
H.~Bai \textit{et al.} [Lattice Parton Collaboration (LPC)],
``Hybrid renormalization for distribution amplitude of a light baryon in large momentum effective theory,''
Phys. Rev. D \textbf{112}, no.11, 114515 (2025)
doi:10.1103/rqmb-x9x8
[arXiv:2508.08971 [hep-lat]].

\bibitem{LPC:2026mvw}
M.~H.~Zhang \textit{et al.} [LPC],
``Baryon Light-Cone Distribution Amplitudes from Lattice QCD: Formalism, Renormalization, Extrapolation, and Matching,''
[arXiv:2606.30387 [hep-lat]].

\bibitem{LPC:2026lcj}
M.~H.~Zhang \textit{et al.} [LPC],
``Complete Access to Leading-Twist $\Lambda$-Baryon Light-Cone Distribution Amplitudes from Lattice QCD,''
[arXiv:2606.29597 [hep-lat]].

\bibitem{Braun:1999te}
V.~M.~Braun, S.~E.~Derkachov, G.~P.~Korchemsky and A.~N.~Manashov,
``Baryon distribution amplitudes in QCD,''
Nucl. Phys. B \textbf{553}, 355-426 (1999)
doi:10.1016/S0550-3213(99)00265-5
[arXiv:hep-ph/9902375 [hep-ph]].

\bibitem{Braun:2008ia}
V.~M.~Braun, A.~N.~Manashov and J.~Rohrwild,
``Baryon Operators of Higher Twist in QCD and Nucleon Distribution Amplitudes,''
Nucl. Phys. B \textbf{807}, 89-137 (2009)
doi:10.1016/j.nuclphysb.2008.08.012
[arXiv:0806.2531 [hep-ph]].

\bibitem{Chen:2016utp}
J.~W.~Chen, S.~D.~Cohen, X.~Ji, H.~W.~Lin and J.~H.~Zhang,
``Nucleon Helicity and Transversity Parton Distributions from Lattice QCD,''
Nucl. Phys. B \textbf{911}, 246-273 (2016)
doi:10.1016/j.nuclphysb.2016.07.033
[arXiv:1603.06664 [hep-ph]].

\bibitem{Zhang:2017bzy}
J.~H.~Zhang, J.~W.~Chen, X.~Ji, L.~Jin and H.~W.~Lin,
``Pion Distribution Amplitude from Lattice QCD,''
Phys. Rev. D \textbf{95}, no.9, 094514 (2017)
doi:10.1103/PhysRevD.95.094514
[arXiv:1702.00008 [hep-lat]].

\bibitem{Musch:2010ka}
B.~U.~Musch, P.~Hagler, J.~W.~Negele and A.~Schafer,
``Exploring quark transverse momentum distributions with lattice QCD,''
Phys. Rev. D \textbf{83}, 094507 (2011)
doi:10.1103/PhysRevD.83.094507
[arXiv:1011.1213 [hep-lat]].

\bibitem{Green:2020xco}
J.~R.~Green, K.~Jansen and F.~Steffens,
``Improvement, generalization, and scheme conversion of Wilson-line operators on the lattice in the auxiliary field approach,''
Phys. Rev. D \textbf{101}, no.7, 074509 (2020)
doi:10.1103/PhysRevD.101.074509
[arXiv:2002.09408 [hep-lat]].

\bibitem{Schaefer:2010hu}
S.~Schaefer \textit{et al.} [ALPHA],
``Critical slowing down and error analysis in lattice QCD simulations,''
Nucl. Phys. B \textbf{845}, 93-119 (2011)
doi:10.1016/j.nuclphysb.2010.11.020
[arXiv:1009.5228 [hep-lat]].

\bibitem{Dreiner:2008tw}
H.~K.~Dreiner, H.~E.~Haber and S.~P.~Martin,
``Two-component spinor techniques and Feynman rules for quantum field theory and supersymmetry,''
Phys. Rept. \textbf{494}, 1-196 (2010)
doi:10.1016/j.physrep.2010.05.002
[arXiv:0812.1594 [hep-ph]].

\end{thebibliography}
\end{document}